\documentclass{statsoc}
\usepackage{CJKutf8}
\usepackage{amsfonts}
\usepackage{mathrsfs}
\usepackage{amsmath,amssymb}
\usepackage{mathtools}
\DeclarePairedDelimiter{\ceil}{\lceil}{\rceil}

\usepackage{bm}
\usepackage{latexsym}
\usepackage[utf8]{inputenc}
\usepackage[T1]{fontenc}
\usepackage{CJK}
\usepackage{float}
\usepackage[a4paper]{geometry}
\usepackage{graphicx}
\usepackage{subfigure}
\usepackage{multirow}
\usepackage{multicol}
\usepackage{dcolumn}
\usepackage{indentfirst}
\usepackage{fancyhdr}
\usepackage{threeparttable}
\usepackage{booktabs}
\usepackage{tabularx}
\usepackage{epstopdf}
\usepackage{color}
\usepackage{rotating} 
\usepackage{threeparttable}
\usepackage{natbib}
\usepackage{algorithm}
\usepackage{algorithmic}
\usepackage{enumitem}
\usepackage{microtype} 
\makeatletter
    
    \newcommand{\Rmnum}[1]{\expandafter\@slowromancap\romannumeral #1@}
  \makeatother                                                      
\allowdisplaybreaks

\newtheorem{assumption}{Assumption}

\newtheorem{theory}{Theorem} 
\newtheorem{remark}{Remark}

\usepackage[colorlinks,
            linkcolor=blue,
            anchorcolor=blue,
            citecolor=blue]{hyperref}
\usepackage{enumitem}
\usepackage{url}
\usepackage{caption}

\newcommand{\figref}[1]{\figurename~\ref{#1}}
\newcommand{\tabref}[1]{\tablename~\ref{#1}}

\usepackage{tikz}
\usepackage{hyperref}
\usetikzlibrary{shapes.geometric, arrows}
\tikzstyle{node} = [rectangle, rounded corners, minimum width=0.5cm, minimum height=0.5cm,text centered, draw=black, fill=white!30]
\tikzstyle{arrow} = [thick,->,>=stealth]

\usepackage{xr}
\title[Functional L-Optimality Subsampling for Massive Data 
]{Functional L-Optimality Subsampling for Massive Data}
\author{\textbf{Hua Liu}}
\address{Department of Statistics and Management, Shanghai University of Finance and Economics,
Shanghai,
China.}
\email{liuhua\_sufe@163.com}
\author{\textbf{Jinhong You}}
\address{Department of Statistics and Management, Shanghai University of Finance and Economics,
Shanghai,
China.}
\email{johnyou07@163.com}
\author[Liu, You and Cao]{\textbf{Jiguo Cao}}
\address{Department of Statistics and Actuarial Science, Simon Fraser University, BC, Canada.}
\email{jiguo\_cao@sfu.ca}

\begin{document}
\begin{CJK}{UTF8}{gbsn}

\begin{abstract}
Massive data bring the big challenges of memory and computation for analysis. These challenges can be tackled by taking subsamples from the full data as a surrogate. For functional data, it is common to collect multiple measurements over their domains, which require even more memory and computation time when the sample size is large. The computation would be much more intensive when statistical inference is required through bootstrap samples. To the best of our knowledge, this article is the first attempt to study the subsampling method for the functional linear model. We propose an optimal subsampling method based on the functional L-optimality criterion. When the response is a discrete or categorical variable, we further extend our proposed functional L-optimality subsampling (FLoS) method to the functional generalized linear model. We establish the asymptotic properties of the estimators by the FLoS method. 
The finite sample performance of our proposed FLoS method is investigated by extensive simulation studies. The FLoS method is further demonstrated by analyzing two large-scale datasets: the global climate data and the kidney transplant data. The analysis results on these data show that the FLoS method is much better than the uniform subsampling approach and can well approximate the results based on the full data while dramatically reducing the computation time and memory. 

\end{abstract}


\newpage

\section{Introduction}
In the past decade, the volume of data increases exponentially with the development of science and technology, which provides researchers more information. At the same time, despite the rapid development of computational resources, the extraordinary amount of data also brings some challenges to researchers in conducting data analysis. One challenge is that fitting a model using massive data needs 
too much memory to this end. Unfortunately, it
often exceeds the available computational resources.
Moreover, the computing time based on the full data may be too long to obtain the results, which means high-performance computing is usually necessary. However, high-performance computing is often a limitation in practice.
To tackle these challenges, an effective way is to take random subsamples from the massive data as a surrogate.

The existing literature about subsampling mainly focuses on the models with scalar variables. For a linear regression, \cite{ma2015statistical} used the probabilities based on statistical leverage scores to randomly subsample data and established the asymptotic properties of the resultant estimators. A method named information-based optimal subdata selection (IBOSS) proposed by \cite{wang2019information} selects subsample data deterministically without involving random sampling. For a logistic regression, \cite{wang2018optimal} proposed a subsampling method based on the A-optimality criterion. \cite{wang2019more} introduced a Poisson subsampling method  \citep{kiefer1959optimum} based on the subsamples obtained by the optimal subsampling probabilities developed in \cite{wang2018optimal}. \cite{cheng2020information} used the IBOSS method to make subsampling for logistic regression. \cite{ai2018optimal}  investigated the optimal subsampling method under the A-optimality
criterion (OSMAC) for generalized linear models. A Poisson subsampling method based on the A-optimality or L-optimality criterion was used for maximum quasi-likelihood estimation in  \cite{yu2020optimal}. \cite{wang2020optimal}, \cite{fan2021optimal} and \cite{ai2020optimal} used the subsampling method for quantile regressions. \cite{wood2017generalized} developed scalable methods for estimating generalized additive models with the marginal discretization of model predictors to reduce memory footprint. We refer the readers to \cite{yao2020review} for a recent review of optimal subsampling methods of massive data when both of the response and predictors are scalar.

It is worth mentioning that there is almost no work of subsampling in the field of functional data analysis (FDA).
In applications, especially in the clinical, biometrical, epidemiological, social and economic fields, many variables are measured or observed at multiple times or spatial locations. This kind of variables is called a functional variable because these variables can be viewed as functions of time or spatial locations, and the data for these variable are called functional data \citep{Ramsay02, Morris15}. The functional data are usually defined on a space that is intrinsically infinite-dimensional. In some applications, we may have massive functional data. 
For example, the global climate data from NASA (\url{https://ds.nccs.nasa.gov/thredds/catalog/NEX-GDDP/IND/BCSD/catalog.html}) records the temperature and precipitation from January 1st, 1950 to December 30th, 2100 of all $1,036,800$ spatial grids in the globe. The data can be used to analyze the changes in the global climate during the past few decades and study the future trend of climate change. The data size of the
whole global climate data during 1950-2100 is 
beyond the terabyte (TB) regime,
which is too large to be stored in personal computers.
Thus, the statistical analysis based on the full data is difficult.
Another example is about the kidney transplant data from the Organ Procurement Transplant Network/United Network for Organ Sharing (Optn/UNOS, \url{https://optn.transplant.hrsa.gov/} ). This data set collects the information of $478,380$ recipients during the follow-up period after the kidney transplant, which can be used to check whether the transplant is successful. For the above classification problem, we may need to use an iterative optimization procedure to obtain the estimator. Then the computation takes too long time to attain the results when the full data is used. When the sample size of functional data is extremely large, we have to face a more serious challenge of the volume than mentioned above. 

One of the most important FDA tools is the functional linear model, which describes the relationship of some functional predictors and  scalar responses 
\citep{cardot2003spline, hall2007methodology, hilgert2013minimax, jiang2011functional,reiss2017methods,reiss2017penalized,jiang2020,li2020inference}. 
Estimating the functional linear model requires the computational time $O(n(K+p+1)+n(K+p+1)^2)$ when using the penalized B-splines method, where $n$ is the number of functional data, $K$ is the number of knots, and $p$ is the degree of the B-splines \citep{cardot2003spline, claeskens2009asymptotic,xiao2019asymptotic}. Usually, the number of knots, $K$, is chosen to be relatively large to capture the local features of the functional coefficients. We also need to select the optimal smoothing parameter by the Bayesian information criterion (BIC), which may take a long computing time when the number of functional data is excessively large.
 
We propose an idea of subsampling to solve the challenges in the computation with functional predictors. The simplest subsampling method is to draw the sample uniformly at random, which will perform poorly when the leverage scores are non-uniform. Moreover, in order to make the B-spline approximation asymptotically unbiased, a relatively large $K$ is usually chosen. Simultaneously, a roughness penalty is used to ensure the smoothness of the estimator, which 
results in the variance of the subsample estimator being more complicated and not as concise as IBOSS in \cite{wang2019information} and \cite{cheng2020information}.
In addition, IBOSS is based on the order statistics of each scalar predictor variable. The functional predictor variable in the functional linear model is a 
curve and is difficult to be ordered.
As a result, IBOSS is not suitable for the subsampling with functional predictors. 

In this paper, we first estimate the functional coefficient using the subsampling data, and derive the asymptotic distribution of the general subsampling estimator. Then, we obtain the optimal subsampling probabilities by minimizing the asymptotic integrated mean squared errors (IMSE) and propose the functional L-optimality criterion. Lastly, we attain the optimal subsampling estimator based on the optimal subdata drawn according to the optimal probability calculated above. Our proposed method is called the functional L-optimality subsampling (FLoS) method in this article. We extend the FLoS method to the functional generalized linear model which has a discrete or categorical response variable and a functional predictor.
Moreover, we establish the asymptotic results of the FLoS estimators for the functional linear model and functional generalized linear model.
In addition, an R package \texttt{SubsamplingFunPredictors} has been developed for implementing the FLoS method.

To the best of our knowledge, this is the first attempt to introduce the subsampling method to the functional data analysis. The FLoS method has several advantages. 
(1) The computing time for this method is $O(n(K+p+1)+L(K+p+1)^2)$, where $L$ is the subsample size. It is significantly faster than $O(n(K+p+1)+n(K+p+1)^2)$ when using the full data. (2) The integrated mean square errors (IMSEs) of the estimators using the FLoS method are smaller than those using the uniform subsampling method; (3) the distributed parallel computing can be adapted based on the FLoS method. We can calculate the subsampling probabilities on each subset independently. (4) One by-product of the FLoS method is to make statistical inference using multiple subsampling datasets, which has a more obvious advantage in reducing computing time. 

The rest of this article is organized as follows. In Section \ref{sec::full}, we briefly introduce the functional linear regression and give the estimation and asymptotic properties of the estimators based on the full data. Section \ref{sec::sub} derives the optimal subsampling strategy and the optimal subsampling algorithm based on the functional L-optimality criteria for the estimator of the coefficient function.  The asymptotic behaviours for the optimal subsampling estimator are also investigated in this section. In Section \ref{sec::quasi}, we extend the optimal subsampling method to the functional generalized linear model. The evaluation of the numerical performance of our proposed estimator via simulation studies is presented in  Section \ref{sec::sim}. We also illustrate our method by the analysis of two real data sets in Section \ref{sec::app}. Some conclusions and discussions are provided in Section \ref{sec::discuss}. 


\section{Preliminary}
\label{sec::full}
\subsection{Functional Linear Model}
\label{sec::model}

In this paper, we consider a scalar-on-function linear regression model:
\begin{equation}\label{eq1}
  y_i = \alpha+\int_{a}^{b} x_i(t)\beta(t) dt +\varepsilon_i,
\end{equation}
where the functional predictor $x_i(t), i=1,\ldots,n,$
is independent realizations of an unknown process $X(t)$ defined on a domain $[a,b]$,
$\alpha$ is the intercept, $\beta(t)$ is the slope function, $y_i$ is the continuous scalar response, the noise term $\varepsilon_i$ is i.i.d, and $\varepsilon_i$ is independent of $x_i(t)$ with $\text{E}(\varepsilon_i|x_i(t))=0$ and $\text{Var}(\varepsilon_i|x_i(t))=\sigma^2$.

Without loss of generality, the model (\ref{eq1}) can be expressed as a centered model without the intercept:
\begin{equation}\label{nointer1}
    y_i^c = \int_{a}^{b} x^c_i(t)\beta(t) dt +\varepsilon^c_i,
\end{equation}
where $y_i^c=y_i-\overline{y}$, $x_i^c(t)=x_i(t)-\overline{x}(t)$ and $\varepsilon^c_i=\varepsilon_i-\overline{\varepsilon}$ are the centered response, pointwise centered predictor curves and centered noise term, respectively. Once we get an estimate $\widehat{\beta}(t)$, the intercept can be estimated as $\widehat{\alpha} = \overline{y}-\int_{a}^{b} \overline{x}(t)\widehat{\beta}(t) dt$.

To ease the notation, we drop the superscript $c$ in (\ref{nointer1}) from now and focus on the estimation of the functional coefficient $\beta(t)$ in the following model 
 \begin{equation*}\label{nointer}
    y_i = \int_{a}^{b} x_i(t)\beta(t) dt +\varepsilon_i.
\end{equation*}

\subsection{Estimating $\beta(t)$ from Full Data}
\label{sec::est}
We utilize the B-spline basis functions \citep{deBoor} to approximate the functional coefficient $\beta(t)$. For $p\geq 1$, let $\mathcal{S}(p+1;k)=\{s(\cdot)\in \mathcal{C}^{p+1}[a,b]: \text{$s$ is a degree $p$ polynomial on} \\
\text{each}\ [k_j,k_{j+1}]\}$ be the space of polynomial splines of degree $p$, implying that the order equals $p+1$. On the domain $[a,b]$, we define a knot sequence with $K$ interior knots $a =k_0<k_1<\ldots<k_K<k_{K+1}=b$. In addition, define the additional knots: $k_{-p}=k_{-p+1}=\cdots=k_{-1} =k_0$, and 
  $k_{K+1}=k_{K+2}=\cdots=k_{K+p+1}$.
According to the definition of B-spline basis functions, the total number of basis functions with degree $p$ and $K$ interior knots is $K+p+1$. Denote the $p$th degree B-spline basis for $\mathcal{S}(p+1;k)$ as $\bm{N}(t)=(N_{j,p+1}(t):-p\leq j\leq K)^{T}$ \citep{schumaker1981spline}.

We denote by $s_{\beta} (t) = \bm{N}^{T}(t)\bm{c} \in \mathcal{S}(p + 1, κ)$ the best $\mathcal{L}_{\infty}$ approximation to the functional coefficient $\beta(t)$\citep{claeskens2009asymptotic}, where $\bm{N}^{T}(t)$ denotes the transpose of $\bm{N}(t)$. The corresponding smoothing estimator for $\beta(t)$ is defined as $\widehat{\beta}(t)=\bm{N}^{T}(t)\widehat{\bm{c}}$,
where $\widehat{\bm{c}}=(c_{-p},\ldots,c_{K})^{T}$ minimizes the penalized least squares
\begin{equation}\label{eq3}
  L(\bm{c};\lambda,K)=\sum_{i=1}^n \left\{y_i - \int_{a}^b x_i(t)\bm{N}^{T}(t)dt \cdot \bm{c}\right\}^2+\lambda \int_{a}^b \left\{\bigg(\bm{N}^{(q)}(t)\bigg)^{T}\cdot \bm{c}\right\}^2dt,
\end{equation}
with the nonnegative smoothing parameter $\lambda$. In the above criterion, the first term is the ordinary least squares error, and the second term is the roughness penalty that aims to enforce smoothness of $\hat{\beta}(t)$. It is a natural choice to have $q\leq p$. Let $\bm{D}_q=\int_a^b [\bm{N}^{(q)}(t)][\bm{N}^{(q)}(t)]^{T} dt$
, $\bm{y}=(y_1,\ldots,y_n)^{T}$, $\bm{X}(t)=(x_1(t),\ldots,x_n(t))^{T}$, and $\bm{N}=\int_{a}^b\bm{X}(t)\bm{N}^{T}(t)dt=(\bm{N}_1,\ldots,\bm{N}_n)^{T}$, where $N_i=\int_a^b x_i(t)\bm{N}(t)dt$, then the estimators of $\bm{c}$ and $\beta(t)$ are given by
\begin{equation}\label{eq4}
  \widehat{\bm{c}} = (\bm{N}^{T}\bm{N}+\lambda \bm{D}_q)^{-1}\bm{N}^{T}\bm{y} \ \ \mbox{and}\ \ \ 
  \widehat{\beta}(t) = \bm{N}^{T}(t)(\bm{N}^{T}\bm{N}+\lambda \bm{D}_q)^{-1}\bm{N}^{T}\bm{y}.
\end{equation}

\subsection{Asymptotic Results of $\widehat{\beta}(t)$}
\label{sec::asy}
In this section, we introduce the asymptotic results of the estimator $\widehat{\beta}(t)$ based on the full data, which is useful to derive the asymptotic distribution of the estimators using the subsampling method. Before we present some assumptions used in the following theorems, we first define some notations. If $0<m<\infty$, $\mathcal{L}^m$ is defined as the space of functions $f(t)$ over the interval $[a,b]$ such that $\int_{a}^{b}|f(t)|^mdt<\infty$. With this convention, $\mathcal{L}^m$ is treated as a Banach space with the norm $\|f\|_m=(\int_{a}^{b}|f(t)|^mdt)^{1/m}$. When $m = 2$, we obtain the Hilbert space $\mathcal{L}^2$ with the inner product $\langle f,g\rangle=\int_{a}^bf(t)g(t)dt$ and the $\mathcal{L}_2$ norm $\|f\|_2$. And $\mathbb{R}^{m^*}$ is also a Hilbert space for a positive integer $m^*$. We also define $\langle \bm{u},\bm{v}\rangle=\bm{u}^{T}\bm{v}$ and $\|\bm{u}\|_2=(\bm{u}^{T}\bm{u})^{1/2}$ as the inner product and the norm of vector $\bm{u}$ and $\bm{v}$, respectively. For a real argument $a$, $\ceil{a}$ means the least integer greater than or equal to $a$.

\begin{assumption}\label{a1}
Let $\upsilon$ be a nonnegative integer, and $\kappa\in(0,1]$ such that $d = \upsilon+\kappa\geq p+1$. We assume the unknown slope function $\beta(\cdot)\in $  $\mathcal{H}^{(d)}([a,b])$, which is the class of function $f$ on $[a,b]$ whose $\upsilon$th deriative exists and satisfies a Lipschitz condition of order $\kappa$: $|f^{(\upsilon)}(t)-f^{(\upsilon)}(s)|\leq C_{\upsilon}|s-t|^{\kappa}$, for $s,t\in[a,b]$ and some constant $C_{\upsilon}>0$.
\end{assumption}

\begin{assumption}\label{s1}
For the functional predictor $X(t)$, it holds that 
$\texttt{E}(\|X\|_4^4)<\infty$.
In addition, the error $\varepsilon_i$  term satisfies that $\mathrm{E}(\varepsilon_i^4)<\infty$.
\end{assumption}

\begin{assumption}\label{a6}
For the roughness penalty, we assume 
tuning parameter $\lambda$ satisfies that  $\lambda=o(n^{1/2}K^{1/2-2q}$). Besides, we assume $q\leq p$.
\end{assumption}

\begin{assumption}\label{a3}
Let $\delta_j=k_{j+1}-k_j$ and $\delta = \max_{0\leq j\leq K} (k_{j+1}-k_j)$. There exists a constant $M > 0$, such that
\begin{equation}\label{eq5}
  \delta/ \min_{0\leq j\leq K} (k_{j+1}-k_j)\leq M,\ \ \max_{0\leq j\leq K-1}|\delta_{j+1}-\delta_j|= o(K^{-1}).
\end{equation}
In addition, let $\bm{G}_{k,n}=\bm{N}^{T}\bm{N}/n$ and $\bm{H}_{k,n}=(\bm{G}_{k,n}+{\lambda}/{n}\bm{D}_q)$. The smallest eigenvalue of $\bm{G}_{k,n}$ is greater than $c_{G}/K$, where $c_{G}$ is a positive constant. 
\end{assumption}

\begin{assumption}\label{a5}
The number of knots $K=o(\sqrt{n})$ and $K = \omega(n^{1/(2d+1)})$, where $K = \omega(n^{1/(2d+1)})$ means $K/n^{1/(2d+1)}\rightarrow\infty$ as $n\rightarrow\infty$.
\end{assumption}

\begin{remark}
Assumption \ref{a1} is about the smoothness of the slope function, which has been widely used in the literature of nonparametric estimation \citep{liu2013oracally,kim2020generalized,yu2020estimation}. Assumption \ref{s1} gives some moment conditions noise term and functional predictor. 
Combing $\|\bm{D}_q\|_{\infty}=O(K^{2q-1})$ with Assumption \ref{a6}, we can get $\|\lambda\bm{D}_q\|_{\infty}=o(n^{1/2} K^{-1/2})$. Thus, we can get $\|\bm{H}_{k,n}\|_{\infty}=O(1/K)$.
Note that (\ref{eq5}) in Assumption \ref{a3} implies that $\delta \sim K^{-1}$, i.e., $\delta$ and $K^{-1}$ are rate-wise
equivalent. 
The second condition in Assumption \ref{a3} implies that the functional predictor $X(t)$ is away from zero in every small area of the 
domain $[a,b]$, which is reasonable to make the coefficient function $\beta(t)$ estimable in the whole domain $[a,b]$.
\end{remark}

\begin{theory}\label{th1}
For any given $t\in[a,b]$, (i) under Assumptions \ref{a1}-\ref{a3}, we can get
\begin{equation*}
  \begin{split}
     &E\{\widehat{\beta}(t)-\beta(t)|X(t)\} = b_a(t)+b_\lambda(t)+o(K^{-d})+o(\lambda n^{-1}K^{2q}),\\
     &Var\{\widehat{\beta}(t)|X(t)\}=\frac{\sigma^2}{n}\bm{N}^{T}(t)\bm{H}_{k,n}^{-1}\bm{G}_{k,n}\bm{H}_{k,n}^{-1}\bm{N}(t).
  \end{split}
     \end{equation*}
The spline approximation bias is
$ b_a(t) = -\frac{\beta^{d}(t)\delta_j^{d}}{d!}B_{d}(\frac{t-t_j}{\delta_j})=O(K^{-d})$, 
where $B_{d}(\cdot)$ is the $d$th Bernoulli polynomial. The shrinkage bias is defined as $ b_{\lambda}=\lambda/n\bm{N}^{T}(t)\bm{H}_{k,n}^{-1}\bm{D}_q\bm{c}=O(\lambda K^{2q}/n)$.
And, the order of the conditional variance is $Var\{\widehat{\beta}(t)|X(t)\}=O(K/n)$.
(ii) Under Assumptions \ref{a1}-\ref{a5}, as
$n\rightarrow \infty$, we have
\begin{equation*}\label{eq7}
  \sqrt{\mathrm{Var}^{-1}\{\widehat{\beta}(t)|X(t)\}} (\widehat{\beta}(t)-\beta(t))\overset{D}\rightarrow \mathrm{N}(0,1).
\end{equation*}
\end{theory}

\begin{remark}
From Assumption \ref{a6}, we can get $b_{\lambda}=o(\sqrt{K/n})$, so the shrinkage bias is negligible. And to make the approxiamtion bias negligible, the Assumption \ref{a5} ensures that the order of $K$ is $n^{\nu}$, where $\nu\geq 1/(2d+1)$.
\end{remark}




\section{The FLoS Method}
\label{sec::sub}
\subsection{Subsample Estimator}
\label{sec::subest}
Denotes $\mathcal {F}_n= \{(x_i(t),y_i),i=1,\ldots,n;t\in[a,b]\}$ be the full data. Let $\eta_i$ be the indicator variable that signifies whether $(x_i(t),y_i;t\in[a,b])$ is included in the subdata, that is
\begin{equation*}
   \eta_i=\left\{
  \begin{split}
     &1&, &\ \ (x_i(t),y_i;t\in[a,b]) \ \ \text{is included},\\
     &0 &, &\ \ \text{otherwise},
  \end{split}
  \right.
\end{equation*}
and $\eta_i\sim \text{Bernoulli}(p_i)$ with $\sum_{i=1}^np_i=1$. Thus, the subsample estimator, denoted as $\widetilde{\bm{c}}$ is the minimizer of
\begin{equation}\label{eq8}
  L^*(\bm{c};\lambda,K)=\sum_{i=1}^n \frac{R_i}{Lp_i}\left\{y_i - \int_{a}^b x_i(t)\bm{N}^{T}(t)dt \bm{c}\right\}^2+\lambda \int_{a}^b \left\{\bigg(\bm{N}^{(q)}(t)\bigg)^{T}\cdot \bm{c}\right\}^2dt,
\end{equation}
where $R_i=\sum_{l=1}^L\eta_{il}$ denotes the total number of times that $i$-th observation is selected into the sample out of the $L$ sampling steps and $R_i\sim \text{Binomial}(L,p_i)$. We weigh the objective function based on the sampling probabilities $p_i$.

To establish the asymptotic result of the subsample estimator, we need the following Assumption \ref{a7}. As mentioned in \cite{ai2018optimal}, Assumption \ref{a7} restricts the weights in the estimation equation (\ref{eq8}) and ensures the order of the extremely small subsampling probabilities. Besides, this assumption gives the order of the subsampling size $L$.

\begin{assumption}\label{a7}
We assume $\max_{1\le i \le n} (np_i)^{-1}=o_p(\sqrt{L})$ and $L=o(K^2)$.
\end{assumption}

The following theorem presents the asymptotic normality of the subsample estimator.

\begin{theory}\label{th5}
Under Assumptions \ref{a1}-\ref{a7}, for any given $t$, as
$L, n\rightarrow \infty$, we have
\begin{equation*}\label{eq15}
  \left\{\bm{N}^{T}(t)\bm{H}_{k,n}^{-1}\bm{W}_p\bm{H}_{k,n}^{-1}\bm{N}(t)\right\}^{-1/2}
  \sqrt{L}(\widetilde{\bm{\beta}}(t)-\beta(t))\rightarrow N(0,1).
\end{equation*}
where $\widetilde{\beta}(t)=\bm{N}^{T}(t)\widetilde{\bm{c}}$ and
\begin{equation}\label{eq16}
  \bm{W}_p = \frac{1}{n^2}\sum_{i=1}^n\frac{\text{E}\{(y_i-\bm{N}_i^{T}\bm{c})^2\}\bm{N}_i\bm{N}_i^{T}}{p_i}.
\end{equation}
\end{theory}
\begin{remark}
In (\ref{eq16}), the term $\text{E}\{(y_i-\bm{N}_i^{T}\bm{c})^2\}=\sigma^2+(\langle x_i,\beta\rangle-\bm{N}^{T}\bm{c})^2$, where $\sigma^2$, $\beta(t)$ and $\bm{c}$ are all unknown, so the optimal subsampling probabilities is not directly implementable based the asymptotic variance of $\widetilde{\beta}(t)$. To practically implement the optimal subsampling probabilities, we establish the asymptotically normality of $\widetilde{\beta}(t)-\widehat{\beta}(t)$.
\end{remark}

\begin{theory}\label{th2}
Under Assumptions \ref{a1}-\ref{a7}, for any given $t$, as $L, n\rightarrow \infty$, conditional on $\mathcal{F}_n$ in probability, 
\begin{equation*}\label{eq9}
  \left\{\bm{N}^{T}(t)\bm{H}_{k,n}^{-1}\bm{V}_p\bm{H}_{k,n}^{-1}\bm{N}(t)\right\}^{-1/2}
  \sqrt{L}
  (\widetilde{\beta}(t)-\widehat{\beta}(t))\rightarrow N(0,1),
\end{equation*}
in distribution, where 
\begin{equation*}\label{eq10}
  \bm{V}_p = \frac{1}{n^2}\sum_{i=1}^n\frac{\{y_i-\bm{N}_i^{T}\widehat{\bm{c}}\}^2\bm{N}_i\bm{N}_i^{T}}{p_i}.
\end{equation*}
\end{theory}

\subsection{Optimal Subsampling Probabilities}
Theorem \ref{th1} and \ref{th5} show that $\widehat{\beta}(t)$ and $\widetilde{\beta}(t)$
are both asymptotically unbiased under some conditions. We aim to find the optimal subsampling probabilities that minimize the asymptotic integrated mean squared error (IMSE) of $\widetilde{\beta}$
in approximating $\widehat{\beta}$. 
The IMSE is defined as follows,
\begin{equation}\label{eq11}
  \text{IMSE}(\widetilde{\beta}-\widehat{\beta}) = \int_a^b \frac{ \bm{N}^{T}(t)\bm{H}_{k,n}^{-1}\bm{V}_p\bm{H}_{k,n}^{-1}\bm{N}(t)}{L}dt.
\end{equation}
In (\ref{eq11}), $L^{-1}\bm{H}_{k,n}^{-1}\bm{V}_p\bm{H}_{k,n}^{-1}$ is the asymptotic covariance matrix of $\widetilde{\bm{c}}-\widehat{\bm{c}}$, where $\bm{H}_{k,n}$ depends on the chosen smoothing parameter $\lambda$. In addition, from (\ref{eq11}), we can see that only $\bm{V}_p$ depends on the sampling probability $p_i$ and the integral $\int_a^b \bm{N}^{T}(t)\bm{H}_{k,n}^{-1}\bm{V}_p\bm{H}_{k,n}^{-1}\bm{N}(t)dt\leq\int_a^b \bm{N}^{T}(t)\bm{H}_{k,n}^{-1}\bm{V}_{p^*}\bm{H}_{k,n}^{-1}\bm{N}(t)dt$ if
$\bm{V}_p\leq\bm{V}_{p^*}$. We propose to obtain the optimal subsampling probability by minimizing $\bm{V}_p$. Several criteria exist for minimizing the matrix. Here we choose to minimize the trace of the matrix $\bm{V}_p$. 
Note that $L^{-1}\bm{V}_p$ is the asymptotic covariance matrix of $\bm{H}_{k,n}^{-1}(\widetilde{\bm{c}}-\widehat{\bm{c}})$, where $\bm{H}_{k,n}^{-1}(\widetilde{\bm{c}}-\widehat{\bm{c}})$ is a linear transformation of the estimator $\widetilde{\bm{c}}-\widehat{\bm{c}}$. Thus,
minimizing $\mathrm{tr}(\bm{V}_p)$ to obtain the optimal subsampling probability is termed the functional L-optimality criterion, which is the functional version of the L-optimality defined in \cite{pukelsheim2006optimal} and \cite{atkinson2007optimum}.

\begin{theory}\label{th3}
If the subsampling probabilities $p_i,i=1,\ldots,n$, are chosen as
\begin{equation}\label{eq12}
  p_i^{\mathrm{FLoS}} = \frac{|y_i-\bm{N}_i^{T}\widehat{\bm{c}}|\|\bm{N}_i\|_{2}}{\sum_{i=1}^n|y_i-\bm{N}_i^{T}\widehat{\bm{c}}|\|\bm{N}_i\|_{2}},
\end{equation}
then $\mathrm{tr}(\bm{V}_p)$ attains its minimum, where the superscript ``FLoS'' indicates that this probability is calculated based on the functional L-optimality criterion.
\end{theory}
\begin{remark}
In (\ref{eq12}), $p_i^{\mathrm{FLoS}}$ not only directly depends on predictors but also on the residual
. For the predictors, the term $\|\bm{N}_i\|_2=\|\int_{a}^bx_i(t)\bm{N}(t)dt\|_2$ describes the structure information of the functional predictors, which is similar to statistical leverage score in linear model. The term $|y_i-\bm{N}_i^{T}\widehat{\bm{c}}|$ represents the effect of the residual.
It will 
more likely 
select samples with larger values of $|y_i-\bm{N}_i^{T}\widehat{\bm{c}}|$  to improve the robustness of the subsample estimator.
\end{remark}

Note that the calculation of $\widehat{\bm{c}}$ in (\ref{eq12}) uses full data and takes $O(n(K+p+1)^2)$. Therefore, we need to replace $\widehat{\bm{c}}$ by a pilot estimator, say $\widehat{\bm{c}}_0$, which can be obtained by a uniform subsample with the sample size $L$
. In addition, 
we need to choose the smoothing parameter $\lambda$, the degree $p$ of the B-spline basis, and the number of knots $K$. In penalized spline method, the choice of $K$ is not crucial \citep{cardot2003spline}, as the roughness of the estimator
is controlled by a roughness penalty, rather than the number
of knots. 
Usually, in practice, we choose $p=3$ and $K$ is chosen to be relatively
large so that
local features of $\beta(t)$ can be captured. 
Once $K$ and $p$ are fixed, 
we can select the smoothing parameter $\lambda$ by minimizing the Bayesian information criterion (BIC):
\begin{equation*}
  \text{BIC}(\lambda) = n\text{log}(\|\bm{y}-\bm{N}\widehat{\bm{c}}(\lambda)\|_2^2/n)+\text{log}(n)\mathrm{df}(\lambda),
\end{equation*}
where $\mathrm{df}(\lambda)=\mathrm{tr}(\bm{N}(\bm{N}^{T}\bm{N}+\lambda\bm{D}_q)^{-1}\bm{N}^{T})$.
Using full data to select the optimal $\lambda$ is computationally expensive. Therefore, we need to select the tuning parameter by BIC using the optimal subsample data. Algorithm \ref{al2} describes the subsampling procedure for estimating the functional linear model in details.
\begin{algorithm}[htp]
\caption{The FLoS Algorithm for Estimating the Functional Linear Model}\label{al2}
\begin{algorithmic}
\STATE
\begin{itemize}
\item \textbf{Step 1:} 
Calculate $\bm{N}_i=\int_{a}^b x_i(t)\bm{N}(t)dt$ and the new data is $(\bm{N}_i,y_i;i=1,\ldots,n)$.
  \item \textbf{Step 2:}
  Draw a subsample of size $L$ 
  using the uniform sampling probabilities $p_i^0=1/n$, and use the subsample data to obtain the pilot estimator $\widehat{\bm{c}}^0$ with $\lambda=0$.
  
  \item \textbf{Step 3:} Using $\widehat{\bm{c}}^0$, we can get the approxiamte optimal subsampling probabilities $p_i^{\mathrm{FLoS},\widehat{\bm{c}}^0}$：
      \begin{equation*}
  p_i^{\mathrm{FLoS},\widehat{\bm{c}}^0} = \frac{|y_i-\bm{N}_i^{T}\widehat{\bm{c}}^0|\|\bm{N}_i\|_{2}}{\sum_{i=1}^n|y_i-\bm{N}_i^{T}\widehat{\bm{c}}^0|\|\bm{N}_i\|_{2}}.
\end{equation*}
Using the subsampling probabilities $p_i^{\mathrm{FLoS},\widehat{\bm{c}}^0}$ to draw a random subsample with replacement of size $L$. Denote the subsample as $(\bm{N}^*_i,y_i^*;t\in[a,b])$, with associated subsampling probabilities $p_i^{*\mathrm{FLoS},\widehat{\bm{c}}^0}$.
  \item \textbf{Step 4:}
Given $\lambda$, we can obtain the estimate $\breve{\bm{c}}_{\mathrm{FLoS}}(\lambda)$ through minimizing
      \begin{equation*}\label{eq17}
  L_{\mathrm{FLoS}}^*(\bm{c};\lambda,K)=\sum_{i=1}^L \frac{1}{Lp_i^{*\mathrm{FLoS},\widehat{\bm{c}}^0}}(y_i^* - \bm{N}^{*T}_i \bm{c})^2+\lambda \int_{a}^b \left\{\bigg(\bm{N}^{(q)}(t)\bigg)^{T}\cdot\bm{c})\right\}^2dt.
\end{equation*}
BIC can be approximated by
\begin{equation*}
 \text{BIC}_{\mathrm{FLoS}}(\lambda) = L\text{log}(\|\bm{y}^*-\bm{N}^*\widetilde{\bm{c}}_{\mathrm{FLoS}}(\lambda)\|_2^2/L)+\text{log}(L)\mathrm{df}(\lambda).
\end{equation*}
The optimal $\lambda$ is selected to minimize $\text{BIC}_{\mathrm{FLoS}}(\lambda)$. Once we get the optimal $\lambda$, we can get the estimator $\breve{\beta}_{\mathrm{FLoS}}(t)=\bm{N}^{T}(t)\breve{\bm{c}}_{\mathrm{FLoS}}$.
\end{itemize}
\end{algorithmic}
\end{algorithm}

Recall in Assumption \ref{a5}, the number of knots is required to satisfy that $K=o(\sqrt{n})$ and  $K/n^{1/(2d+1)}\rightarrow\infty$ as $n\rightarrow\infty$. 
Suppose we let the order of $K$ be $n^{1/(2d)}$ in practice.
In Algorithm \ref{al2}, the used subsample size is $L<<n$ and the computing time of the algorithm is
$O(nK+LK^2)$.
And, if the full data size $n$ is very large, the time complexity 
$O(nK)$
of this subsampling algorithm is much smaller than the computing time 
$O(nK^2)$
based on the full data.
Thus, the Algorithm \ref{al2} can reduce computing time dramatically.
Algorithm \ref{al2} is also naturally suited for distributed storage and parallel computing. We can divide the full data into several subsets, simultaneously compute the $\bm{N}_i$ and optimal subsample probabilities $p_i^{\mathrm{FLoS},\widehat{\bm{c}}^0}$ on each subset. Combining the optimal subsample probabilities of each subset, we can get the indices of a random subsample in the full data and use these indices to extract the corresponding data on each subset.

\begin{remark}
Our proposed method can be extended to the following functional linear model with multiple functional predictors:
\begin{equation*}\label{eqmul}
y_i = \sum_{m=1}^M\int_a^b x_{im}(t)\beta_m(t) dt +\varepsilon_i=\int_a^b \bm{x}_i^{T}(t)\bm{\beta}(t)dt +\varepsilon_i,
\end{equation*}
where $y_i$ is the scalar response , $\bm{x}_{i}(t)=(x_{i1}(t),\cdots,x_{iM}(t))^{T}$ is a functional predictor vector defined on domain $[a,b]$, $\bm{\beta}(t)=(\beta_1(t),\cdots,\beta_M(t))^{T}$ and $\varepsilon_i$ is the noise. And, the smoothing estimator for $\beta_m(t)$ is defined as
\begin{equation*}\label{meq2}
\widehat{\beta}_m(t)=\bm{N}^{T}(t)\widehat{\bm{c}}_m,
\end{equation*}
where $\widehat{\bm{c}}=(\bm{c}_1^{T},\cdots,\bm{c}_M^{T})^{T}$ minimizes the penalized least squares
\begin{equation*}\label{eqmul2}
L(\bm{c};\lambda,K)=\sum_{i=1}^n (y_i - \sum_{m=1}^M\int_{a}^b x_{im}(t)\bm{N}^{T}(t)dt \bm{c}_m)^2+\sum_{m=1}^M\lambda_m \int_{a}^b \left\{\bigg(\bm{N}^{(q)}(t)\bigg)^{T}\bm{c}_m)\right\}^2dt.
\end{equation*}
Let $\bm{X}_m(t)=(x_{1m}(t),\cdots,x_{nm}(t))^{T}$, for each predictor $x_m(t), m=1,\ldots, M$, we compute a matrix $\bm{N}_m= \int_{a}^b \bm{X}_m(t)\bm{N}^{T}(t)dt$. Denote $\bm{N}=(\bm{N}_1,\cdots,\bm{N}_M)$ be the column catenation of $\bm{N}_1,\cdots,\bm{N}_M$ and corresponding set $\bm{D} = \text{diag}(\bm{D}_q,\cdots,\bm{D}_q)$, where $\bm{D}$ is the matrix with M blocks $\bm{D}_q$ in its main diagonal and zeros elsewhere. After replacing $\bm{N}$ and $\bm{D}_q$ by new defined $\bm{N}$ and $\bm{D}$, respectively, the estimations and algorithms described in Section \ref{sec::full} and Section \ref{sec::sub} can be carried out to estimate $\beta_1(t),\ldots,\beta_M(t)$ simultaneously. It is worth mentioned that we can simultaneously compute all matrix $\bm{N}_1,\ldots,\bm{N}_M$.
\end{remark}

\subsection{Asymptotic Results of $\breve{\beta}_{\mathrm{FLoS}}(t)$}
Next theorem shows the asymptotic property of the estimator $\breve{\beta}_{\mathrm{FLoS}}(t)$ obtained from Algorithm \ref{al2}.
\begin{theory}\label{th6}
	If Assumptions \ref{a1}-\ref{a7} hold, for any given $t$, as 
	$L\rightarrow \infty$ and $n\rightarrow \infty$, conditionally on $\mathcal{F}_n$ in probability, 
	\begin{equation*}\label{eql15}
	\left\{\bm{N}^{T}(t)\bm{H}_{k,n}^{-1}\bm{V}_{\mathrm{FLoS}}\bm{H}_{k,n}^{-1}\bm{N}(t)\right\}^{-1/2}
	\sqrt{L}(\breve{\beta}_{\mathrm{FLoS}}(t)-\widehat{\beta}(t))\rightarrow N(0,1),
	\end{equation*}
	in distribution, where $\bm{V}_{\mathrm{FLoS}}$ has the minimum trace, and it has the explicit expression
	\begin{equation*}\label{eq18}
	\bm{V}_{\mathrm{FLoS}} = \frac{1}{n}\sum_{i=1}^n\frac{|y_i-\bm{N}_i^{T}\widehat{\bm{c}}|\bm{N}_i\bm{N}_i^{T}}{\|\bm{N}_1\|_2}\times \frac{1}{n}\sum_{i=1}^n|y_i-\bm{N}_i^{T}\widehat{\bm{c}}|\|\bm{N}_i\|_2.
	\end{equation*}
\end{theory}

\section{Extension to Functional Generalized Linear Models}
\label{sec::quasi}
In most applications with a discrete response, the functional linear model may not be appropriate to fit the data. To describe the relationship between the functional predictors and the scalar response from an exponential family distribution (e.g.  the Binomial distribution and Poisson distribution), we consider a functional generalized linear model, namely, FGLM. FGLM was first proposed by 
\cite{james2002generalized}. \cite{muller2005generalized} approximated the functional predictor with a truncated Karhunen-Lo{\`e}ve expansion and gets the estimators through maximizing a functional quasi-likelihood. \cite{ cardot2005estimation}, \cite{yao2005functional}, \cite{crainiceanu2009generalized},
\cite{li2010generalized},
\cite{mclean2014functional} and \cite{li2020inference} also studied the FGLM and extended the FGLM to semi-parametric FGLM.
The basic FGLM can be expressed as:
\begin{equation*}\label{fglm}
E(Y|X)= \psi\left(\alpha+\int_{a}^b Z(t)\beta(t)dt\right),
\end{equation*}
where $\alpha$ is the intercept, $\psi(\cdot)$ is a twice continuously
differentiable function and the function $\psi^{-1}(\cdot)$ is called the link function. For example, in the case of logistic functional regression, $\psi(\cdot) = exp(\cdot)/(1+exp(\cdot))$.

\subsection{Full Data Estimation}
\label{subsec::model}
The intercept $\alpha$ can be represented by the constant basis function $1(t)$, the value of the constant basis $1(t)$ is one everywhere, as follows: $\alpha = 1(t)\alpha=\int_a^b1(t)/(b-a)dt \alpha$.
Denote 
$\bm{Z}^*(t) = (1(t)/(b-a),Z(t))^{T}$
and $\bm{\beta}(t)=(\alpha,\beta(t))^{T}$, then, 
$\alpha+\int_{a}^b Z(t)\beta(t)=\int_{a}^b \bm{Z}^{*T} (t)\bm{\beta}(t)dt$.
Suppose the data $(y_i,\bm{z}_i^*(t)),i=1,\ldots,n$ are i.i.d. copies of $(Y,\bm{Z}^*(t))$. 
In this section, 
we rewrite $\bm{N}$ and $\bm{N}_i$ as $\bm{N} = \int_a^b \bm{X}(t)\bm{N}^{*T}(t)dt$ and $\bm{N}_i = \int_a^b \bm{N}^*(t)\bm{z}_i^*(t)dt$, where $\bm{X}(t) = (\bm{z}_1^*(t),\cdots,\bm{z}_n^*(t))^{T}$ and 
\begin{equation*}
    \bm{N}^*(t) = \begin{pmatrix}
1(t)&0\\
\bm{0}_{K+p+1}&\bm{N}(t)
\end{pmatrix},
\end{equation*}
with $\bm{0}_{K+p+1}$ be a $(K+p+1)\times 1$ vector with $0$s.

Combining the maximum quasi-likelihood estimator in the generalized linear model \citep{chen1999strong,muller2005generalized} and the penalized B-splines, we can obtain the penalized quasi-likelihood estimator $\widehat{\bm{\beta}}_{\mathrm{PQL}}(t)=\bm{N}^{T}(t)\widehat{\bm{c}}_{\mathrm{PQL}}$, where $\widehat{\bm{c}}_{\mathrm{PQL}}$ can be inferred by solving the following equation:
\begin{equation}\label{qeq1}
Q_{\mathrm{PQL}}(\bm{c})=\sum_{i=1}^n\{y_i-\psi(\bm{N}_i^{T}\bm{c})\}\bm{N}_i-\lambda\bm{D}_q\bm{c}=0,
\end{equation}
and $\bm{D}_q$ is rewritten as $\bm{D}_q= \begin{pmatrix}
0 & \bm{0}_{q}^{T}\\
\bm{0}_{q} & \bm{D}_q
\end{pmatrix}$.

\begin{assumption}\label{a8}
Let $\dot{Q}_{\mathrm{PQL}}(\gamma,y)$ be the first order derivative of $Q_{\mathrm{PQL}}(\gamma,y)$ with respect to $\gamma$. The function  $\dot{Q}_{\mathrm{PQL}}(\gamma,y)<0$ for $\eta\in\mathbb{R}$ and $y$ in the range of the response variable. The functions $\psi(\cdot)$, and the first order
derivative of $\psi(\cdot)$ are continuous. There exist positive constants $c_{Q}$ and $C_{Q}$ such that $c_{Q}\leq \dot{Q}_{\mathrm{PQL}}(\gamma,y)\leq C_{Q}$. In addition, the assumptions of $Z(t)$ is same as those of $X(t)$ in Section \ref{sec::asy}.
And 
for each $\bm{z}$, $\mathrm{Var}(Y|\bm{Z}^* = \bm{z})$ and $\psi^{-1}(\int_{a}^b \bm{z}^{T}(t)\bm{\beta}(t)dt)$ are 
nonzero.
\end{assumption}
The above assumption is a common assumptions used under the quasi likelihood frame work \citep{carroll1997generalized,wang2011estimation,liu2013oracally,wang2018efficient,kim2020generalized,yu2020estimation}. And $\dot{Q}_{\mathrm{PQL}}(\eta,y)<0$ ensures the uniqueness of the solution (\ref{qeq1}). 

Denote $\bm{\Psi} = \text{Diag}(\dot{\psi}(\bm{N}^{T}_1\bm{c}),\cdots,\dot{\psi}(\bm{N}^{T}_n\bm{c}))$, $\bm{G}_{k,n}^{\psi}=\frac{1}{n}\bm{N}^{T}\bm{\Psi}\bm{N}$ and $\bm{H}_{k,n}^{\psi} = \bm{G}_{k,n}^{\psi}+\lambda\bm{D}_q$, where $\dot{\psi}(\cdot)$ is the first order deriative of $\psi(\cdot)$. The asymptotic property of $\widehat{\beta}_{\mathrm{PQL}}(t)$ is given in the next theorem.
\begin{theory}\label{th7}
	Under Assumptions \ref{a1}-\ref{a5} and \ref{a8}, for any given $t$, as $n\rightarrow\infty$, we have
	\begin{equation*}\label{qeq2}
	\sqrt{\mathrm{Cov}^{-1}\{\widehat{\bm{\beta}}_{\mathrm{PQL}}(t)|\bm{X}(t)\}} (\widehat{\bm{\beta}}_{\mathrm{PQL}}(t)-\bm{\beta}(t))\overset{D}\rightarrow \mathbb{N}(\bm{0}_2,\bm{I}_2),
	\end{equation*}
	where 
	\begin{equation*}
	\mathrm{Cov}\{\widehat{\bm{\beta}}_{\mathrm{PQL}}(t)|\bm{X}(t)\}=\frac{\sigma^2}{n}\bm{N}^{T}(t)\bm{H}_{k,n}^{\psi,-1}\bm{G}_{k,n}\bm{H}_{k,n}^{\psi-1}\bm{N}(t).
	\end{equation*}
\end{theory}

\subsection{Subsampling  Based Estimation}
The subsample penalized quasi-likelihood estimator, denoted as $\widetilde{\beta}_{\mathrm{PQL}}(t)$ is given by $\widetilde{\bm{\beta}}_{\mathrm{PQL}}(t) = \bm{N}^{T}(t)\widetilde{\bm{c}}_{\mathrm{PQL}}$,
where $\widetilde{\bm{c}}_{\mathrm{PQL}}$ can be obtained through the equation
\begin{equation*}\label{qeq4}
Q^*_{\mathrm{PQL}}(\bm{c}) := \sum_{i=1}^{n}\frac{R_i}{Lp_i}\{y_i-\psi(\bm{N}_i^{T}\bm{c})\}\bm{N}_i-\lambda\bm{D}_q\bm{c}=0.
\end{equation*}

\begin{theory}\label{th8}
	Under Assumptions \ref{a1}-\ref{a8}, for any given $t$, as 
	$L,n\rightarrow\infty$, we have
	\begin{equation*}\label{qeq5}
	\left\{\bm{N}^{T}(t)\bm{H}_{k,n}^{\psi,-1}\bm{W}_p^{\psi}\bm{H}_{k,n}^{\psi,-1}\bm{N}(t)\right\}^{-1/2}
	\sqrt{L}(\widetilde{\bm{\beta}}_{\mathrm{PQL}}(t)-\bm{\beta}(t))\rightarrow \mathbb{N}(\bm{0}_2,\bm{I}_2).
	\end{equation*}
	where 
	\begin{equation}\label{qeq6}
	\bm{W}_p^{\psi} = \frac{1}{n^2}\sum_{i=1}^n\frac{\text{E}\left\{\left(y_i-\psi(\bm{N}_i^{T}\bm{c})\right)^2\right\}\bm{N}_i\bm{N}_i^{T}}{p_i}.
	\end{equation}
\end{theory}
\begin{remark}
	In (\ref{qeq6}), the term $\text{E}\left\{\left(y_i-\psi(\bm{N}_i^{T}\bm{c})\right)^2\right\}$ is unknown, so the optimal subsampling probabilities is not directly implementable based the asymptotic variance of $\widetilde{\bm{\beta}}_{\mathrm{PQL}}(t)$. Similar with Section \ref{sec::subest}, we establish the asymptotically normality of estimator 
	$\widetilde{\bm{\beta}}_{\mathrm{PQL}}(t)$ in approximating the full data estimator $\widehat{\bm{\beta}}_{\mathrm{PQL}}(t)$ to obtain the optimal subsampling probabilities.
\end{remark}

\begin{theory}\label{th9}
	Under Assumptions \ref{a1}-\ref{a8}, for any given $t$, as 
	$L,n\rightarrow\infty$, conditionally on $\mathcal{F}_n$ in probability, 
	\begin{equation*}\label{qeq7}
	\left\{\bm{N}^{T}(t)\bm{H}_{k,n}^{\psi,-1}\bm{V}_p^{\psi}\bm{H}_{k,n}^{\psi,-1}\bm{N}(t)\right\}^{-1/2}
	\sqrt{L}
	(\widetilde{\bm{\beta}}_{\mathrm{PQL}}(t)-\widehat{\bm{\beta}}_{\mathrm{PQL}}(t))\rightarrow \mathbb{N}(\bm{0}_2,\bm{I}_2),
	\end{equation*}
	where 
	\begin{equation*}\label{qeq8}
	\bm{V}_p^{\psi} = \frac{1}{n^2}\sum_{i=1}^n\frac{\{y_i-\psi(\bm{N}_i^{T}\widehat{\bm{c}}_{\mathrm{PQL}})\}^2\bm{N}_i\bm{N}_i^{T}}{p_i}.
	\end{equation*}
\end{theory}
\subsection{Optimal Subsampling Probabilities}
   Under some conditions, $\widehat{\bm{\beta}}_{\mathrm{PQL}}(t)$ and $\widetilde{\bm{\beta}}_{\mathrm{PQL}}(t)$
	are both asymptotically unbiased. 
	We want to find the optimal subsampling probabilities that minimizing IMSE of $\widetilde{\bm{\beta}}_{\mathrm{PQL}}$
	in approximating $\widehat{\bm{\beta}}_{\mathrm{PQL}}$,
    where 
	the IMSE is defined as follows,
	\begin{equation}\label{qeq12}
	\text{IMSE}(\widetilde{\bm{\beta}}_{\mathrm{PQL}}-\widehat{\bm{\beta}}_{\mathrm{PQL}}) = \int_a^b \frac{ \bm{N}^{T}(t)\bm{H}_{k,n}^{\psi,-1}\bm{V}_p^{\psi}\bm{H}_{k,n}^{\psi,-1}\bm{N}(t)}{L}.
	\end{equation}
	From (\ref{qeq12}), it is clear that only $\bm{V}_p^{\psi}$ depends on $p_i$'s, therefore, similar to the subsampling method in the functional linear, we use the functional L-optimality criterion that is 
	 minimizing the $\mathrm{tr}(\bm{V}_p^{\psi})$ to get the optimal subsampling probabilities.

\begin{theory}\label{th10}
	If the subsampling probabilities $p_i,i=1,\ldots,n$, are chosen as
	\begin{equation}\label{qeq13}
	p_{\mathrm{PQL},i}^{\mathrm{FLoS}} = \frac{|y_i-\psi(\bm{N}_i^{T}\widehat{\bm{c}}_{\mathrm{PQL}})|\|\bm{N}_i\|_{2}}{\sum_{i=1}^n|y_i-\psi(\bm{N}_i^{T}\widehat{\bm{c}}_{\mathrm{PQL}})|\|\bm{N}_i\|_{2}},
	\end{equation}
	then $\mathrm{tr}(\bm{V}_p^{\psi})$ attains its minimum.
\end{theory}
\begin{remark}
Analogous to the optimal subsampling probabilities (\ref{eq12}) for the functional linear model, the subsampling probabilities (\ref{qeq13}) are related with the predictors and response. Suppose the response $y_i\in\{0,1\}, i = 1,\ldots,n$, we study the effect of the response on the subsampling probabilities. For these individuals with response $y_i = 1$, a smaller estimated probability $\psi(\bm{N}_i^{T}\widehat{\bm{c}}_{\mathrm{PQL}})$ using full data results in a larger subsampling probability  $p_{\mathrm{PQL},i}^{\mathrm{FLoS}}$. On the contrary, for these samples with $y_i=0$, the subsampling probability $p_{\mathrm{PQL},i}^{\mathrm{FLoS}}$ increases 
as the estimated probability $\psi(\bm{N}_i^{T}\widehat{\bm{c}}_{\mathrm{PQL}})$ increases. In summary, this subsampling method is more likely to select those samples that are more easily misclassified, which means this method improves the robustness of the subsample estimator.
\end{remark}

Akin to the subsampling steps in the functional linear, we give the practical subsampling procedure for FGLM in Algorithm \ref{al3}. 
\begin{algorithm}[htp]
	\caption{FLoS Algorithm for Estimating the Functional Generalized Linear Model}\label{al3}
	\begin{algorithmic}
		\STATE
		\begin{itemize}
			\item \textbf{Step 1:} 
			Calculate $\bm{N}_i=(1,\int_{a}^b z_i(t)\bm{N}^{T}(t)dt)^{T}$ and the new data is $(\bm{N}_i,y_i;i=1,\ldots,n)$.
            \item \textbf{Step 2:} 
			Draw a subsample of size $L$
			using the uniform sampling probabilities $p_i^0=1/n$, and use it to obtain the pilot estimator $\widehat{\bm{c}}^0_{\mathrm{PQL}}$ with $\lambda=0$.
			\item \textbf{Step 3:} Using $\widehat{\bm{c}}^0_{\mathrm{PQL}}$, we can get the approximate optimal subsampling probabilities $p_{\mathrm{PQL},i}^{\mathrm{FLoS},\widehat{\bm{c}}^0}$：
			\begin{equation*}
			p_{\mathrm{PQL},i}^{\mathrm{FLoS},\widehat{\bm{c}}^0_{\mathrm{PQL}}} = \frac{|y_i-\psi(\bm{N}_i^{T}\widehat{\bm{c}}^0_{\mathrm{PQL}})|\|\bm{N}_i\|_{2}}{\sum_{i=1}^n|y_i-\psi(\bm{N}_i^{T}\widehat{\bm{c}}^0_{\mathrm{PQL}})|\|\bm{N}_i\|_{2}},
			\end{equation*}
			Using the subsampling probabilities $p_{\mathrm{PQL},i}^{\mathrm{FLoS},\widehat{\bm{c}}^0_{\mathrm{PQL}}}$ to draw a random subsample with replacement of size $L$. Denote the subsample as $(\bm{N}^*_i,y_i^*)$, with associated subsampling probabilities $p_{\mathrm{PQL},i}^{*\mathrm{FLoS},\widehat{\bm{c}}^0_{\mathrm{PQL}}}$.
			\item \textbf{Step 4:}
			Given $\lambda$, we can obtain the estimate $\breve{\bm{c}}_{\mathrm{FLoS}}^{\mathrm{PQL}}(\lambda)$ through solving
			\begin{equation}\label{eqll17}
			Q_{\mathrm{PQL}}^{*\mathrm{FLoS}}(\bm{c})=\sum_{i=1}^L \frac{1}{Lp_{\mathrm{PQL},i}^{*\mathrm{FLoS},\widehat{\bm{c}}^0_{\mathrm{PQL}}}}(y_i^* - \psi(\bm{N}^{*T}_i \bm{c}))\bm{N}_i-\lambda \bm{D}_q=0,
			\end{equation}
			and based on the optimal subsample data, we can use BIC to choose the optimal tuning parameter $\lambda$. Once we obtain the optimal $\lambda$, we can get the estimator $\breve{\bm{\beta}}_{\mathrm{PQL}}^{\mathrm{FLoS}}(t)=\bm{N}^{T}(t)\breve{\bm{c}}_{\mathrm{FLoS}}^{\mathrm{PQL}}$.
		\end{itemize}
	\end{algorithmic}
\end{algorithm}

As in functional linear model, we suppose the order of $K$ used in the estimation of functional generalized linear model is $n^{1/(2d)}$. In Algorithm \ref{al3}, we need to use an iterative procedure, such as Newton's method, to get the pilot estimator and solve (\ref{eqll17}).  In step 1 \& 3, it takes 
$O(nK)$ computing time to calculate the matrix $\bm{N}$ and the subsampling probabilities. To get the pilot estimator $\widehat{\bm{c}}_{\mathrm{PQL}}^0$ in step 2, the computing time is 
$O(\xi_0 LnK^2)$ 
where $\xi_0$ is the number of iterations. In step 3, for each iteration, it takes 
$O(LnK^2)$ 
computing time and the whole procedure requires 
$O(\xi LnK^2)$
with the number of iterations $\xi$. Thus, when the full data size $n$ is very large, total computing time 
$O(nK+\xi_0 LnK^2+\xi LnK^2)\approx O(nK)$ 
is smaller than the total computing time based on full data 
$O(nK+\xi_{full}nK^2)\approx O(nK^2)$.

The asymptotic result of the estimator obtained from Algorithm \ref{al3} is presented as follows.
\begin{theory}\label{th11}
	Under Assumptions \ref{a1}-\ref{a8}, for any given $t$, as 
	$L\rightarrow\infty$ and $n\rightarrow\infty$, conditionally on $\mathcal{F}_n$ in probability, 
	\begin{equation*}\label{eqll15}
	\left\{\bm{N}^{T}(t)(\bm{H}_{k,n}^{\psi})^{-1}\bm{V}_{\mathrm{FLoS}}^{\psi}(\bm{H}_{k,n}^{\psi})^{-1}\bm{N}(t)\right\}^{-1/2}
	\sqrt{L}(\breve{\bm{\beta}}_{\mathrm{PQL}}^{\mathrm{FLoS}}(t)-\widehat{\bm{\beta}}_{\mathrm{PQL}}(t))\rightarrow \mathbb{N}(\bm{0}_2,\bm{I}_2),
	\end{equation*}
	in distribution, where $\bm{V}^{\psi}_{\mathrm{FLoS}}$ has the minimum trace, and it has the explicit expression
	\begin{equation*}\label{eql18}
	\bm{V}_{\mathrm{FLoS}} ^{\psi}= \frac{1}{n}\sum_{i=1}^n\frac{|y_i-\psi(\bm{N}_i^{T}\widehat{\bm{c}}_{\mathrm{PQL}})|\bm{N}_i\bm{N}_i^{T}}{\|\bm{N}_1\|_2}\times \frac{1}{n}\sum_{i=1}^n|y_i-\psi(\bm{N}_i^{T}\widehat{\bm{c}}_{\mathrm{PQL}})|\|\bm{N}_i\|_2.
	\end{equation*}
\end{theory}


\section{Simulation Studies}
\label{sec::sim}
In this section, we conduct three simulation studies to evaluate the finite sample performance of the functional L-optimality subsampling approach in comparison with the uniform subsampling method and the estimation from the full data.
\subsection{Simulation I}
\label{subsec::sim1}
To investigate the numerical performance of the functional L-optimality subsampling (FLoS) approach described in Algorithm \ref{al2},  we consider synthetic data of 500 runs generated from the following functional linear model
\begin{equation*}
  y_i = \int_0^1x_i(t)\bm{\beta}(t)dt + \varepsilon_i, \qquad i=1,\ldots,n,
\end{equation*}
where $\varepsilon_i\overset{iid} \sim N(0,\sigma^2)$,
$\beta(t)= \text{exp}(-32(t-0.5)^2)+2t-1$, and $\sigma^2=0.1$. The functional predictor $x_i(t)$ is generated by $x_i(t) =\sum a_{ij}B_j(t)$, where $B_j(t)$
are cubic B-spline basis functions defined on 
$[0,1]$. We consider the following three different scenarios to generate the basis coefficients $a_{ij}$:
\begin{itemize}
  \item \textbf{Scenario I.} The coefficient $a_{ij}$ are i.i.d from the standard normal distribution, namely, $a_{ij}\overset{iid}\sim N(0,1)$.
  \item \textbf{Scenario II.} We generate the basis coefficient $a_{ij}$ from the $t$ distribution with 3 degree of freedom and 
  zero mean, namely, 
  $a_{ij}\overset{iid}\sim t_3(0)$.
  \item \textbf{Scenario III.} We generate the coefficient $a_{ij}$ from the $t$ distribution with 2 degree of freedom and zero mean
  , namely, $a_{ij}\overset{iid}\sim t_2(0)$.
\end{itemize}
\begin{figure}[htbp]
  \centering
  \subfigure[Scenario I]{\includegraphics[width=4.8cm]{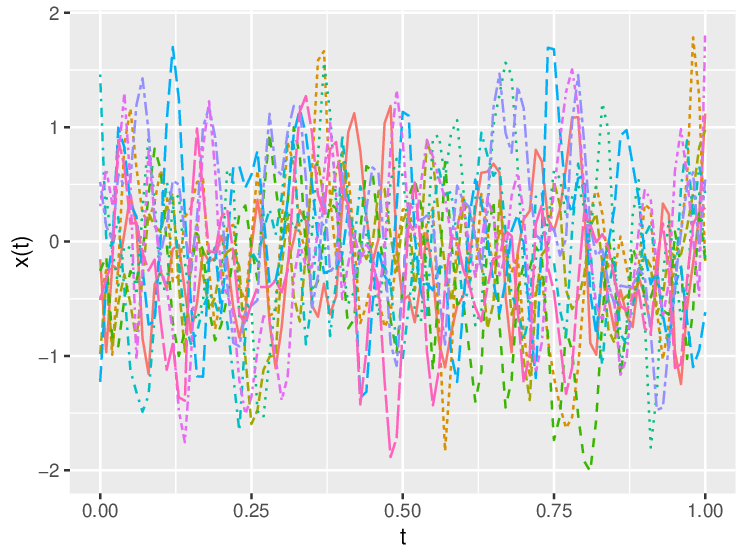}}
  \subfigure[Scenario II]{\includegraphics[width=4.8cm]{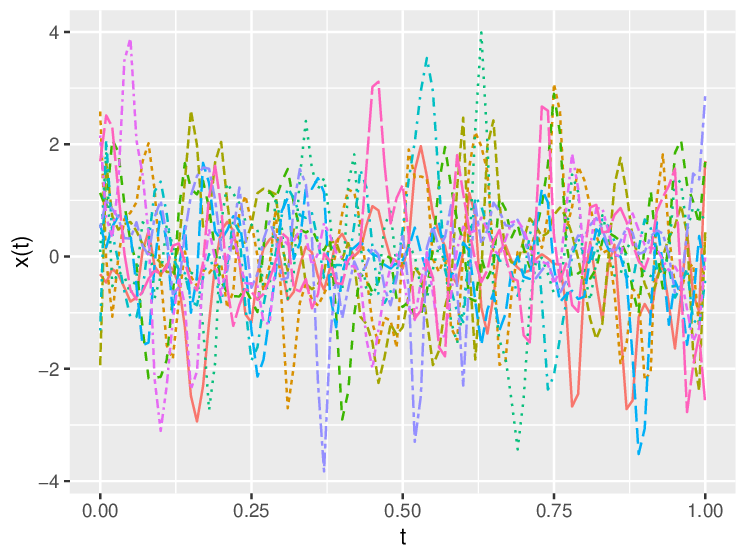}}
  \subfigure[Scenario III]{\includegraphics[width=4.8cm]{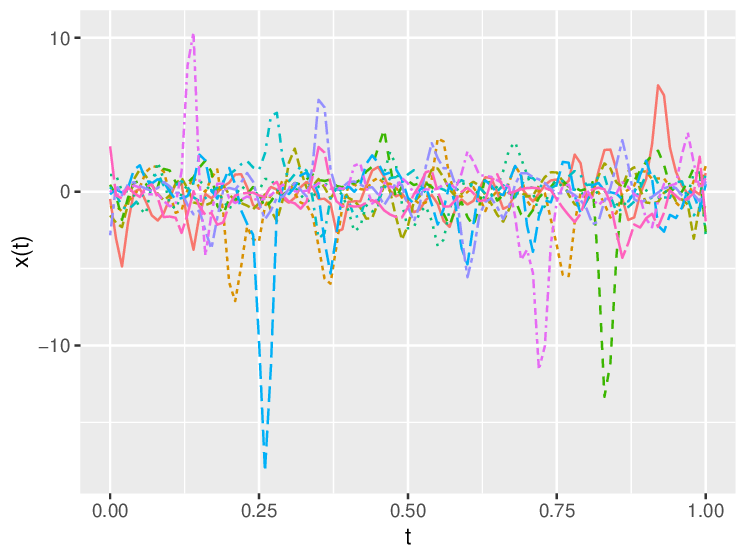}}
  \caption{A random subset of 10 curves for the functional predictor $x_i(t)$ under three scenarios
  when the full sample size is $n=10^5$.}\label{f1}
\end{figure}
\figref{f1} displays a random subset of 10 curves for the functional predictor $x_i(t)$ under three scenarios when the sample size $n = 10^5$. It shows that the variation among the functional predictor $x_i(t)$ is the smallest when $a_{ij}$ is generated from Scenario I, while the variation is the largest when $a_{ij}$ is generated from Scenario III. It means that the data generated under Scenario I is more uniform.

In the following, we want to compare two different approaches: the functional L-optimality subsampling (FLoS) method described in Algorithm \ref{al2} and the uniform subsampling (UNIS) approach. For the fairness of comparison, we use the same basis functions and the same smoothing parameter in the two approaches with the same full data. The integrated mean squared error (IMSE) of the estimated functional coefficient $\widetilde{\beta}$ from 500 replications is defined as follows:
\begin{equation*}
    \mathrm{IMSE} = \frac{1}{500}\sum_{s=1}^{500}\int (\widetilde{\beta}^{(s)}(t)-\beta(t))^2dt.
\end{equation*}


\begin{figure}[htbp]
  \centering
  \subfigure[Scenario I, $n=10^5$]{\includegraphics[width=4.8cm]{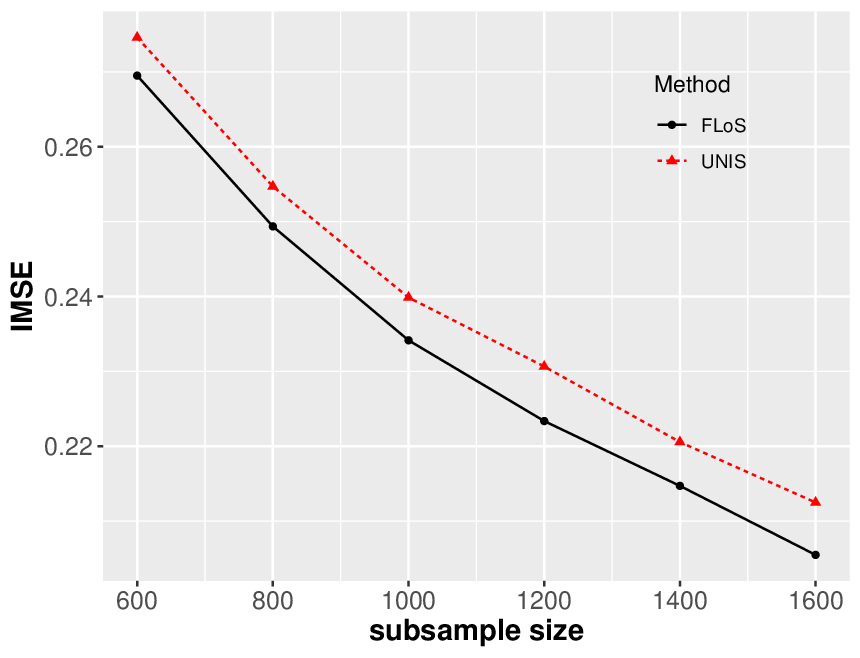}}
  \subfigure[Scenario II, $n=10^5$]{\includegraphics[width=4.8cm]{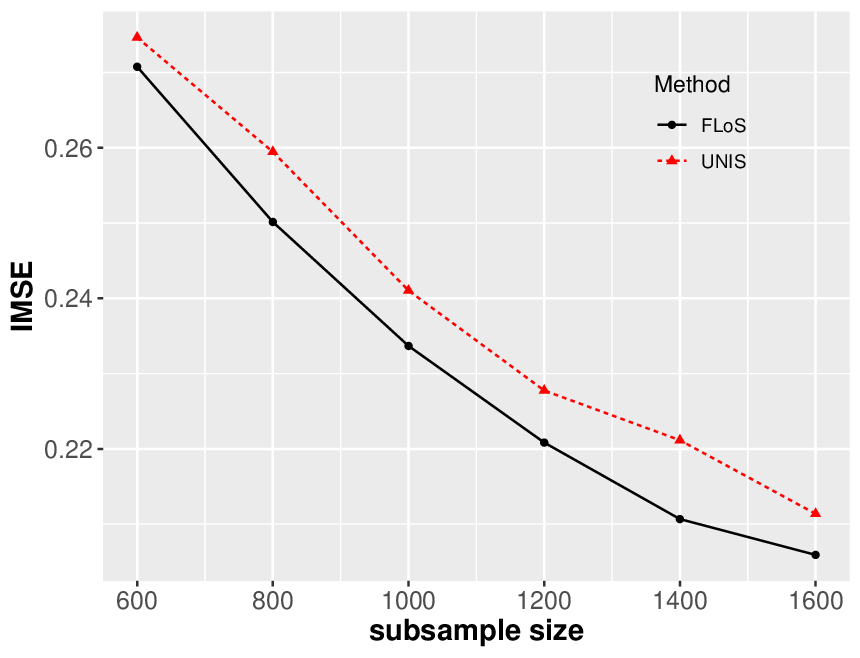}}
  \subfigure[Scenario III, $n=10^5$]{\includegraphics[width=4.8cm]{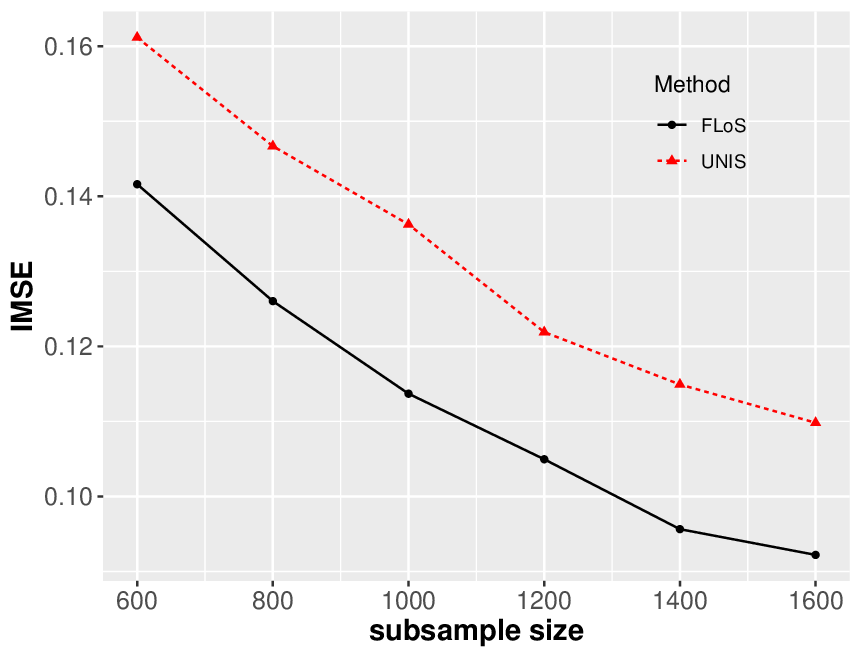}}\\
  \subfigure[Scenario I, $n=10^6$]{\includegraphics[width=4.8cm]{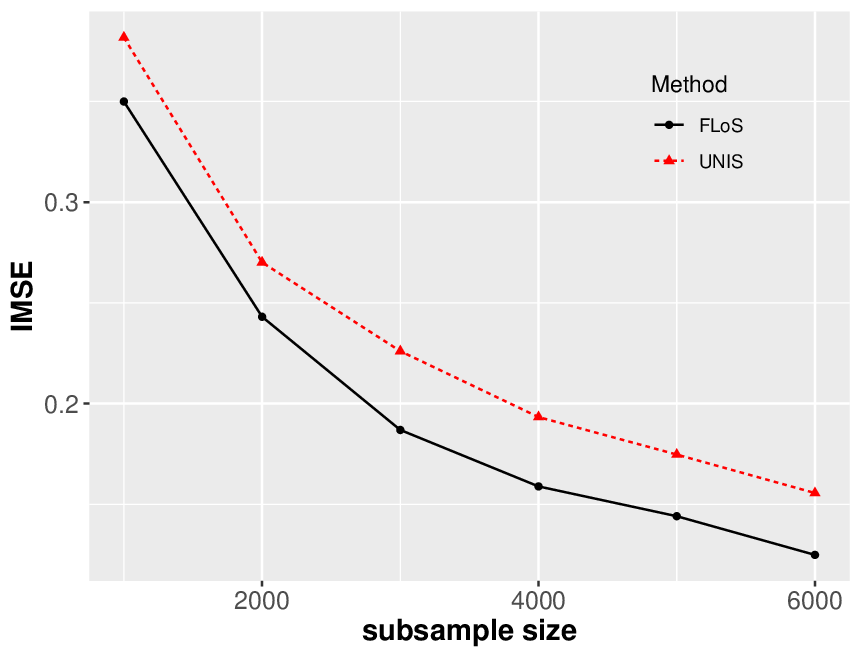}}
  \subfigure[Scenario II, $n=10^6$]{\includegraphics[width=4.8cm]{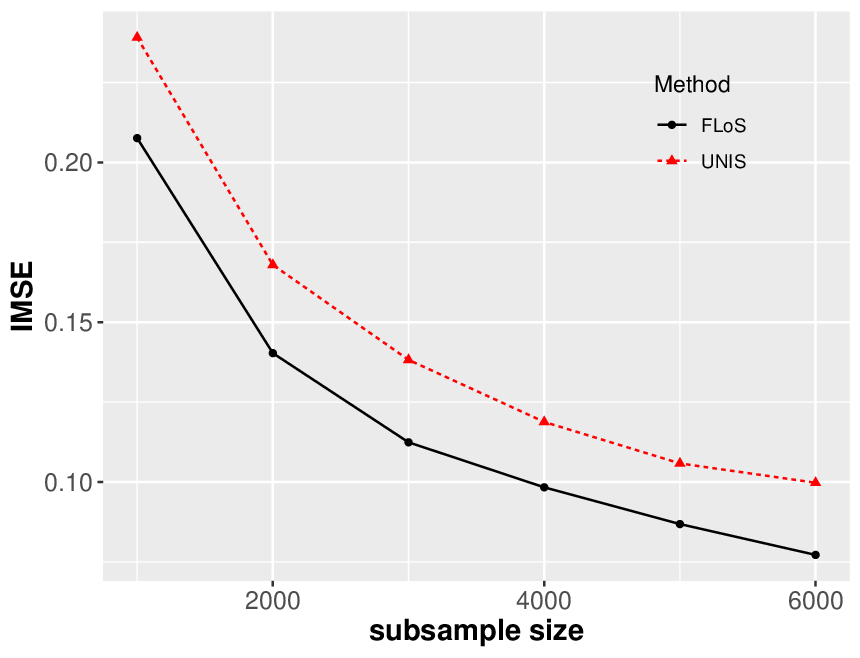}}
  \subfigure[Scenario III, $n=10^6$]{\includegraphics[width=4.8cm]{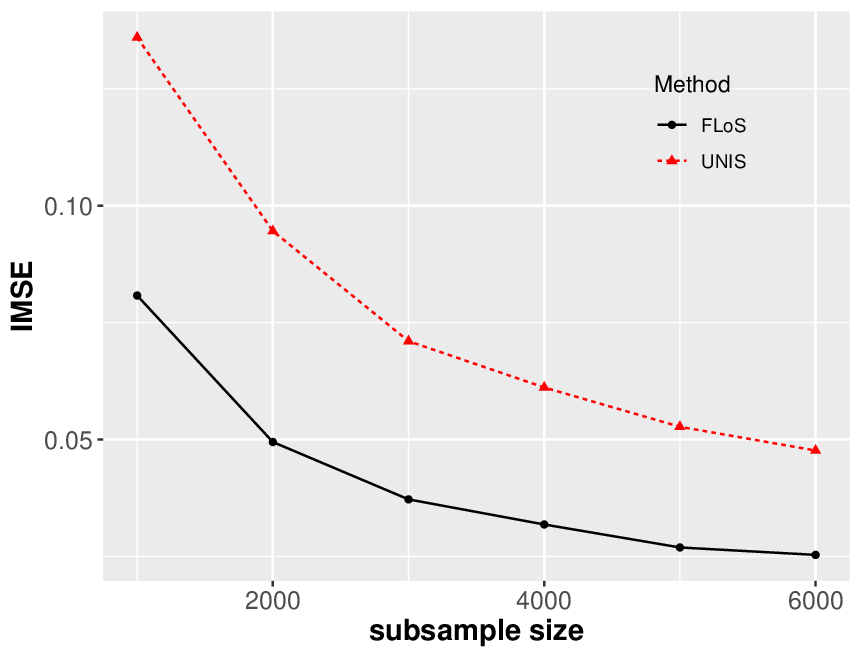}}\\
  \subfigure[Scenario I, $n=5\times 10^6$]{\includegraphics[width=4.8cm]{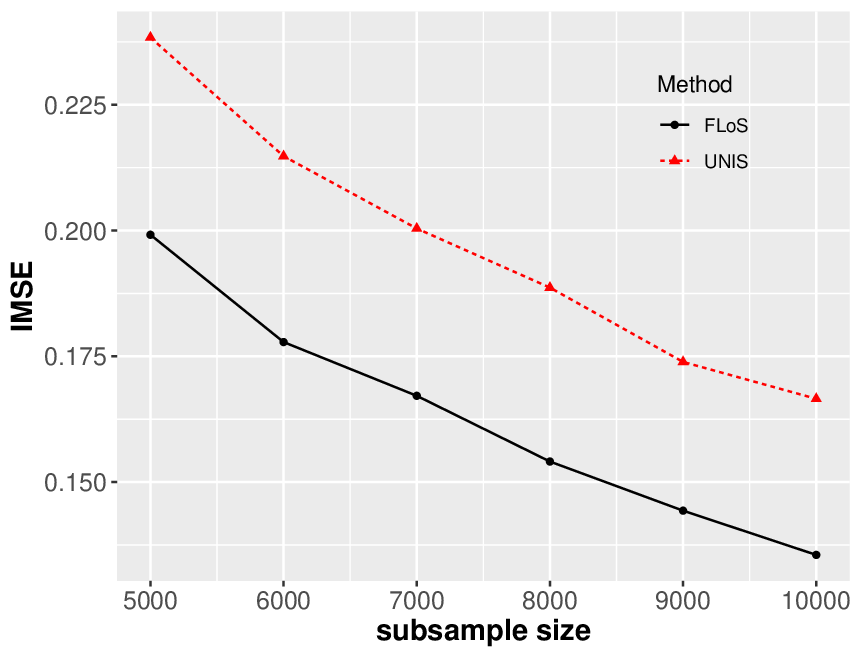}}
  \subfigure[Scenario II, $n=5\times 10^6$]{\includegraphics[width=4.8cm]{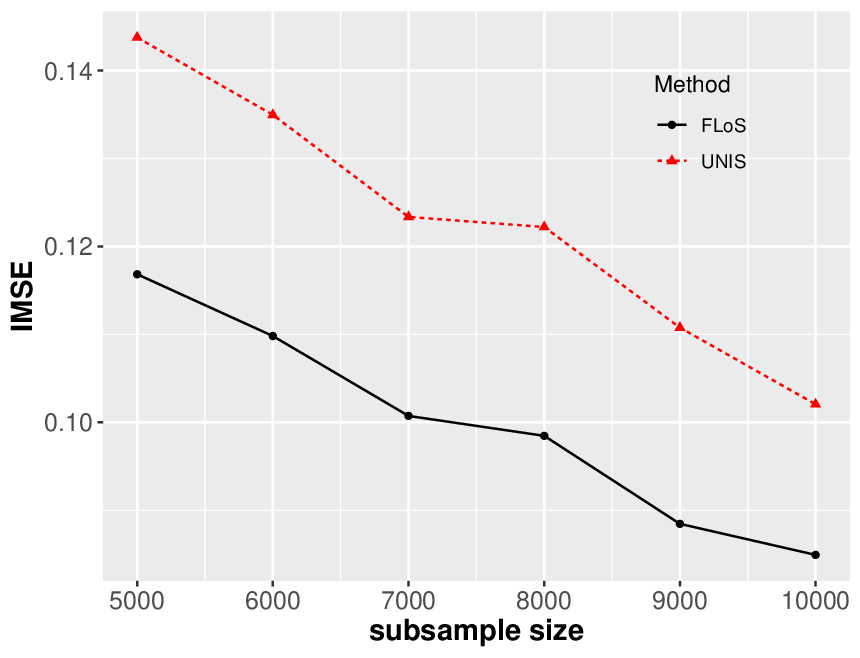}}
  \subfigure[Scenario III, $n=5\times 10^6$]{\includegraphics[width=4.8cm]{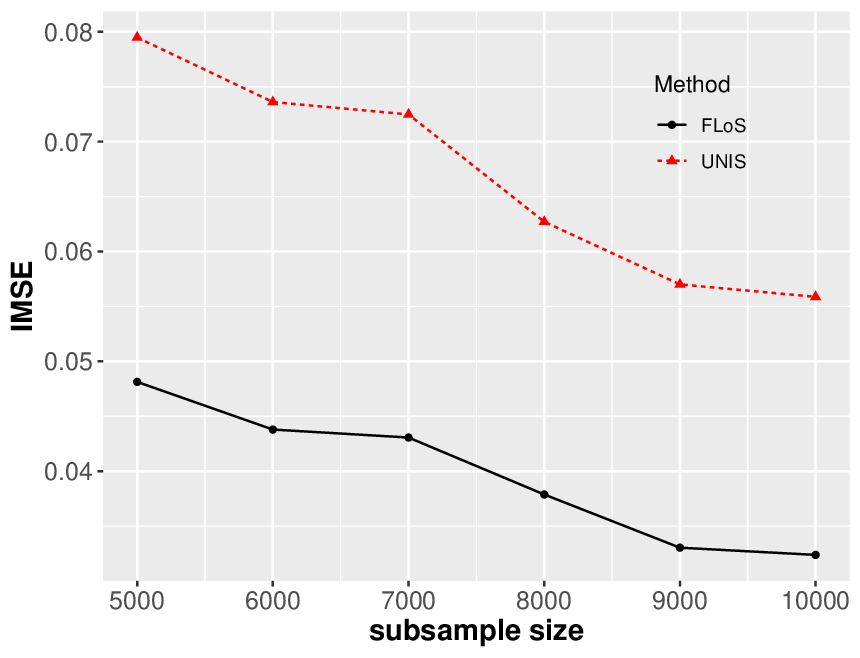}}\\
  \caption{The integrated mean squared error (IMSE) of the estimated functional coefficient $\widetilde{\beta}$ in the functional linear model by using the functional L-optimality subsampling (FLoS) method and the uniform subsampling (UNIS) approach under three scenarios with various subsample sizes $L$ when the full data size $n=10^5$, $10^6$, and $5\times 10^6$ .
   }\label{f2}
\end{figure}

\figref{f2} displays the mean of IMSE with various subsample sizes when the full data size $n=10^5$, $10^6$, and $5\times 10^6$. It shows that for all three scenarios, the functional L-optimality subsampling method always results in smaller IMSEs than the uniform subsampling method, which is consistent with the theoretical results that aim to minimize the IMSE of the estimator. Moreover, the advantage of the functional L-optimality subsampling method is more significant when the distribution tail of the basis coefficients $a_{ij}$ is heavier. It is not surprising to see that the IMSEs from both methods decrease as the subsample size $L$ increases when the full data size $n$ is fixed. In other words, the IMSEs decrease as the ratio $L/n$ increases.

  


To evaluate the computational efficiency of the subsampling strategies, we record the CPU times (in seconds)
of the two subsampling strategies and using the full data.  In this paper, we use the R programming language (enhanced R distribution Microsoft R 4.0.2) to implement each method. 
All computations are carried on a PC running Windows 7 with an 2.20 GHz Intel Core I5 Quad-Core Processor and 12GB memory. \tabref{ta1} displays the computation time for different combinations of the full data size $n$ and the subsample size $L$ under Scenario I. The results under the other two scenarios are similar and thus omitted. \tabref{ta1} shows that the functional L-optimality subsampling method is significantly faster than using the full data. The difference between the functional L-optimality subsampling method and the uniform subsampling method is small. In the implementation, we make the number of knots $K=\ceil{5\times n^{1/4}}$. When the full data size 
$n=5\times10^6$, the size of the basis matrix $\bm{N}$ is about $9$GB and the computing time for using full data exceeds 40 minutes. 
Moreover, 
the basis matrix needs about 21.6 GB memory under the full data size $n=10^7$, which goes beyond the maximum memory of a general PC with a 16G memory, so the estimation using the full data is not feasible. In this case, for the functional L-optimality subsampling method and the uniform subsampling method, we can take advantage of parallel computing to calculate the basis matrix $\bm{N}$ and the subsampling probability $p^{\mathrm{FLoS},\widehat{\bm{c}}^0}$. We then use the optimal subsampling data to estimate the functional linear model.

\renewcommand{\arraystretch}{1.2} 
\begin{table}
\centering
  \fontsize{7}{8}\selectfont
  \begin{threeparttable}
  \caption{The computing time for estimating the functional linear model using the functional L-optimality subsampling method and the uniform sampling method when the full data size $n=10^5$, $10^6$, $5\times 10^6$ and $n=10^7$. 
  When the full data size $n=10^7$, the estimation is beyond the computer's memory and fails when using the full data.}
  \label{ta1}
  \setlength{\tabcolsep}{4mm}{
    \begin{tabular}{cccccccc}
    \toprule
    \multirow{2}{*}{Full data size}&\multirow{2}{*}{Method}&\multicolumn{6}{c}{Subsample size L} \cr
    \cmidrule(lr){3-8}
    &&$1000$&$2000$&$3000$&$4000$&$5000$&$6000$\cr
    \midrule
   \multirow{3}{*}{$n=10^5$}&FLoS & 0.073 & 0.136 & 0.245 & 0.498 & 0.517 & 0.760 \\ 
  & UNIS & 0.027 & 0.081 & 0.183 & 0.275 & 0.498 & 0.746 \\ 
    \cmidrule(lr){2-8}
    & FULL & \multicolumn{6}{c}{\bf{1.243}}\\
    \midrule
  \multirow{3}{*}{$n=10^6$}&FLoS & 0.719 & 0.817 & 0.948 & 1.105 & 1.337 & 1.571\\ 
  &UNIS & 0.042 & 0.118 & 0.235 & 0.391 & 0.594 & 0.887 \\ 
  \cmidrule(lr){2-8}
 &FULL & \multicolumn{6}{c}{
 \bf{ 12.259}}\\
\midrule
\multirow{3}{*}{$n=5\times 10^6$}&FLoS &  11.975&12.068&13.038&15.354&29.633&41.238\\
&UNIS  & 0.344& 2.940&6.372&8.491&8.750&16.910 \\ 
\cmidrule(lr){2-8}
&FULL & \multicolumn{6}{c}{\bf{2518.364}}\\
\midrule
\multirow{3}{*}{$n=10^7$}
&FLoS &48.216&59.020&64.428&77.524&151.485&201.362 \\
&UNIS &2.102&13.518&26.656&31.738&36.216&67.147  \\ 

\cmidrule(lr){2-8}
&FULL & \multicolumn{6}{c}{\textbf{FAIL}}\\

\bottomrule
    \end{tabular}}
    \end{threeparttable}
   \end{table}

\newpage
\subsection{Simulation II}
\label{subsec::sim2}
\begin{figure}[htbp]
  \centering
  \subfigure[Scenario I]{\includegraphics[width=3.5cm]{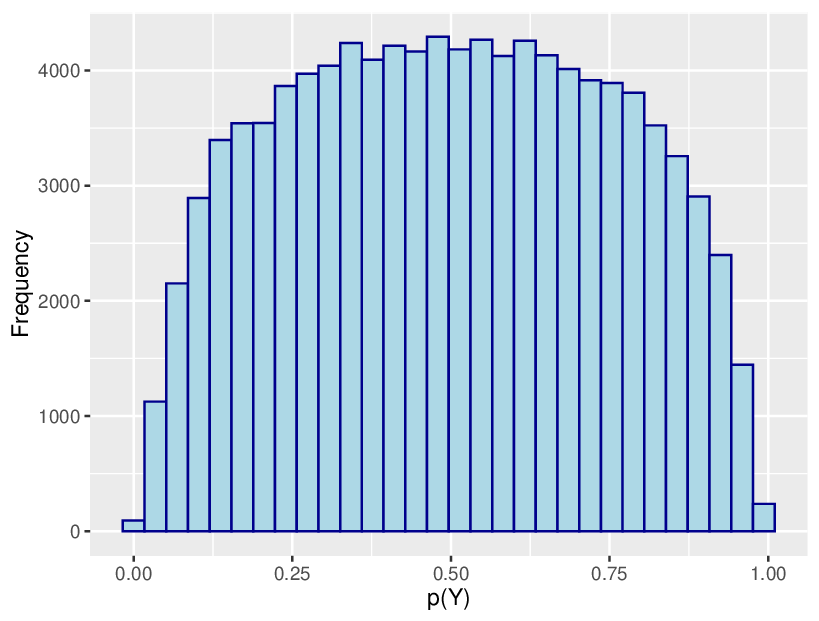}}
  \subfigure[Scenario II]{\includegraphics[width=3.5cm]{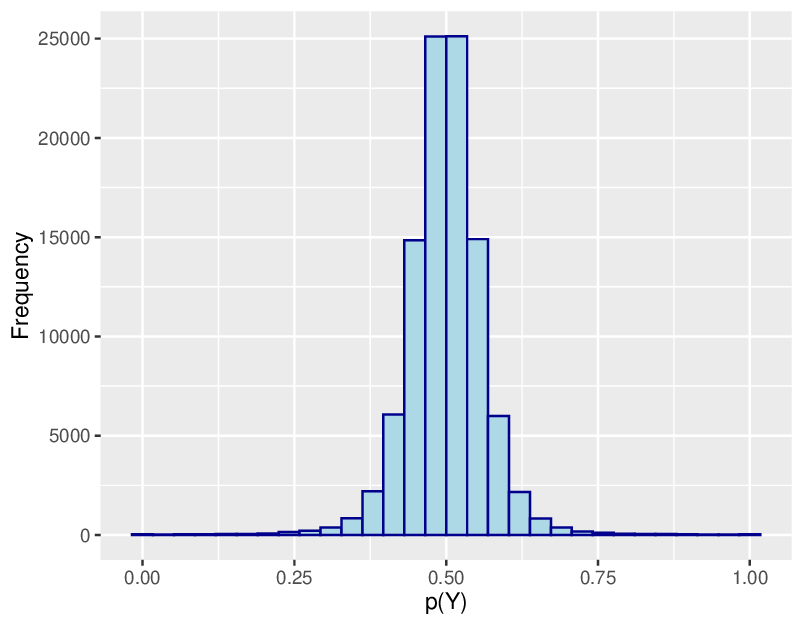}}
  \subfigure[Scenario III]{\includegraphics[width=3.5cm]{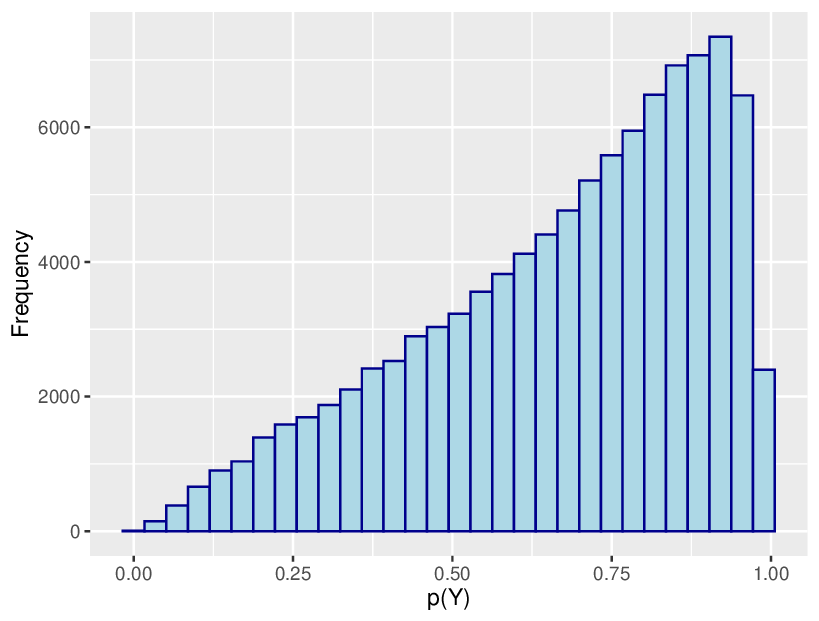}}
  \subfigure[Scenario IV]{\includegraphics[width=3.5cm]{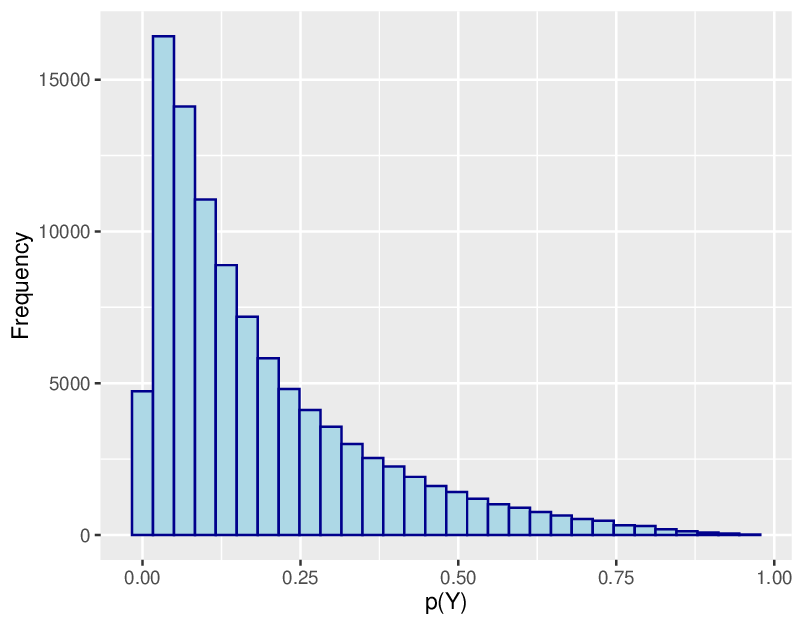}}\\
  \caption{The histogram of $p(x_i)$ under four scenarios
  when the full data size is $n=10^5$. }\label{f9}
\end{figure}

In this section, we evaluate the finite sample performance of the functional L-optimality subsampling method described in Algorithm \ref{al3} for estimating the functional logistic regression in comparison with the uniform subsampling method. We set the true functional coefficient $\beta(t)= \text{sin}(0.5\pi t)$. Denote $\psi(\cdot)= \text{exp}(\cdot)/(1+\text{exp}(\cdot))$ and $p(x_i) = \psi(\int_{0}^1x_i(t)\beta(t)dt)$, then
we generated responses $y(x_i) \sim \text{ Binomial}(1,p(x_i))$ as pseudo-Bernoulli r.v.s
with probability $p(x_i)$. The simulation designs for the functional predictors $x_i(t)$ are the same as in Simulation I, except that  
 we consider the following four different scenarios to generate the basis coefficients $a_{ij}$,
\begin{itemize}
  \item \textbf{Scenario I.} The coefficient $a_{ij}$ are i.i.d from $\sim N(0,15)$. \figref{f9} (a) shows that in the simulated data set under this scenario, the distribution of the probability $p(x_i)$ is symmetric about 0.5 and the number of $1$'s and the number of $0$'s in the responses are roughly equal.
  \item \textbf{Scenario II.} We generate the coefficient $a_{ij}$ from the $t$ distribution with 2 degree of freedom and zero mean, namely, $a_{ij}\overset{iid}\sim t_2$. For this scenario, \figref{f9} (b) shows that the probability $p(x_i)$ is symmetric about 0.5 and is less uniform than those $p(x_i)$ of Scenario I.
  Similar with Scenario I, in the simulated data set under Scenario II, the number of $1$'s and the number of $0$'s in the responses are roughly equal.
  \item \textbf{Scenario III.} Similar with the setting in \cite{wang2018optimal}, we generate the coefficient $a_{ij}$ from $ N(1.5,15)$. In this scenario, the distribution of probability $p(x_i)$ is skewed left and about 67.09\% of  responses are 1, which is shown in \figref{f9} (c). This data set is an imbalanced data.
  \item \textbf{Scenario IV.} We generate the coefficient $a_{ij}$ from $N(-3.0,15)$. The data set generated under this scenario is an example of rare events data with about 18.87\% of responses as 1, which is similar to the rare event data used in \cite{wang2018optimal}. \figref{f9} (d) shows that the distribution of probability $p(x_i)$ is skewed right.
\end{itemize}

\begin{figure}[htbp]
  \centering
  \subfigure[Scenario I]{\includegraphics[width=3.5cm]{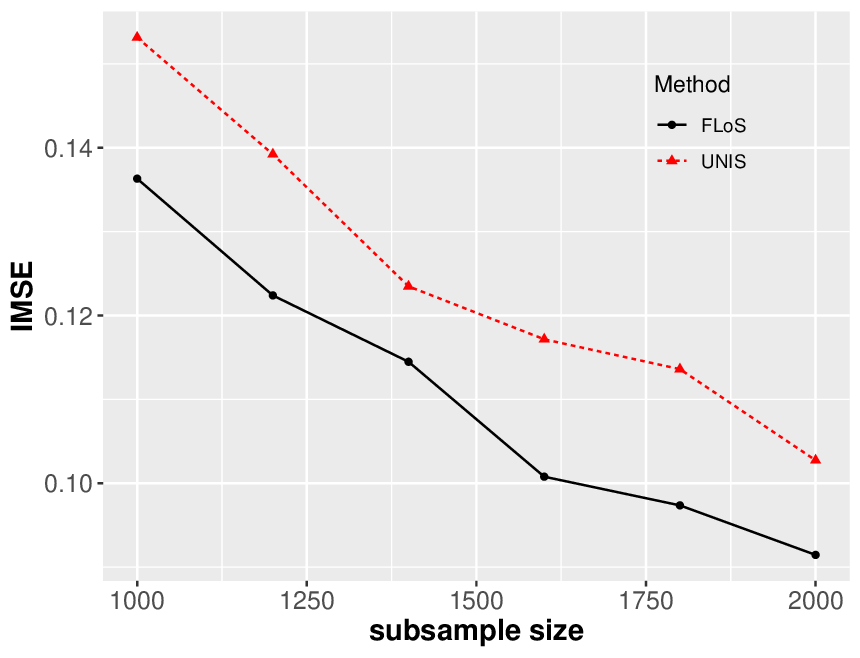}}
  \subfigure[Scenario II]{\includegraphics[width=3.5cm]{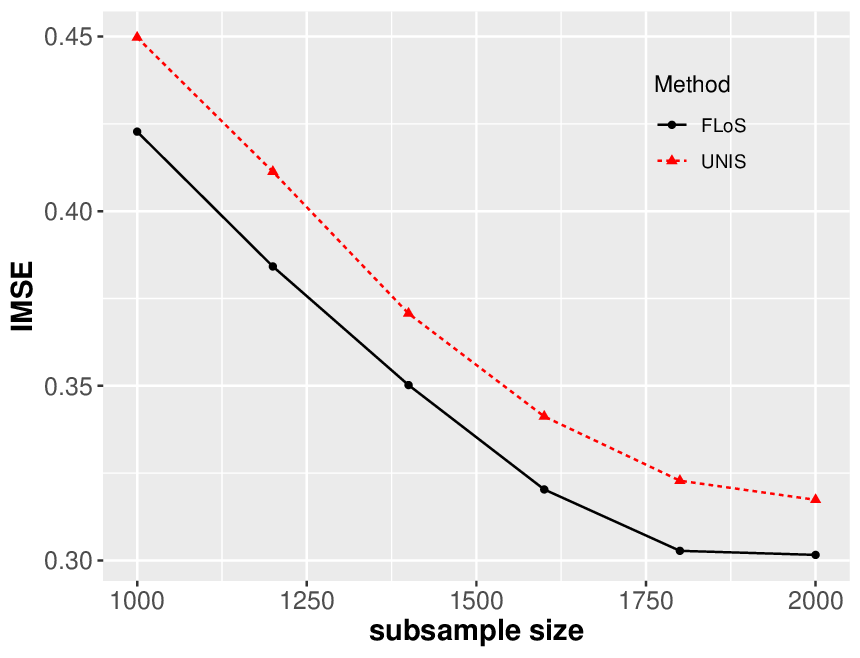}}
  \subfigure[Scenario III]{\includegraphics[width=3.5cm]{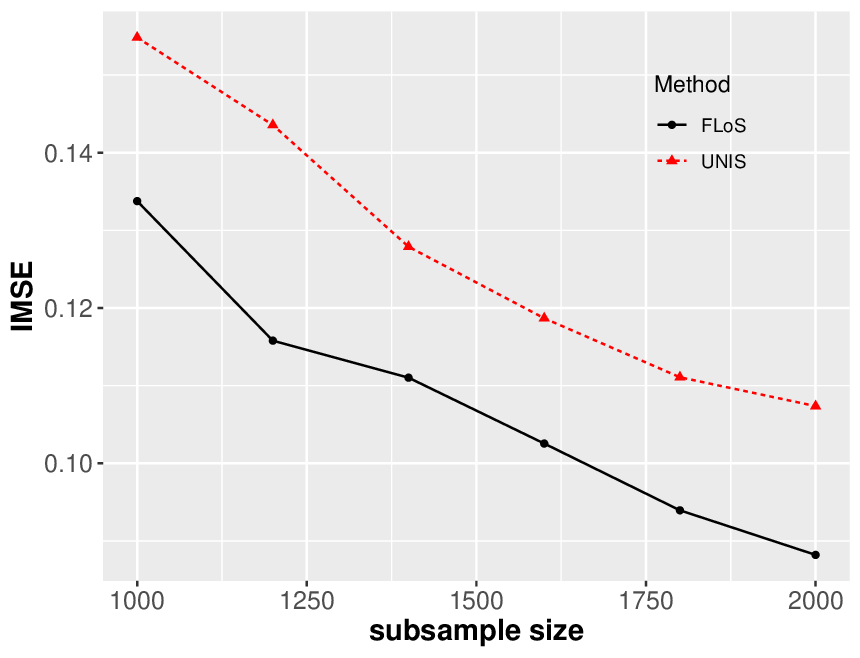}}
  \subfigure[Scenario IV]{\includegraphics[width=3.5cm]{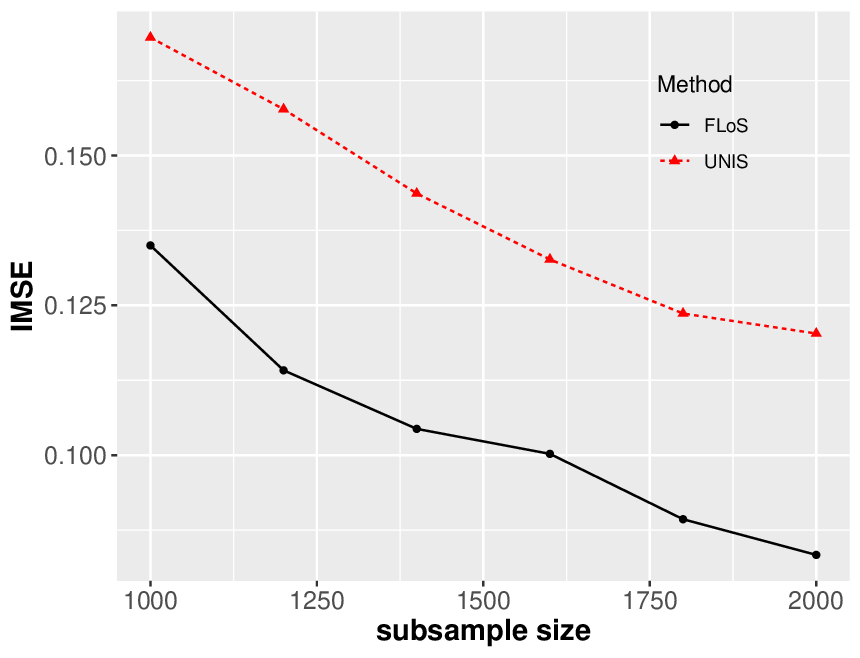}}\\
  \subfigure[Scenario I]{\includegraphics[width=3.5cm]{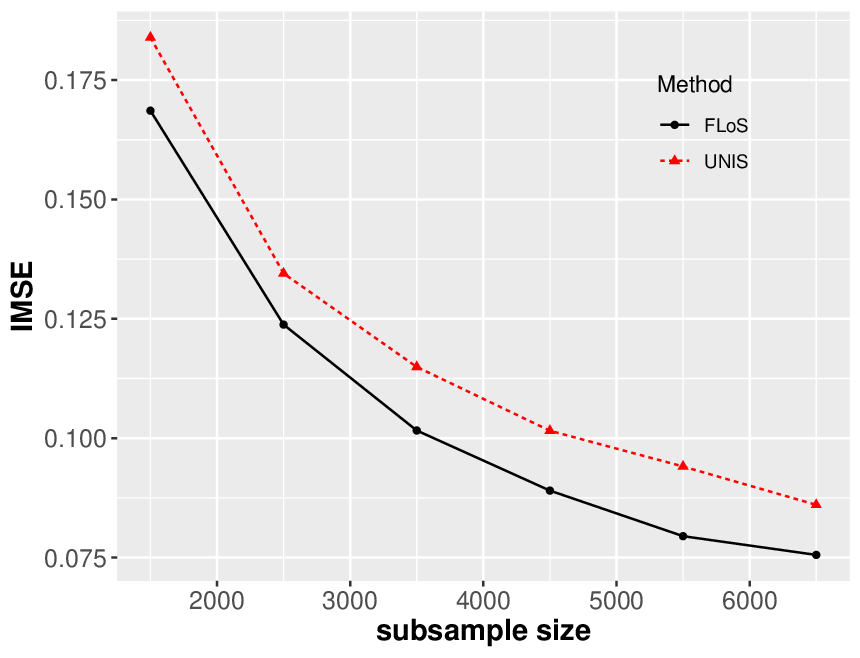}}
  \subfigure[Scenario II]{\includegraphics[width=3.5cm]{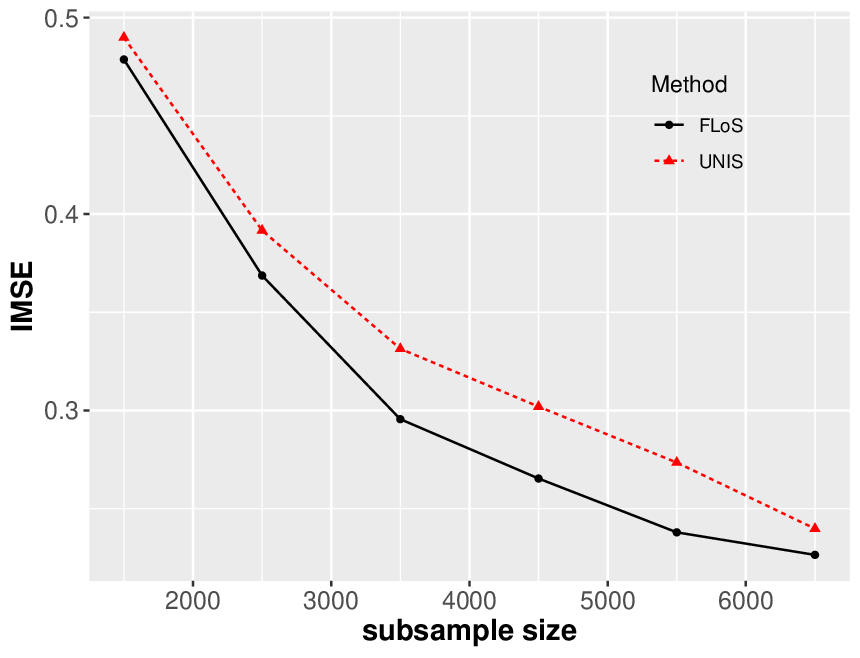}}
  \subfigure[Scenario III]{\includegraphics[width=3.5cm]{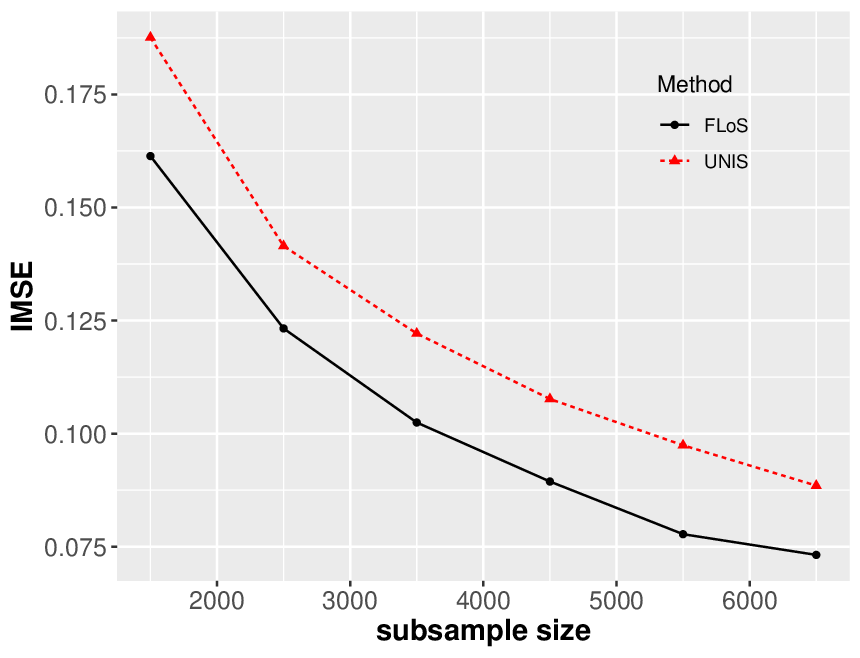}}
  \subfigure[Scenario IV]{\includegraphics[width=3.5cm]{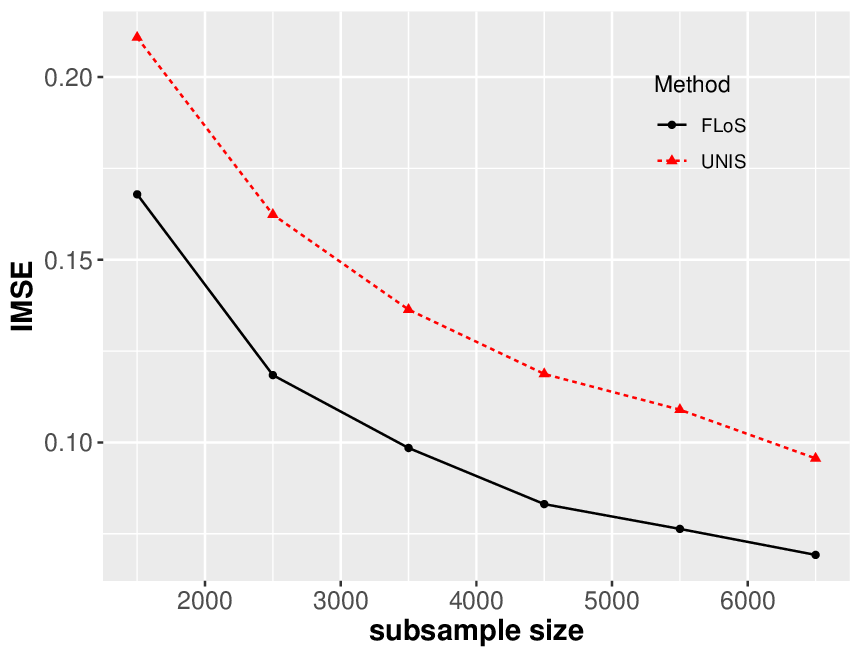}}\\
  \subfigure[Scenario I]{\includegraphics[width=3.5cm]{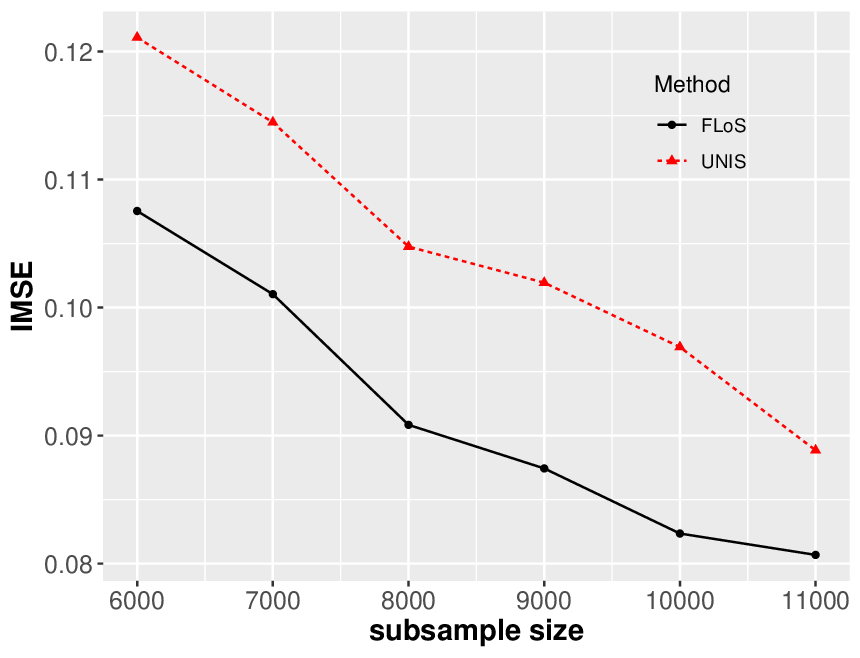}}
  \subfigure[Scenario II]{\includegraphics[width=3.5cm]{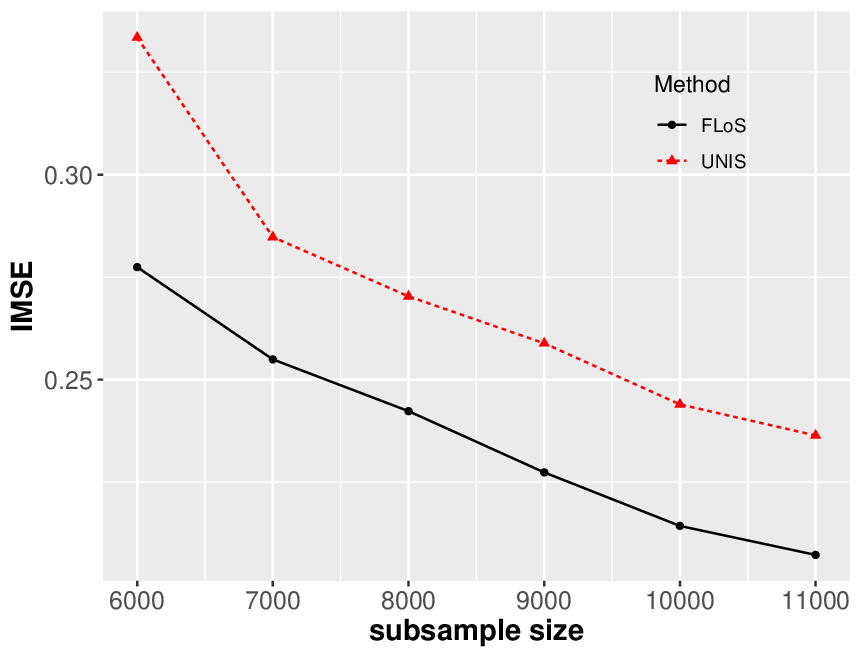}}
  \subfigure[Scenario III]{\includegraphics[width=3.5cm]{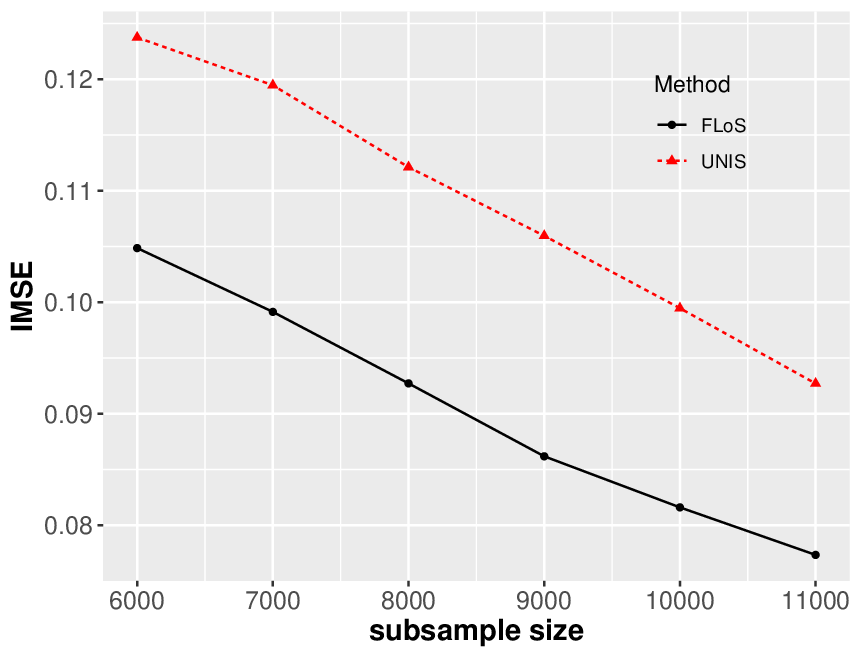}}
  \subfigure[Scenario IV]{\includegraphics[width=3.5cm]{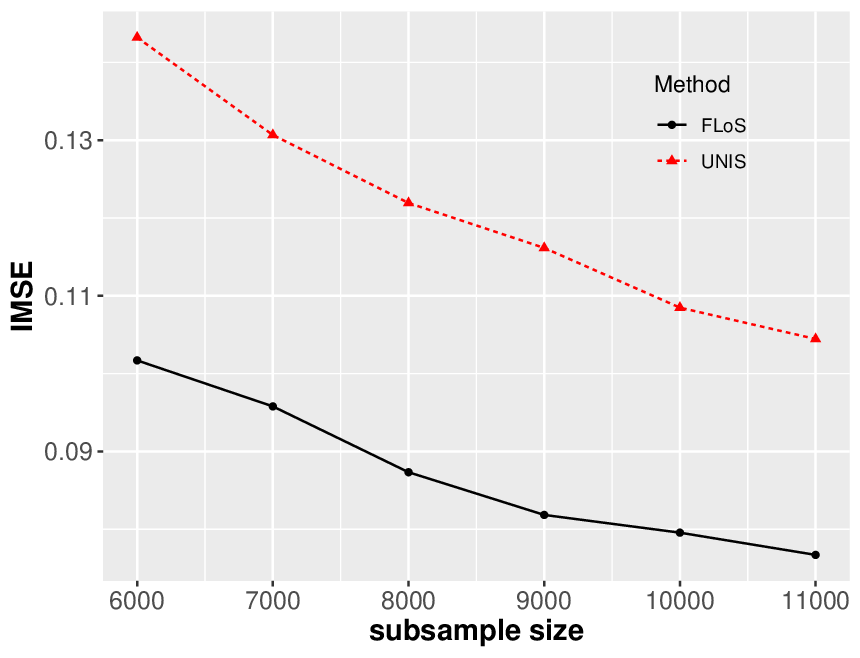}}\\
  \caption{The integrated mean squared error (IMSE) of the estimated functional coefficient $\widetilde{\beta}$ in the functional logistic regression model by using the functional L-optimality subsampling (FLoS) method and the uniform subsampling (UNIS) approach under four scenarios with various subsample sizes $L$ when the full data size $n=10^5$ (Panels (a)-(d)), $10^6$ (Panels (e)-(h)), and $5\times 10^6$ (Panels (i)-(l)).} \label{f11}
\end{figure}

\begin{figure}[htbp]
  \centering
  \subfigure[4.33\% of 1's in the responses]{\includegraphics[width=7cm]{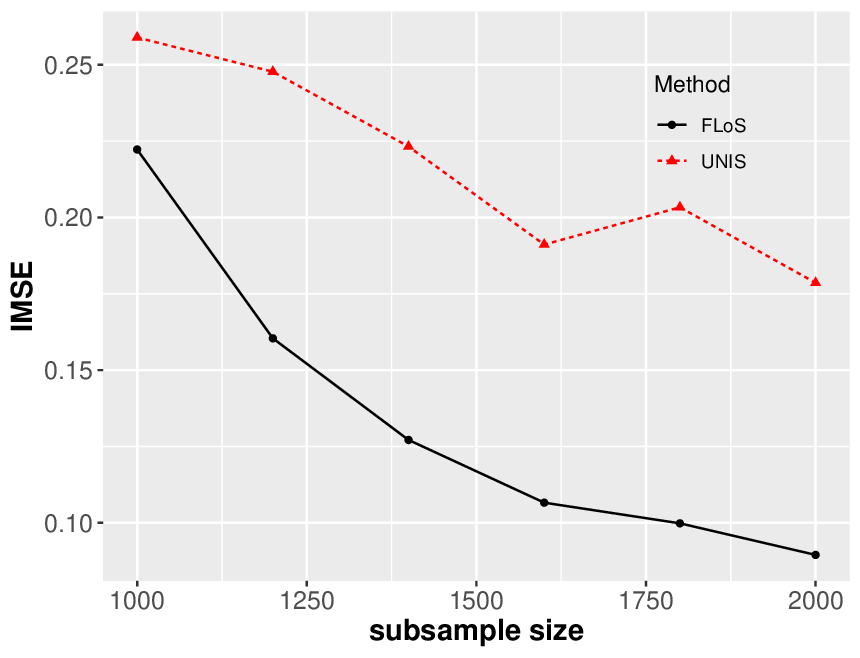}}
  \subfigure[1.34\% of 1's in the responses]{\includegraphics[width=7cm]{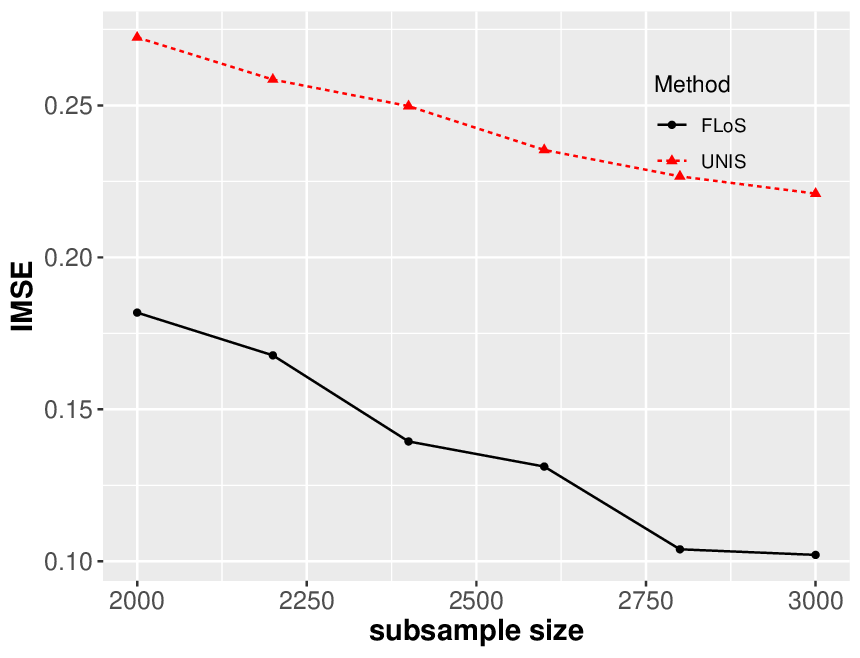}}
  \caption{The integrated mean squared error (IMSE) of the estimated functional coefficient $\widetilde{\beta}$ in the functional logistic regression model by using the functional L-optimality subsampling (FLoS) method and the uniform subsampling (UNIS) approach from the rare event data with 4.33\% or 1.34\% of 1's in the response when the full data size $n=10^5$.}\label{rare}
\end{figure}

\figref{f11} displays the mean of IMSEs when the full data size is $10^5$, $10^6$ and $5\times 10^6$. \figref{f11} shows that the functional L-optimality subsampling method outperforms the uniform subsampling approach for all scenarios and all full data sizes. The IMSEs for both subsampling methods decrease as the subsample increases. When the full data size is fixed, the more imbalanced the data, the greater the advantage of the functional L-optimality subsampling method over the uniform subsampling approach. \figref{rare} shows that our method can still outperform the uniform subsampling approach when the proportion of 1’s in the responses reaches 4.33\% ($a_{ij}\overset{iid}\sim N(-6,15)$) or even 1.34\% ($a_{ij}\overset{iid}\sim N(-8,15)$). On the other hand, when the data is extremely rare data (e.g. 0.02\% of 1's in the responses, that is, $a_{ij}\overset{iid}\sim N(-15,15)$), neither subsampling methods or the method using the full data work well. 
In Scenario II when the variation among functional predictor is larger, \figref{f11} (b), (f) and (j) show that the functional L-optimality subsampling method also dominates the uniform subsampling approach.

\begin{figure}[htbp]
  \centering
  \subfigure[Scenario I]{\includegraphics[width=3.5cm]{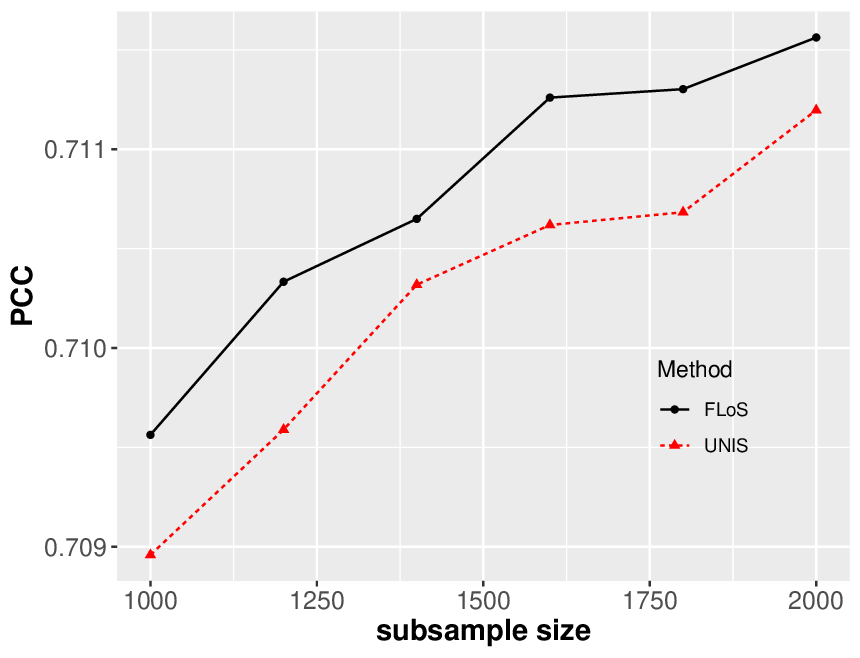}}
  \subfigure[Scenario II]{\includegraphics[width=3.5cm]{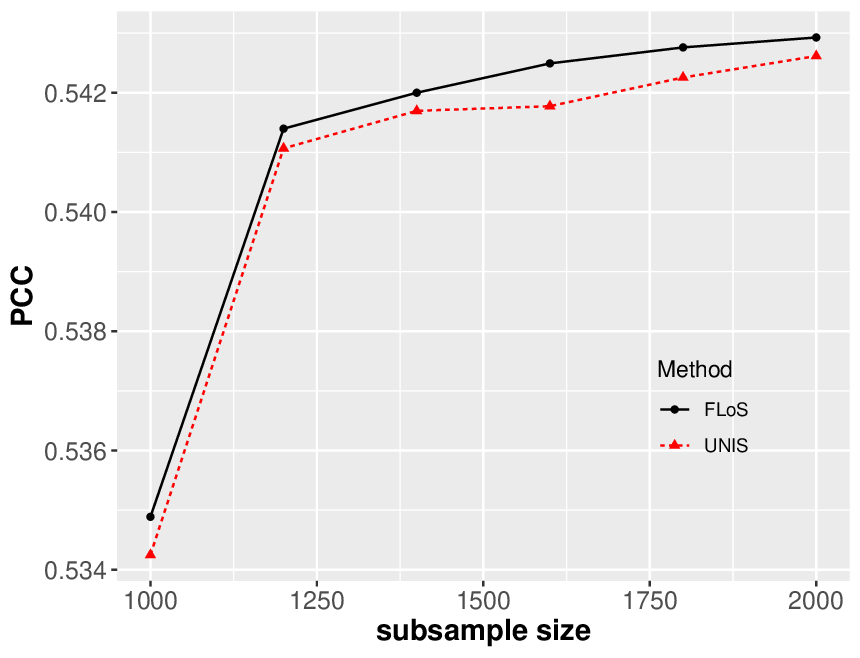}}
  \subfigure[Scenario III]{\includegraphics[width=3.5cm]{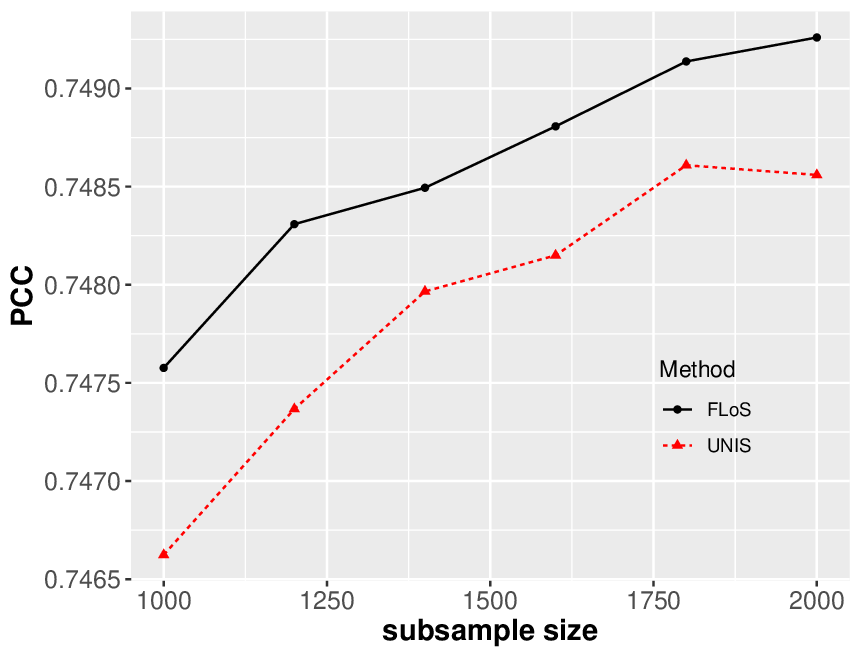}}
  \subfigure[Scenario IV]{\includegraphics[width=3.5cm]{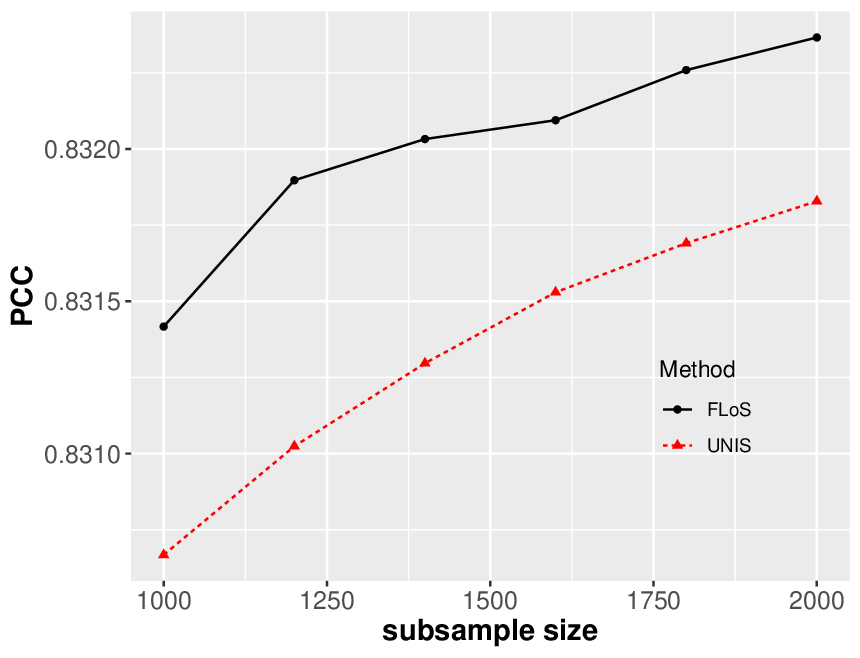}}\\
  \subfigure[Scenario I]{\includegraphics[width=3.5cm]{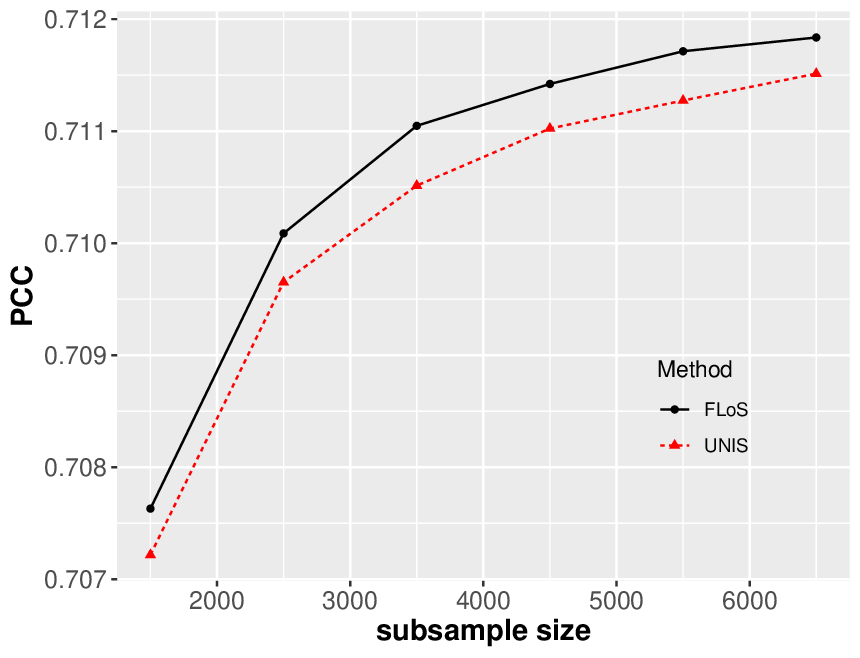}}
  \subfigure[Scenario II]{\includegraphics[width=3.5cm]{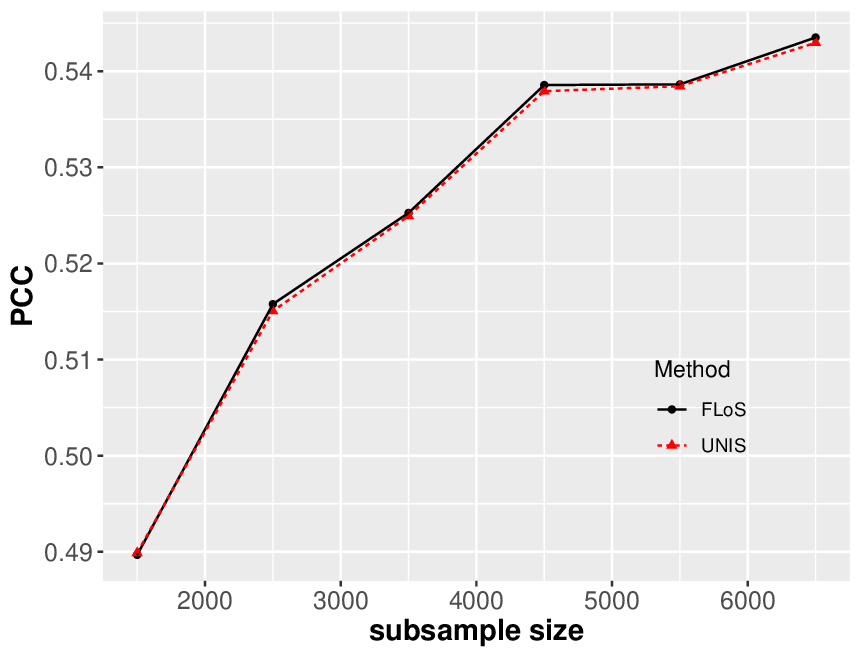}}
  \subfigure[Scenario III]{\includegraphics[width=3.5cm]{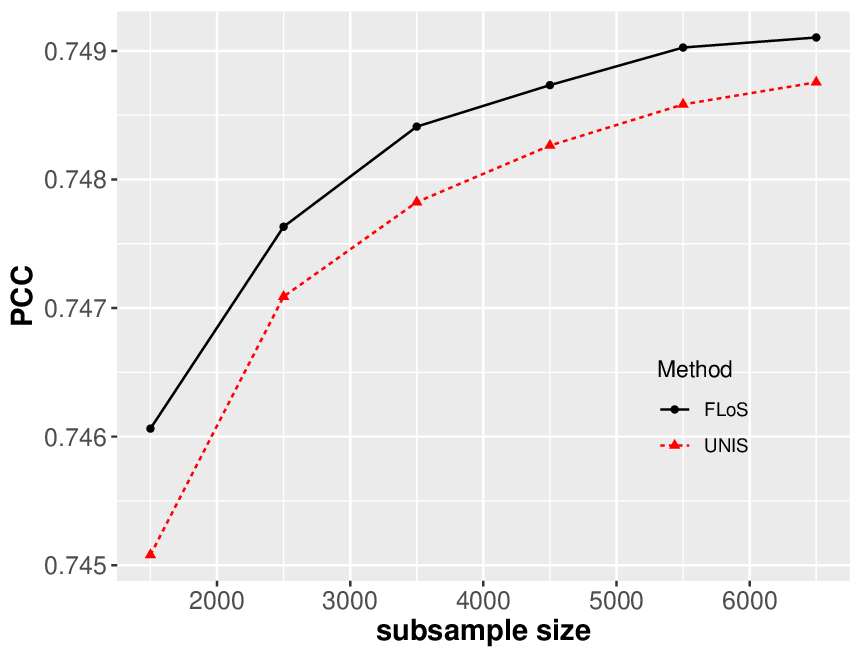}}
  \subfigure[Scenario IV]{\includegraphics[width=3.5cm]{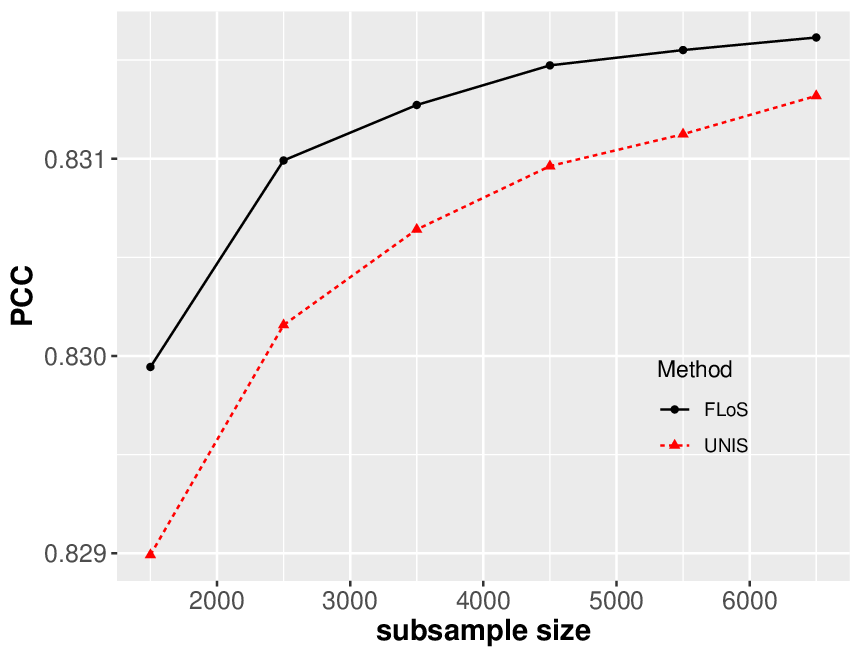}}\\
  \subfigure[Scenario I]{\includegraphics[width=3.5cm]{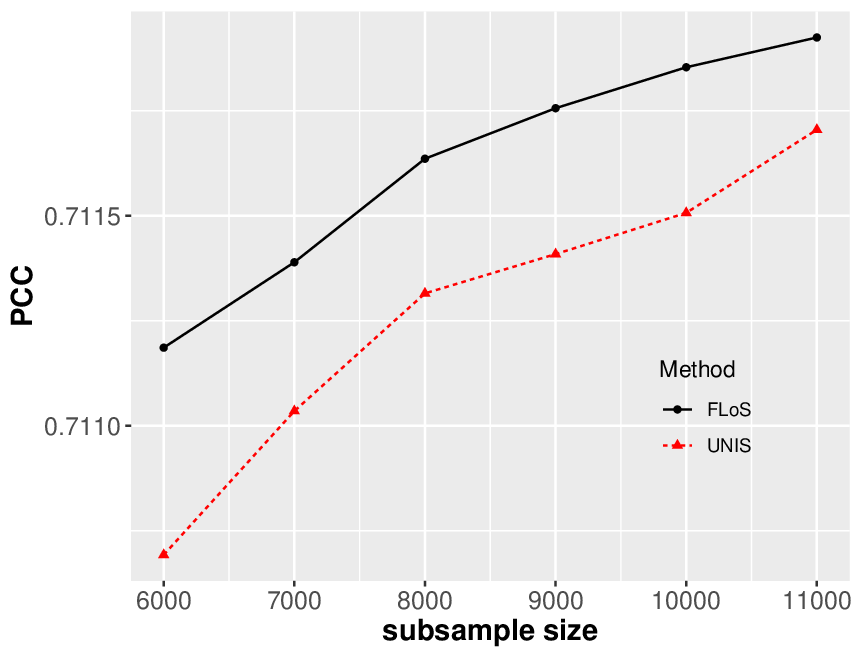}}
  \subfigure[Scenario II]{\includegraphics[width=3.5cm]{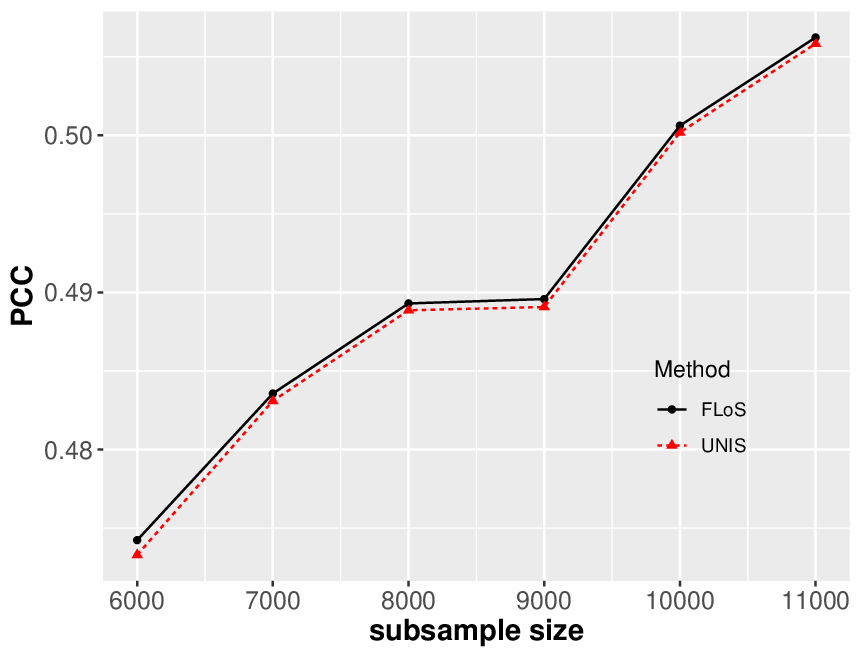}}
  \subfigure[Scenario III]{\includegraphics[width=3.5cm]{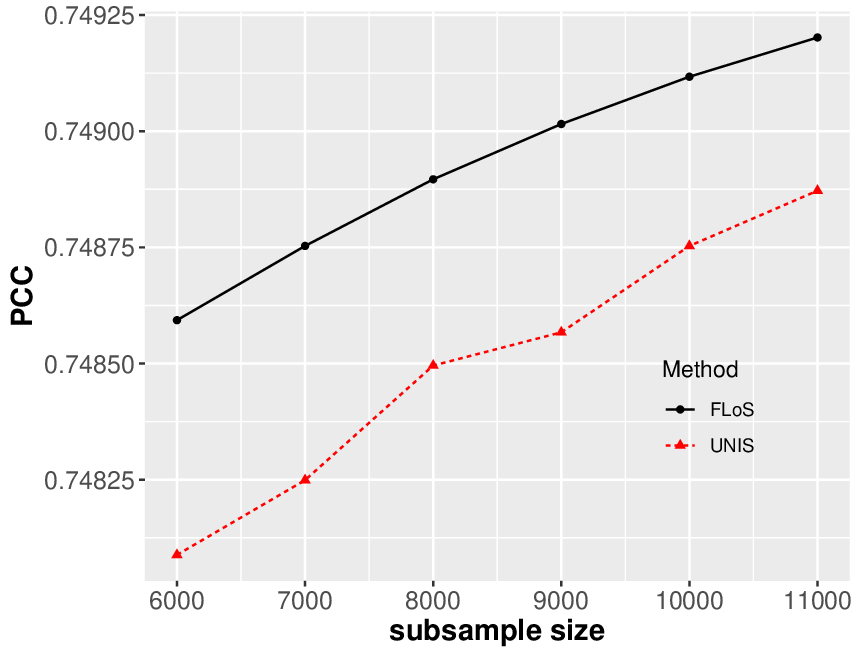}}
  \subfigure[Scenario IV]{\includegraphics[width=3.5cm]{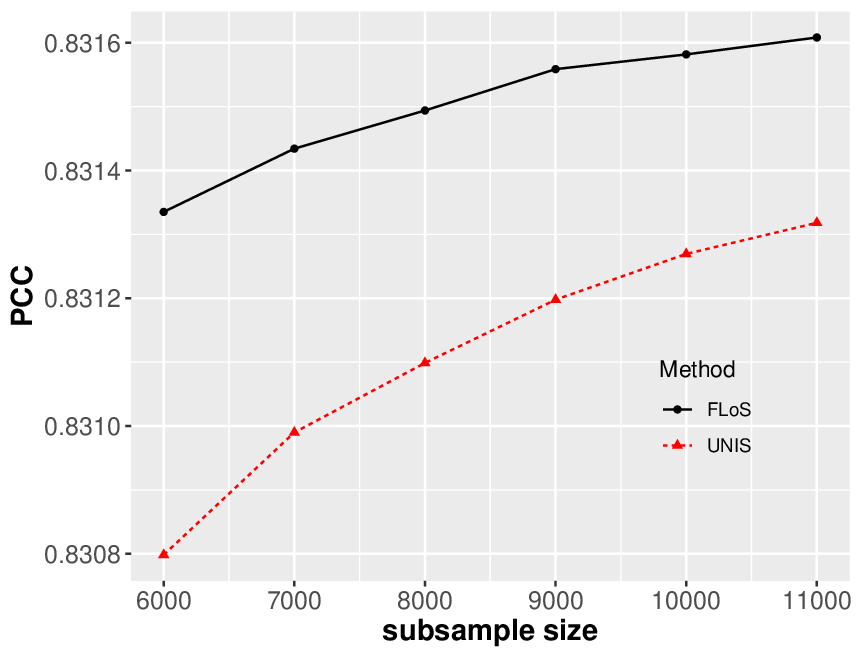}}\\
  \caption{The proportions of correct classifications (PCC) defined in (\ref{PCC}) in the functional logistic regression model by using the functional L-optimality subsampling (FLoS) method and the uniform subsampling (UNIS) approach under four scenarios with various subsample sizes $L$ when the full data size $n=10^5$ (Panels (a)-(d)), $10^6$ (Panels (e)-(h)), and $5\times 10^6$ (Panels (i)-(l)).}\label{f12}
\end{figure}

To compare the performance of the two subsampling methods on the classification accuracy, \figref{f12} displays proportions of correct classifications (PCC), which is defined as:
\begin{equation}\label{PCC}
    \mathrm{PCC} = \frac{\#\{y_i=1 \quad
    \text{and} \quad \psi(\bm{N}_i^{T}\widehat{\bm{c}})>0.5 \}+\# \{y_i=0 \quad \text{and} \quad \psi(\bm{N}_i^{T}\widehat{\bm{c}})\leq 0.5\} }{n}.
\end{equation}
\figref{f12} shows that the functional L-optimality subsampling method performs better than the uniform subsampling approach in all four scenarios. For Scenario II, although the two methods do not perform well, the functional L-optimality subsampling method is still slightly better than the uniform subsampling approach.
We also find that the performance using the full data is not good either under Scenario II.

In summary, regardless of whether the variation among the generated functional predictors is  large or the responses are imbalanced, our proposed functional L-optimality subsampling method is better than the uniform subsampling approach.

\subsection{Simulation III}
\label{subsec::sim3}
\begin{figure}[htbp]
  \centering
  \subfigure[Scenario I]{\includegraphics[width=4cm]{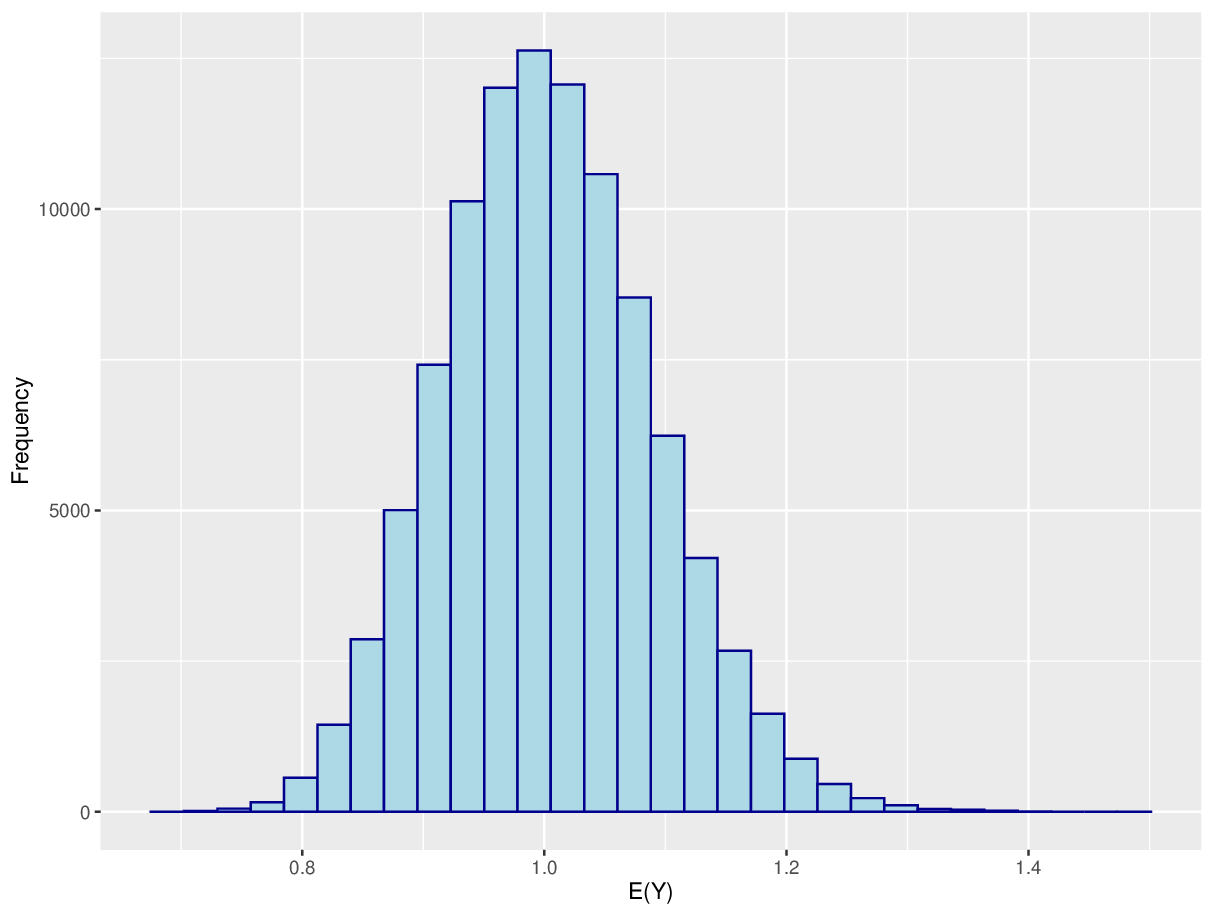}}
  \subfigure[Scenario II]{\includegraphics[width=4cm]{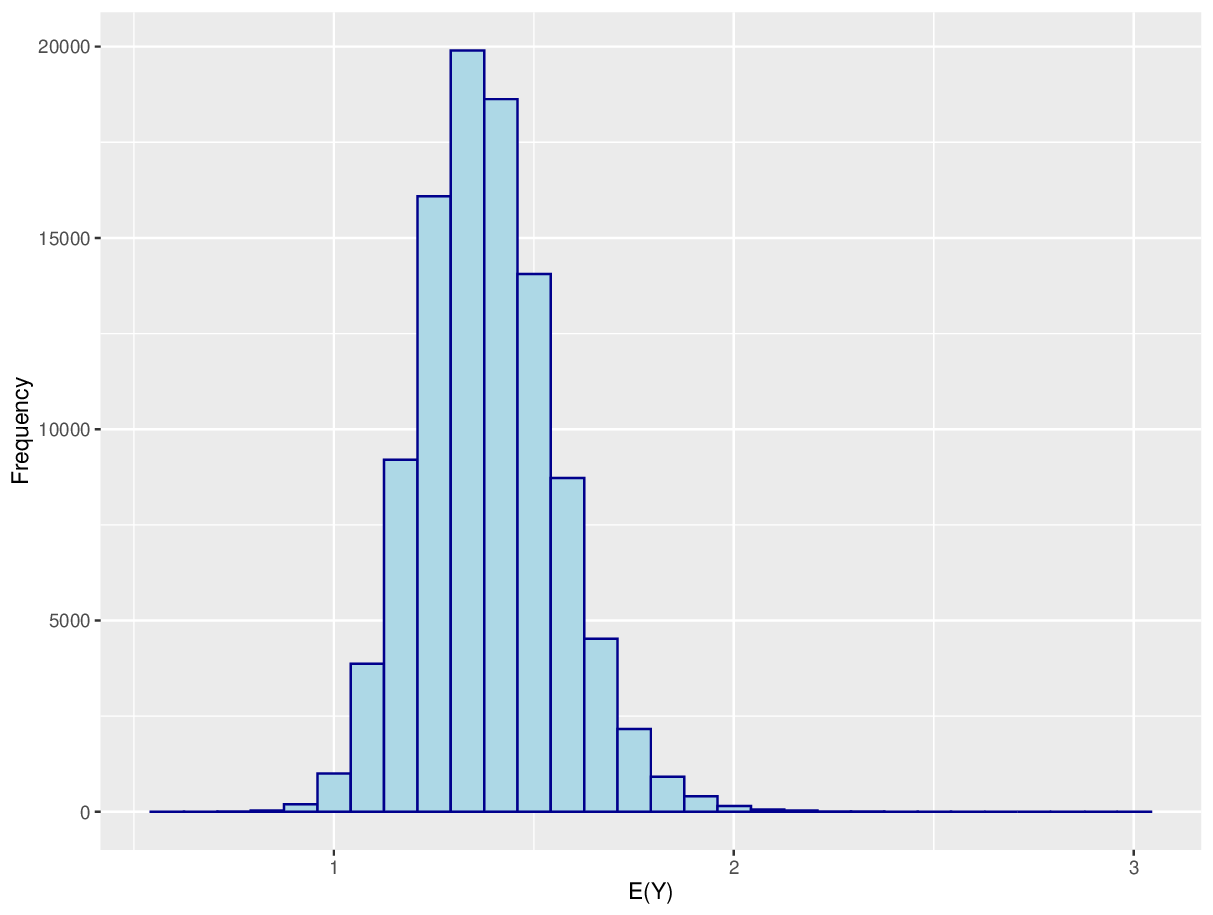}}
  \subfigure[Scenario III]{\includegraphics[width=4cm]{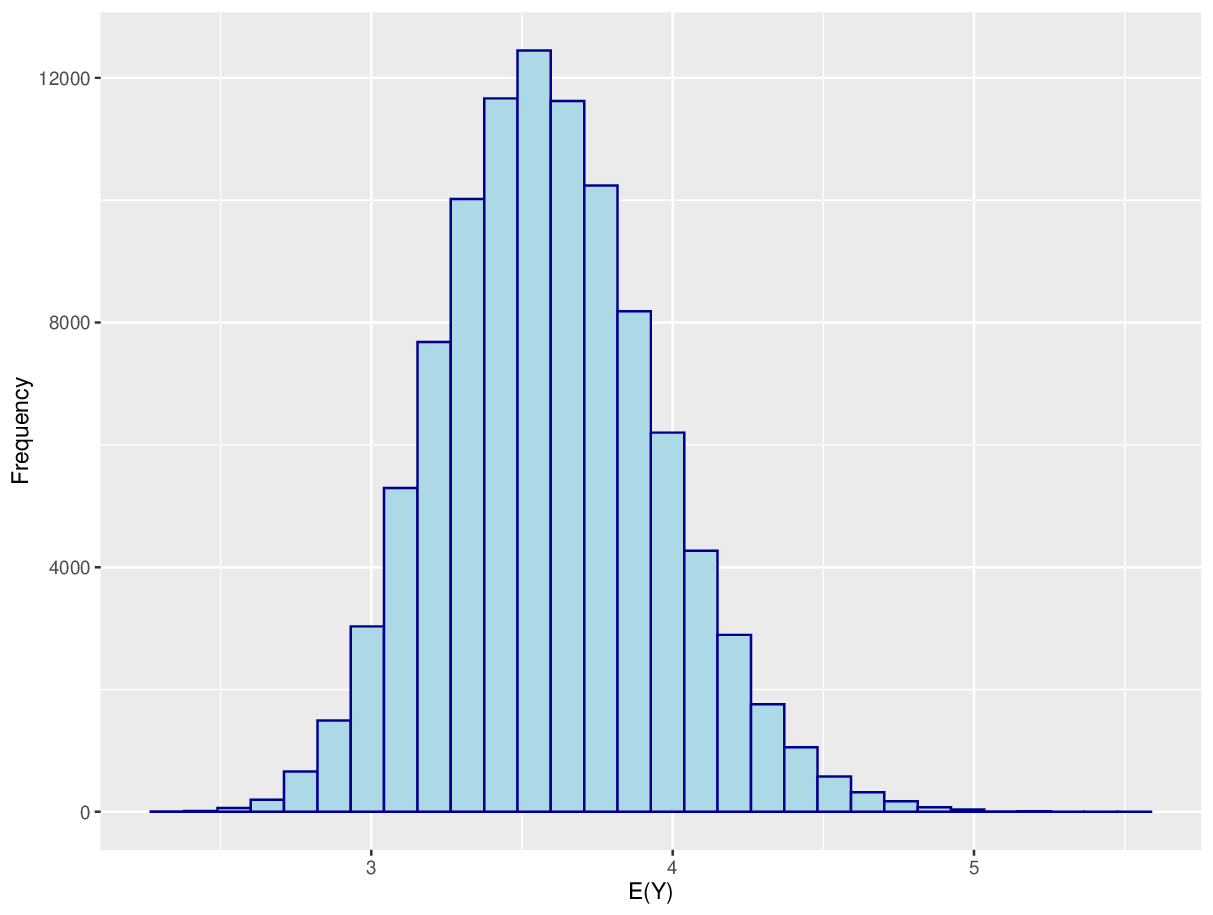}}\\
  \subfigure[Scenario I]{\includegraphics[width=4cm]{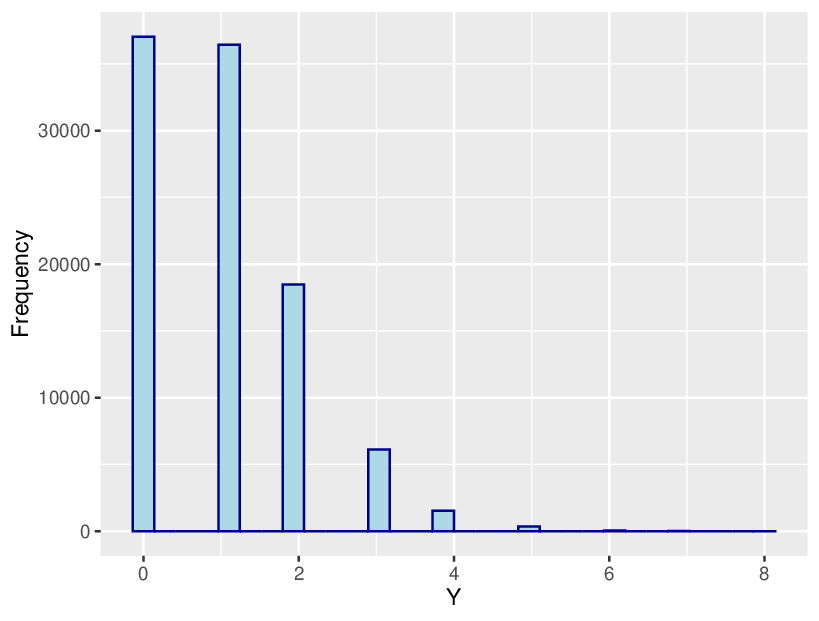}}
  \subfigure[Scenario II]{\includegraphics[width=4cm]{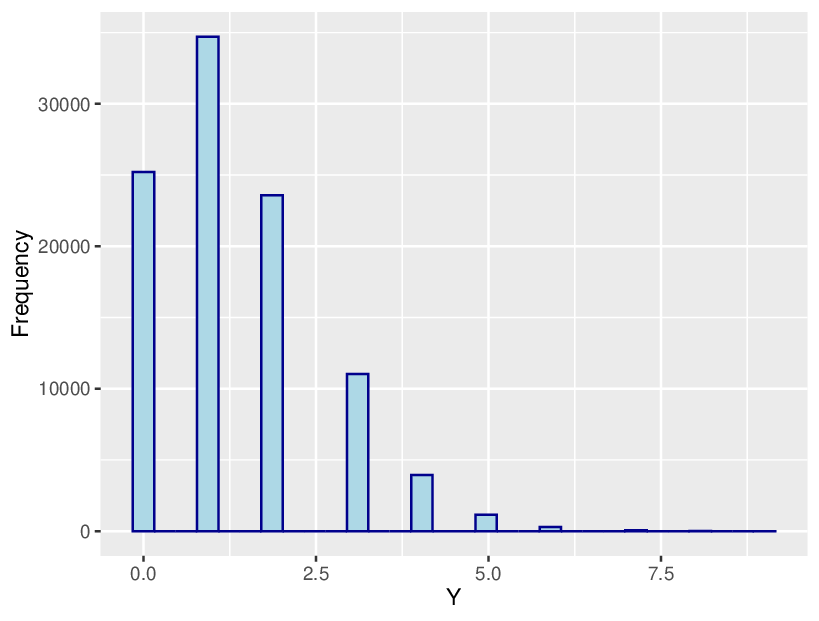}}
  \subfigure[Scenario III]{\includegraphics[width=4cm]{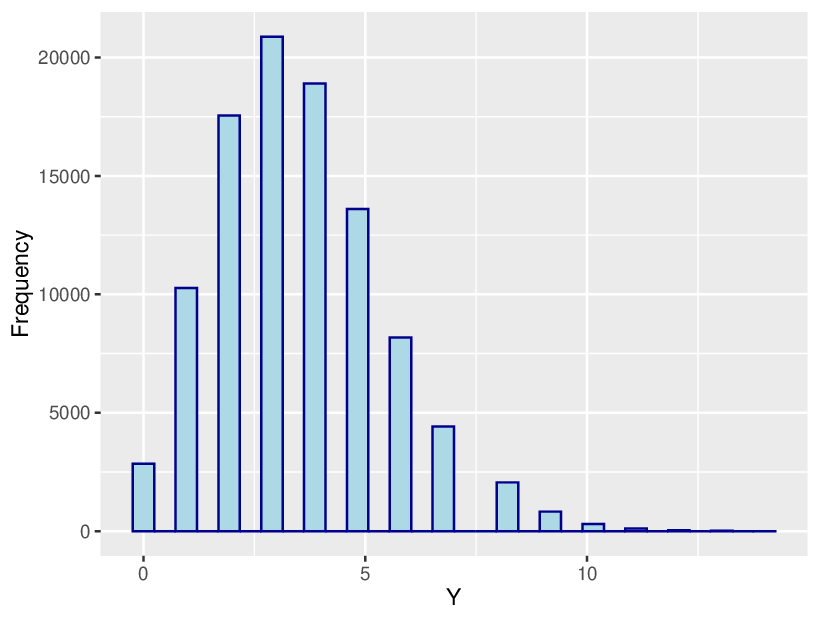}}\\
  \caption{The histogram of $E(y_i) = \lambda(x_i)$ and $y_i$ under three scenarios
  when the full data size is $n=10^5$.
  }\label{f6}
\end{figure}

In this section, we evaluate the finite sample performance of the proposed subsampling method described in Algorithm \ref{al3} for estimating the functional Poisson regression in comparison with the uniform subsampling approach. We set the true functional coefficient $\beta(t)= \text{sin}(0.5\pi t)$. Denote $\psi(\cdot)= \text{exp}(\cdot)$ and $\lambda(x_i) = \psi(\int_{0}^1x_i(t)\beta(t)dt)$, then
we generated responses $y(x_i) \sim \text{Poisson}(\lambda(x_i))$
with the mean $\lambda(x_i)$. The simulation designs of the functional predictors $x_i(t)$ are the same as in Simulation I, except that  
 we consider the following three different scenarios to generate the basis coefficients $a_{ij}$,
\begin{itemize}
  \item \textbf{Scenario I.} The basis coefficient $a_{ij}$ are i.i.d from the standard normal distribution, namely, $a_{ij}\overset{iid}\sim N(0,1)$. \figref{f6} (a) and (d) show that the distribution of the expected value $\lambda(x_i)$ ranges from 0.6 to 1.5 and is approximately symmetric about 1. About 70\% of responses are equal to 0 or 1.
  \item \textbf{Scenario II.} We generate the basis coefficient $a_{ij}$ from the $t$ distribution with 4 degrees of freedom and the variance is 1, namely, $a_{ij}\overset{iid}\sim t_4(0.5,1)$. \figref{f6} (b) and (e) show that the $\lambda(x_i)$ varies from 0.6 to 1.5, and about 80\% of responses lie between 0-2.
  \item \textbf{Scenario III.} We generate the basis coefficient $a_{ij}$ from the uniform distribution between 0 and 4, namely, $a_{ij}\overset{iid}\sim U(0,4)$. \figref{f6} (c) and (f) show that the expected value $\lambda(x_i)$ ranges from 2 to 6 and the distribution of responses is more uniform than Scenario I and II.
\end{itemize}

\begin{figure}[htbp]
  \centering
  \subfigure[Scenario I]{\includegraphics[width=4.8cm]{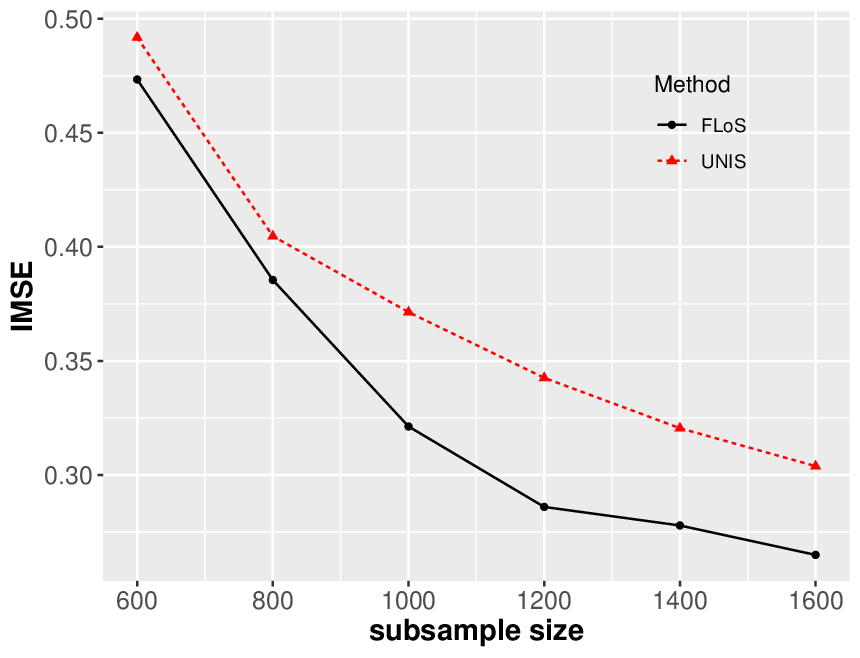}}
  \subfigure[Scenario II]{\includegraphics[width=4.8cm]{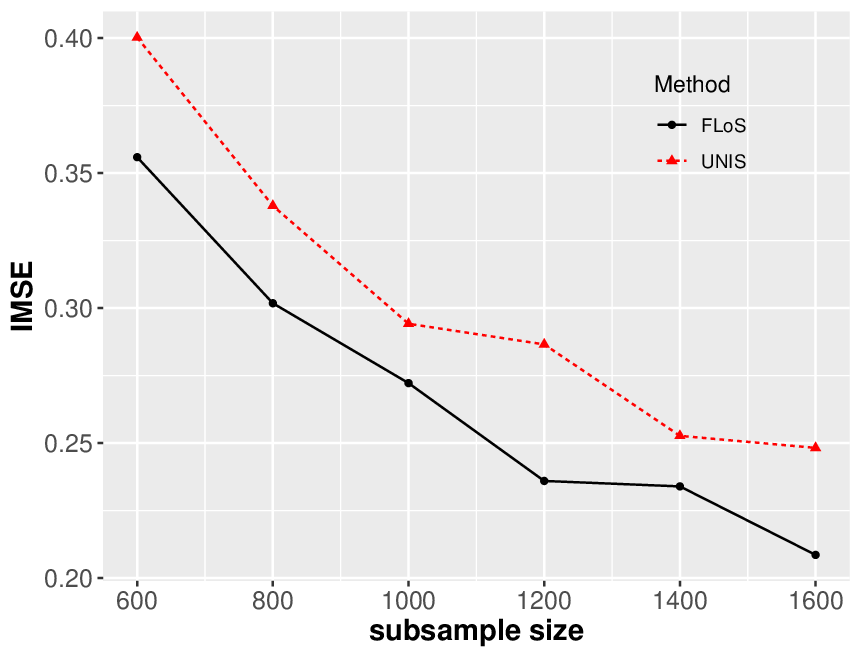}}
  \subfigure[Scenario III]{\includegraphics[width=4.8cm]{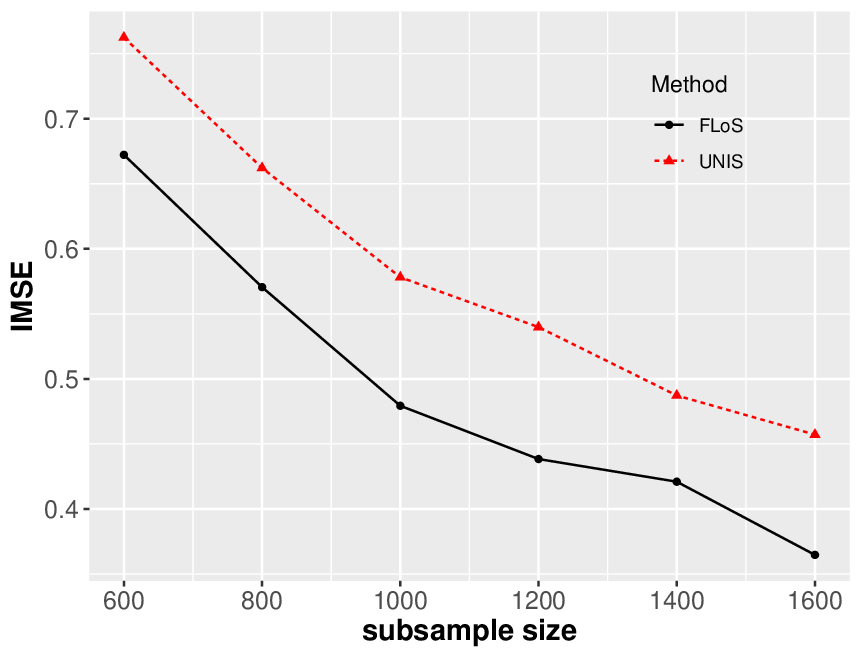}}\\
  \subfigure[Scenario I]{\includegraphics[width=4.8cm]{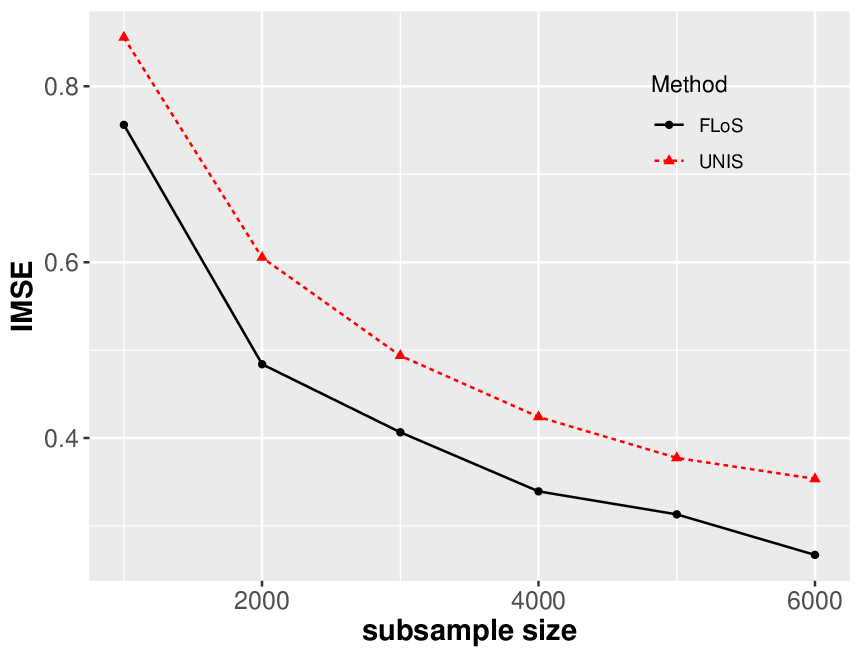}}
  \subfigure[Scenario II]{\includegraphics[width=4.8cm]{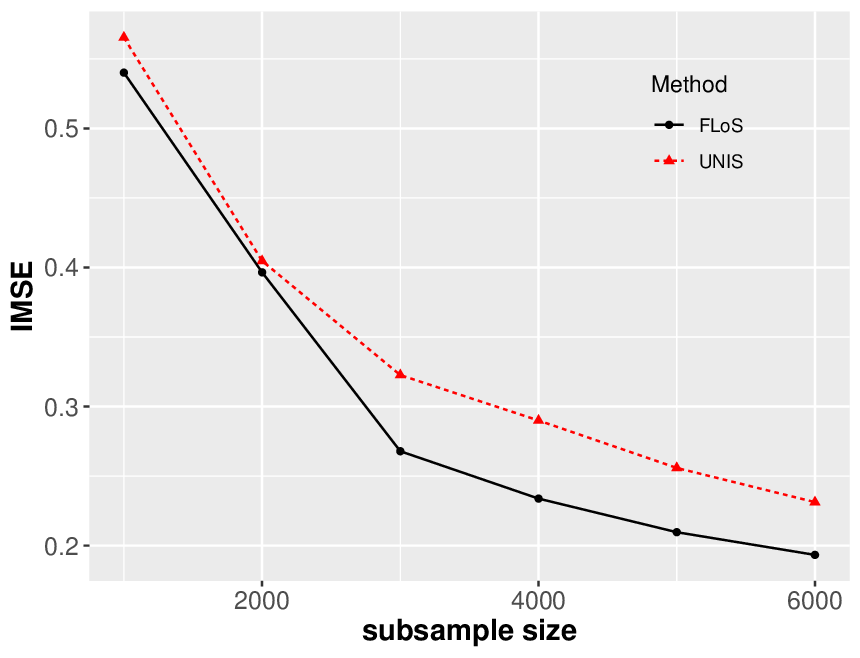}}
  \subfigure[Scenario III]{\includegraphics[width=4.8cm]{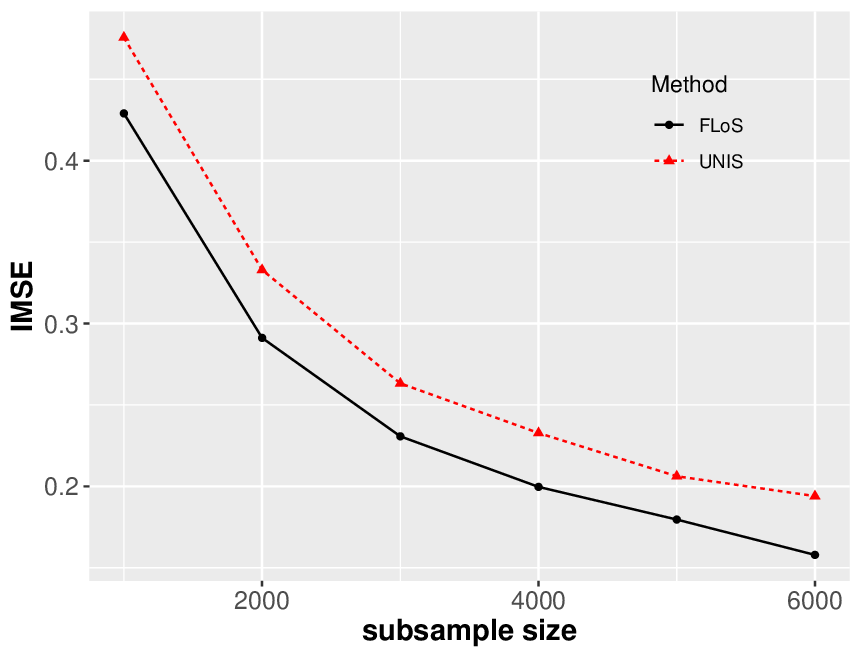}}\\
  \subfigure[Scenario I]{\includegraphics[width=4.8cm]{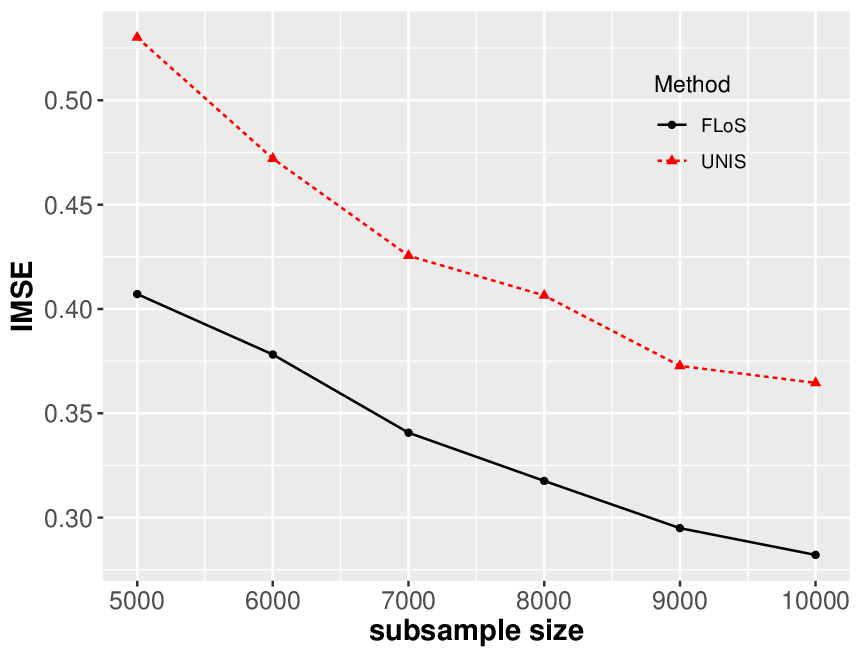}}
  \subfigure[Scenario II]{\includegraphics[width=4.8cm]{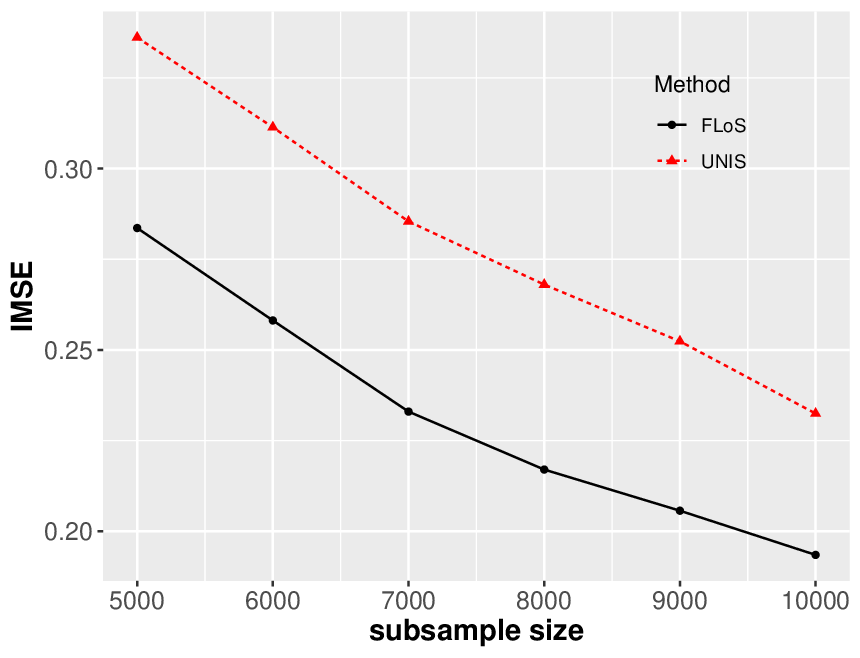}}
  \subfigure[Scenario III]{\includegraphics[width=4.8cm]{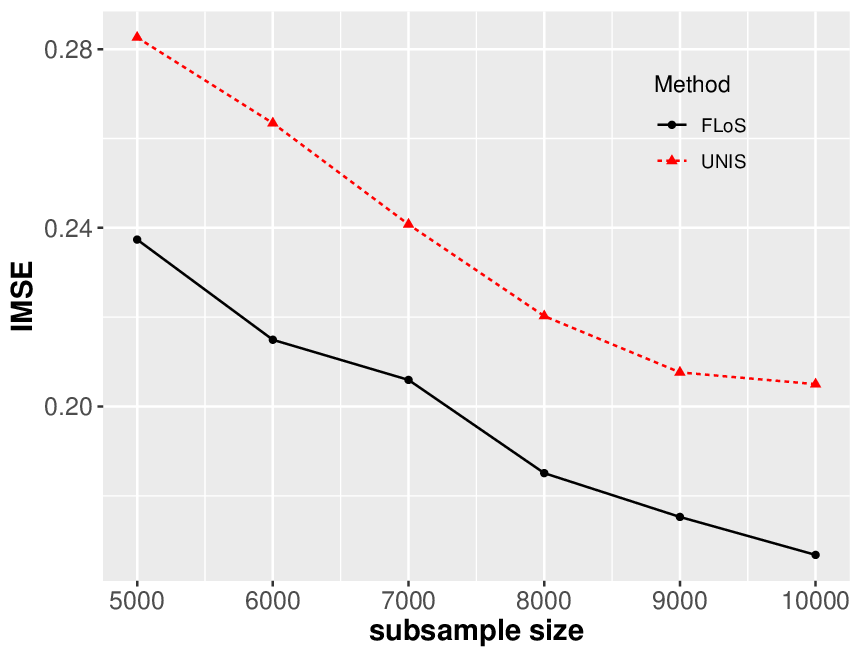}}\\
  \caption{The integrated mean squared error (IMSE) of the estimated functional coefficient $\widetilde{\beta}$ in the functional Poisson regression model by using the functional L-optimality subsampling (FLoS) method and the uniform subsampling (UNIS) approach under three scenarios with various subsample sizes $L$ when the full data size $n=10^5$ (Panels (a)-(c)), $10^6$ (Panels (d)-(f)), and $5\times 10^6$ (Panels (g)-(i)).}\label{f8}
\end{figure}

\figref{f8} displays the mean of IMSEs of the estimated functional coefficient $\widetilde{\beta}$ in the functional Poisson regression model when using the functional L-optimality subsampling method and the uniform subsampling approach under three scenarios when the full data size $n=10^5$, $10^6$, and $5\times 10^6$. \figref{f8} shows that the functional L-optimality subsampling method outperforms the uniform subsampling approach for all three scenarios and all full data sizes. This numerical results are consistent with our theoretical results that the functional L-optimality subsampling method aims to minimize the IMSE of $\widetilde{\beta}$ in approximating the estimator using the full data. Besides, when the full data size is fixed, the IMSEs of $\widetilde{\beta}$ using both methods become smaller as the subsample size $L$ increases.

\newpage
\section{Applications}
\label{sec::app}
In this section, we apply the proposed functional L-optimality subsampling method  to estimate the functional logistic regression model from the kidney transplant data set and to estimate the functional linear model from the global climate data set.

\subsection{Kidney Transplant Data}
The kidneys are a pair of organs in the human body, whose primary function is to remove waste from the body through the production of urine and regulate the chemical (electrolyte) composition of the blood.
Renal failure means that the kidneys can no longer remove wastes and maintain electrolyte balance, which will threaten a human's life. Renal failure can be divided into acute renal failure and chronic renal failure. Regarding the treatment of chronic renal failure, one method is the kidney transplant. A successful kidney transplant can restore normal renal function to the patients and extend their survival time.

After kidney transplantation, kidney transplant recipients still face a high probability of losing transplant function. It is also important to follow up the graft function and predict the patient's expected lifespan after a kidney transplant. 
Creatinine is the waste product of creatine, which the muscles use to make energy. Typically, creatinine travels from the blood to the kidneys where it leaves the body in the urine. A high level of creatinine in the blood indicates that the kidney is not working correctly.
On the other hand, only looking at how much creatinine in the blood is not the best way to check how well the kidneys are working, because the level of creatinine in blood is related to age, race, gender, and body size. In other words, what’s considered “normal” depends on these factors. The best way to know if kidneys are working properly is by looking at glomerular filtration rate (GFR), which considers the creatinine level and the associated factors simultaneously \citep{levey1999more,dongfunctional,keong2016decreasing}. For adults (Age$\geq19$), we use the Chronic Kidney Disease Epidemiology Collaboration (CKD-EPI, \cite{levey2009new}) equation to obtain the estimated glomerular filtration rate (eGFR, mL/min/1.73m$^2$).
For child (Age$\leq18$), we use the Schwartz formula \citep{schwartz1976simple,schwartz1987use,schwartz2009new} to estimate the glomerular filtration rate.



Our objective is to predict whether the kidney transplant recipients can survive over ten years based on their eGFR trajectories in the first six years after kidney transplant. The data resource used in this section is kidney transplant data from the Organ Procurement Transplant Network/United Network for Organ Sharing (Optn/UNOS) as of September 2020, which collect the basic description (e.g. age, race, gender, and height) of the kidney transplant recipients at the time of transplant and the information (e.g. serum creatinine, recipient status and the follow-up time) during the followed-up period. This data is available at
\url{https://optn.transplant.hrsa.gov/} with the permission of OPTN/UNOS.

\begin{figure}[htbp]
  \centering
  \subfigure{\includegraphics[width=10cm]{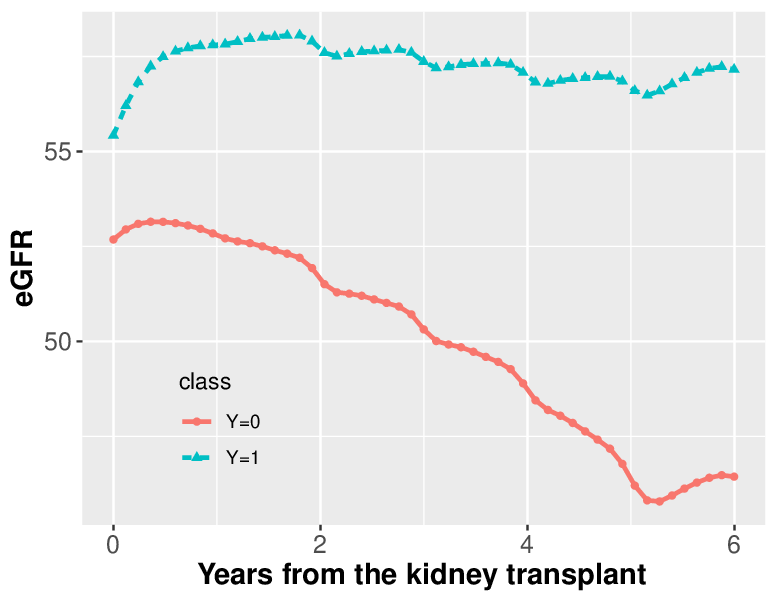}}
  \caption{
  The mean eGFR curves for the group of recipients who die or need to be re-transplanted during the sixth to tenth year after the transplant ($Y=0$) and the group of recipients who have lived for at least ten years after transplant ($Y=1$). 
  }\label{fl23}
\end{figure}

After matching data and deleting missing data, there are $n=130313$ recipients who have lived for at least six years after kidney transplant. We divide these recipients into two categories: the first category is the $30590$（23.3\%) recipients who die or need to be re-transplanted during the sixth to tenth year after the transplant ($Y=0$), and the other category is the $100713$（76.7\%) recipients who have lived for at least ten years after transplant ($Y=1$). \figref{fl23} display the mean eGFR trajectories for these two categories. It shows that the mean eGFR curve of $Y=1$ is higher than that of $Y=0$, which is consistent with the fact that a higher eGFR mean a better renal function. For those recipients who have not lived for ten years after transplant, the eGFR shows a significant downward trend. On the contrary, the eGFR curve remains stable for those recipients who have lived for ten years after transplant.
 
\begin{figure}[htbp]
  \centering
  \subfigure[]{\includegraphics[width=7cm]{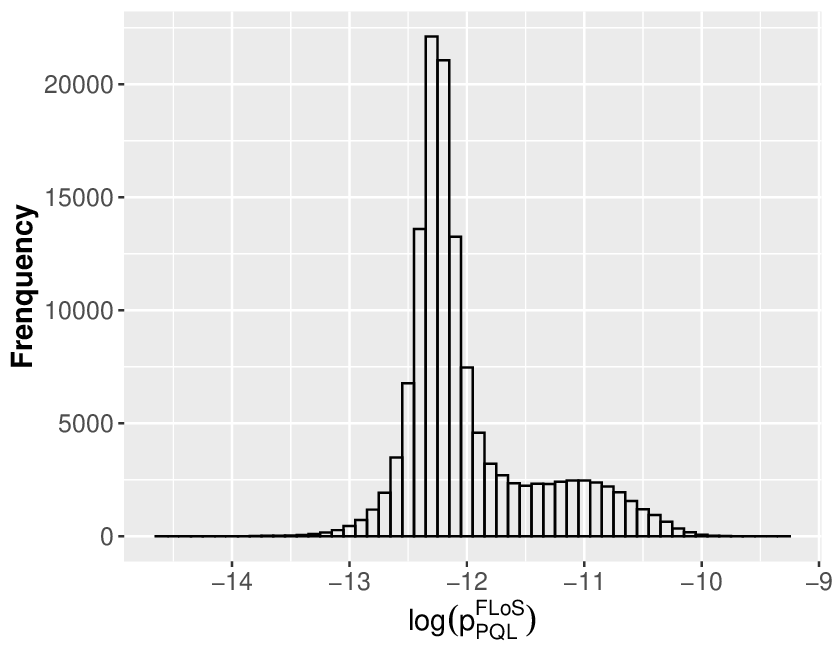}}
  \subfigure[]{\includegraphics[width=7cm]{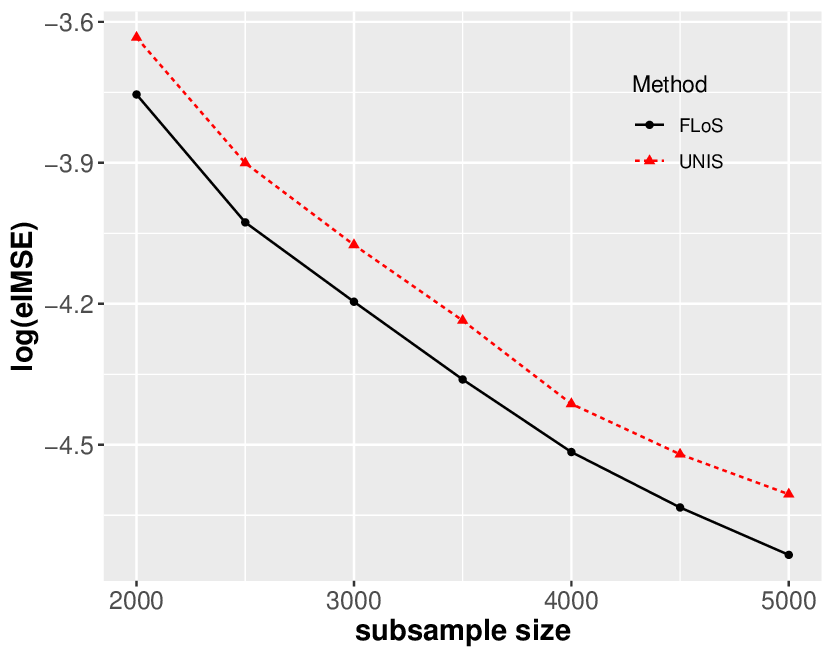}}
  \caption{(a) Histogram of the log of the subsampling probabilities $p_{\mathrm{PQL}}^{\mathrm{FLoS}}$ in the functional L-optimality subsampling method. (b) The logarithm of the empirical integrated mean square error (eIMSE) defined in (\ref{eIMSE}) for the estimated functional coefficient using the functional L-optimality subsampling (FLoS) method and the uniform subsampling (UNIS) approach with different subsample sizes.
  }\label{f24}
\end{figure}

We consider fitting a functional logistic regression model: \begin{equation}\label{GFRmodel}
    E(Y_i|\mathrm{eGFR}_i) = \psi\left(\alpha+\int_{0}^6\mathrm{eGFR}_i(t)\cdot \beta(t)dt\right).
\end{equation}
\figref{f24} (a) displays the histogram of the log of the subsampling probabilities $p^{\mathrm{FLoS}}$ in the 
the functional L-optimality subsampling method. It shows that the subsampling probabilities for different samples are very different. \figref{f24} (b) displays the logarithm of the empirical integrated mean square error (eIMSE) defined in (\ref{eIMSE}) for the estimated functional coefficient using both subsumpling methods. It indicates that the functional L-optimality subsampling method has smaller eIMSEs than the uniform subsampling approach for all subsample sizes.

\begin{figure}[htbp]
  \centering
  \includegraphics[width=9cm]{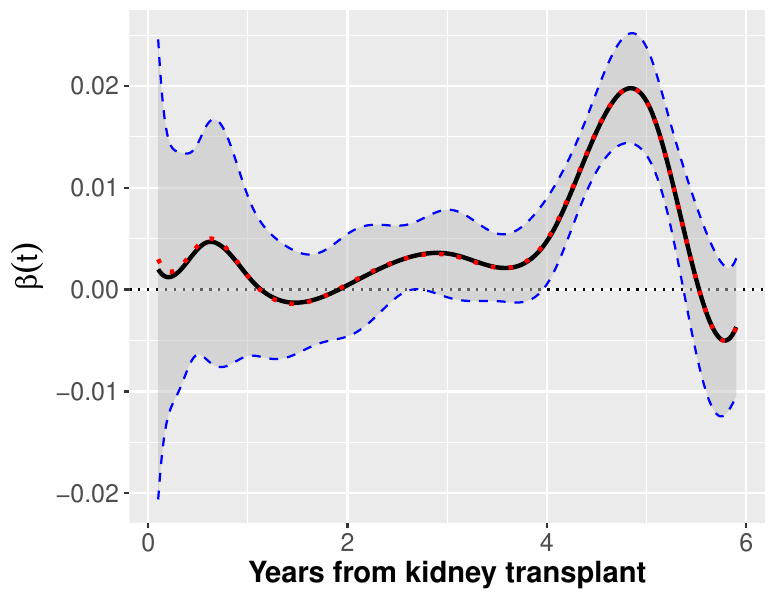}\\
  \caption{The red dotted line is the estimated $\beta(t)$ based on the full fata for predicting whether the recipient can live for at least 10 years after transplant based on the eGFR information in about the first six years. The black solid curve is the averaged estimated $\beta(t)$ using the FLoS method 
  based on 1000 subsampling datasets with subsample size $L=5000$.
  The blue dashed lines are 95\% point-wise confidence limits on the curve based on 1000 subsampling datasets with the subsample size $L=5000$.}\label{f25}
\end{figure}

\figref{f25} displays the estimated functional coefficient for the functional logistic regression model (\ref{GFRmodel}) by using the full data and using $L=5000$ data subsampled with the functional L-optimality subsampling method. The two estimated functional coefficients are almost identifical.
\figref{f25} also provides the corresponding 95\% point-wise confidence interval for the functional coefficient based on 1000 subsampling datasets with the subsample size $L=5000$ by using the functional L-optimality subsampling method. It shows that only the functional coefficientis significantly non-zero only from
the fourth year after transplant. Therefore, the information of eGFR during the 4th to the 5.5th year is more helpful to predict whether a recipient can live beyond ten years.

\subsection{Global Climate Data}
In recent years, climate change has created enormous challenges and costs for societies worldwide. 
For example, climate change is considered very likely to have contributed to the unprecedented extent and severity of the 2019–20 Australian bushfires. Thus, climate change is a global issue that should be addressed.

Rising temperature is the most obvious feature of climate change. According to the Intergovernmental Panel on Climate Change’s (IPCC) fifth assessment report (\url{http://www.climatechange2013.org/images/report/WG1AR5_TS_FINAL.pdf}), it is extremely likely that human activities caused more than half of the observed increase in global average surface temperature from 1951 to 2010. From the National Oceanic and Atmospheric Administration's (NOAA) Global Climate Report - Annual 2020 (\url{https://www.ncdc.noaa.gov/sotc/global/202013}), we can know that (1) the month of December 2020 had a global land and ocean surface temperature departure of 0.78$^{\circ}$C above the 20th-century average—this was the smallest monthly temperature departure during 2020; (2) the month of December 2020 was the eighth warmest December on record; (3) with a slightly cooler end to the year, the year 2020 secured the rank of second warmest year in the 141-year record, with a global land and ocean surface temperature departure from average of $+0.98^{\circ}$C. Besides, Global warming increases the severity of extreme rainfall and snowfall almost everywhere. A warmer world will increase soil evaporation and reduce the snow pack, exacerbating droughts even in the absence of reduced precipitation.

In this section, we use the global climate data set to analyze the relationship of temperature and precipitation in three distinct years: 1950, 2020, and 2100.
RCP4.5 is a pathway labeled after a possible range of radiative forcing values at the end of the 21st century relative to pre-industrial values (+4.5 W per square meter), in which emissions peak in 2040. Because RCP4.5 is a more moderate scenario than RCP8.5 and RCP2.6, we choose to use the global climate data based on RCP4.5 to analyze.

\begin{figure}[htbp]
  \centering
  \subfigure[Year = 1950]{\includegraphics[width=7cm]{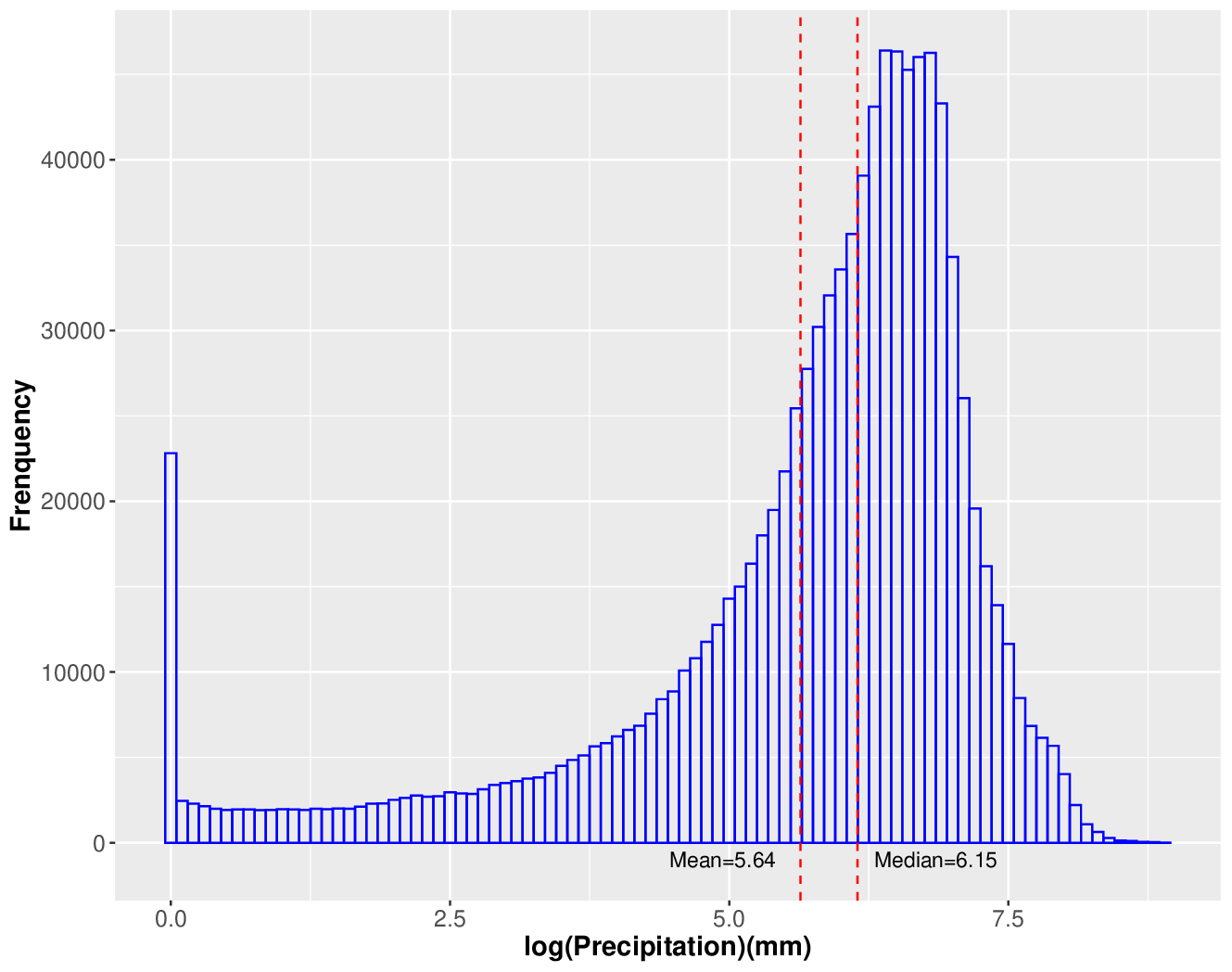}}
  \subfigure[Year = 2020]{\includegraphics[width=7cm]{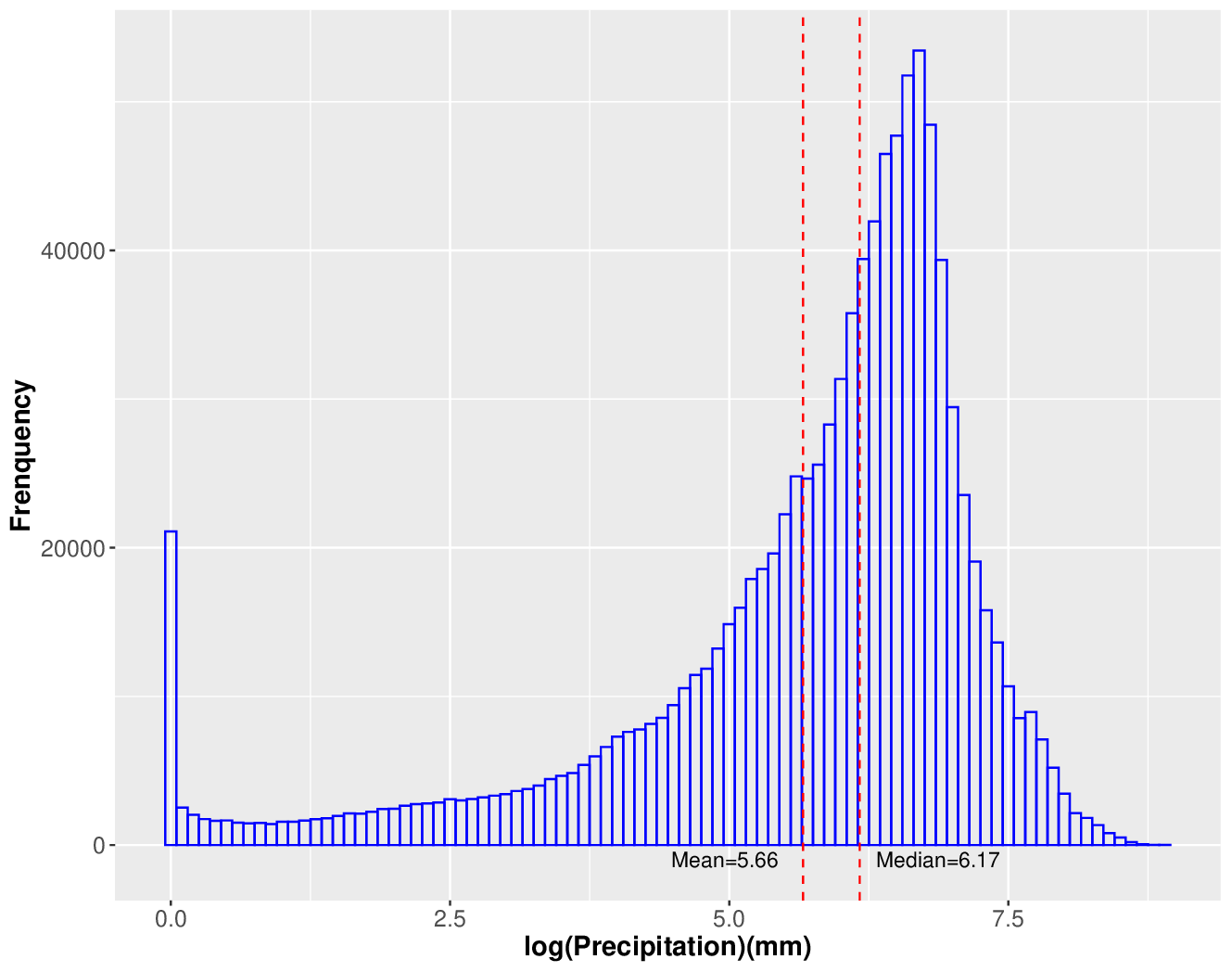}}\\
  \subfigure[Year = 2100]{\includegraphics[width=7cm]{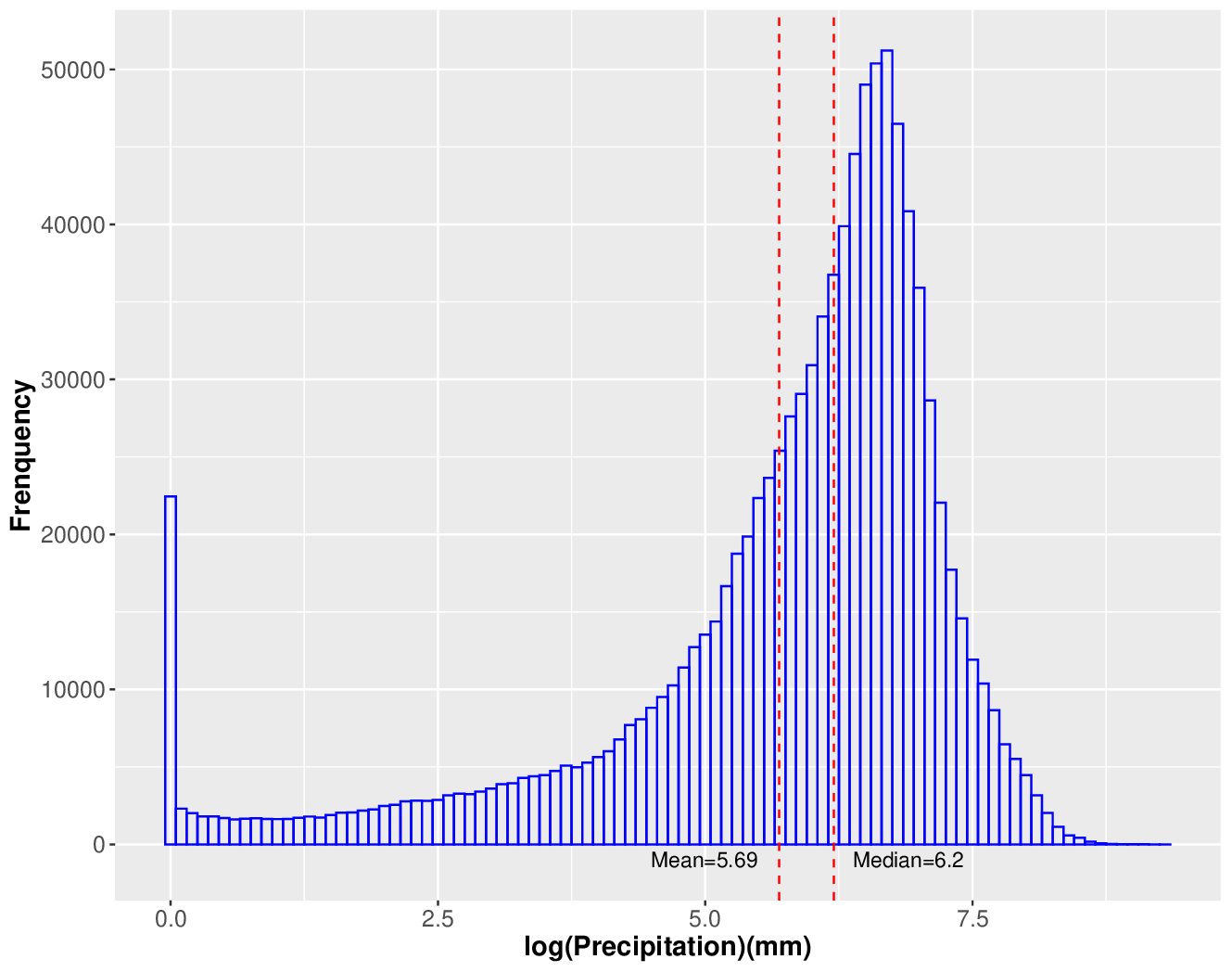}}
  \subfigure[Global daily average temperature]{\includegraphics[width=7cm]{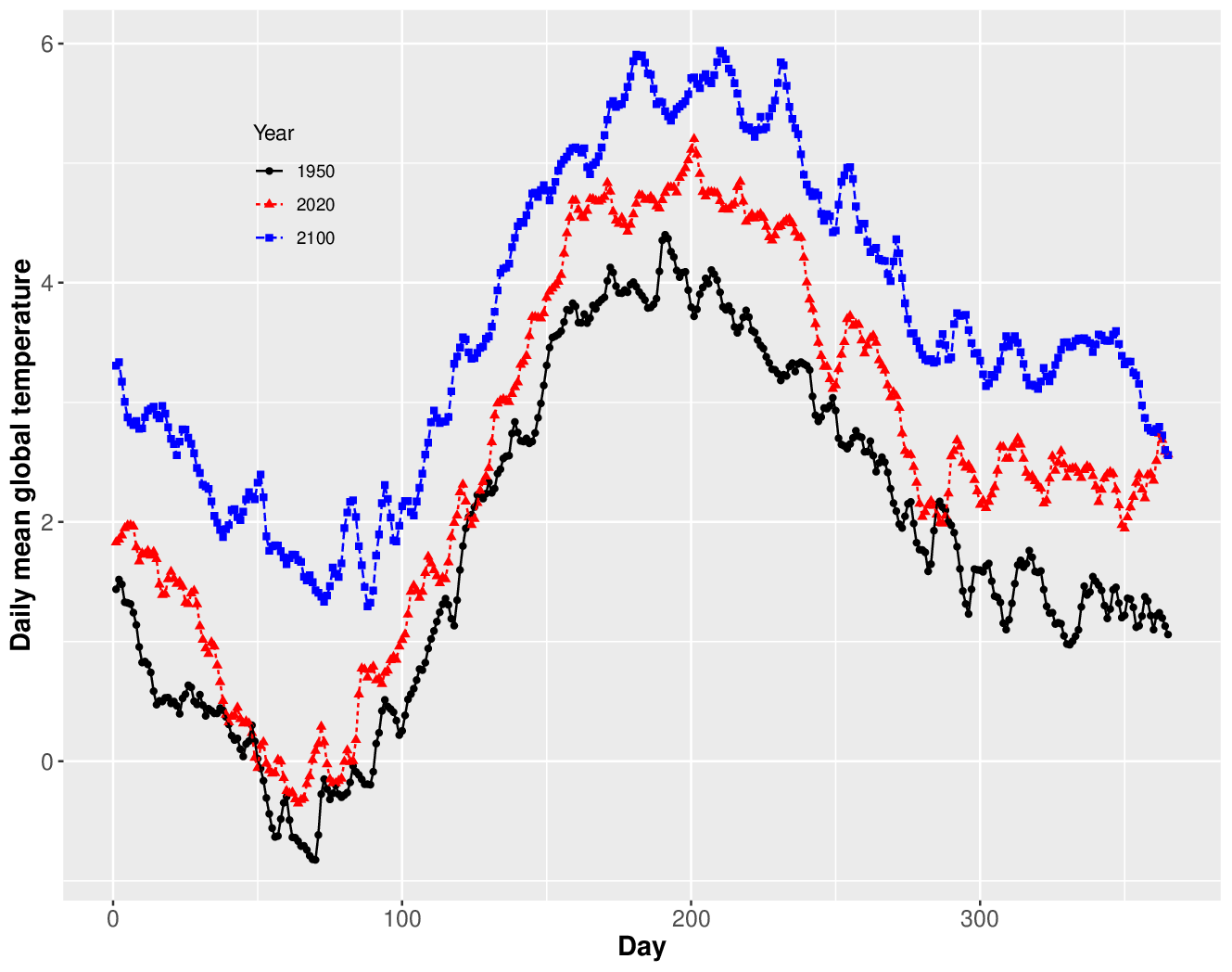}}\\
  \caption{The histogram of the log annual precipitation (mm) in three distinct years: 1950, 2020 and 2100, and the daily mean global temperature ($^{\circ}$C) in the three years. The two vertical lines in Panels (a)-(c) indicate the mean and median of the log annual precipitation (mm) in the globe.}\label{f20}
\end{figure}

The precipitation and temperature data under RCP4.5 are from the NASA Earth Exchange Global Daily Downscaled Projections (NEX-GDDP) data set (\url{https://ds.nccs.nasa.gov/thredds/catalog/NEX-GDDP/IND/BCSD/catalog.html}). In this data set, the globe is divided into $1,036,800$ grids of 0.25 degrees x 0.25 degrees using the Bias-Correction Spatial Disaggregation (BCSD). After deleting missing data, the full data size is $n = 1,028,032$. 
\figref{f20} (a)-(c) display the histograms of the log annual precipitation in 1950, 2020, and 2100, which show that there is no obvious difference in the mean and median precipitation in these three distinct years. \figref{f20} (d) shows that the last century witnessed an increase in daily mean global temperature.

\begin{figure}[htbp]
  \centering
  \subfigure[]
  [The heatmap of 
$\|\bm{N}_i\|_2$
  ]
  {\includegraphics[
  width = 0.475\textwidth,
  height=0.4\textheight
  ]{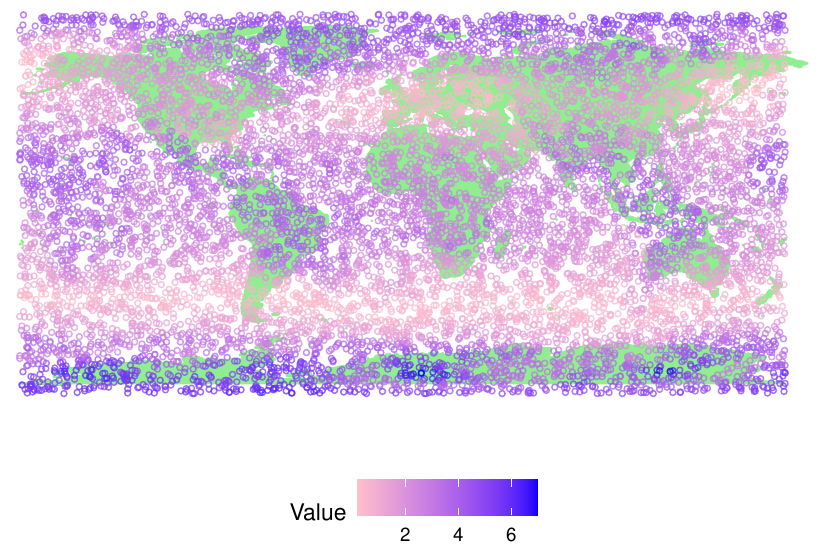}}\hfill
  \subfigure[]
  [The heatmap of the fitted residual $|y_i-\bm{N}_i^{T}\widehat{\bm{c}}|$]
  {\includegraphics[
width = 0.475\textwidth,
height=0.4\textheight]{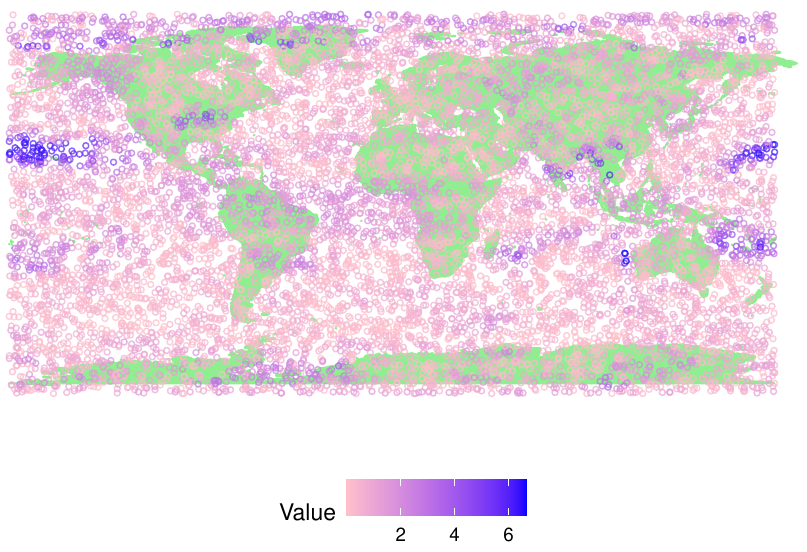}}
  \subfigure
  [Functional L-optimality Subsampling]
  {\includegraphics[
   width= 0.475\textwidth,
   height=0.3\textheight
  ]{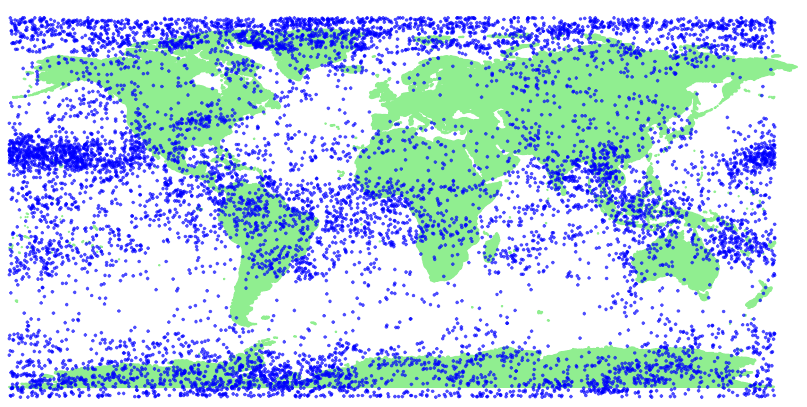}}
  \subfigure
  [Uniform Subsampling]
  {\includegraphics[
   width = 0.475\textwidth,
   height=0.3\textheight
  ]{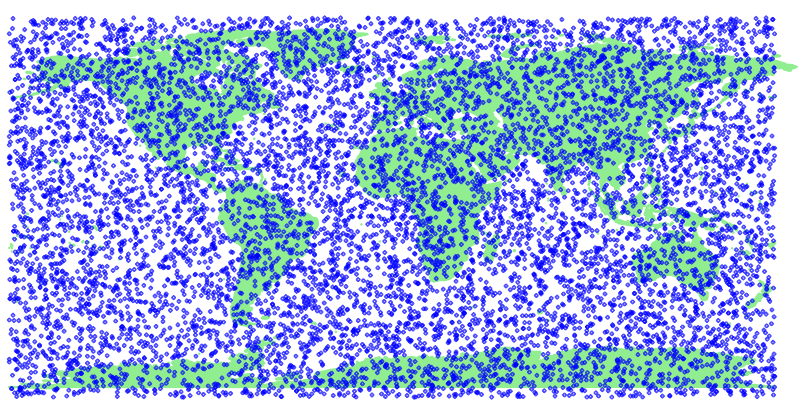}}
  \caption{Panels (a) and (b) show the heatmap of the terms related to
  the optimal subsampling probabilities (\ref{eq12}) in the functional L-optimality subsampling method. Panels (c) and (d) display the selected samples in the year of 2020 using the functional L-optimality subsampling method and the uniform subsampling method, respectively. }\label{f23}
\end{figure}

\begin{figure}[htbp]
  \centering
  \subfigure[Year = 1950]{\includegraphics[width=4.8cm]{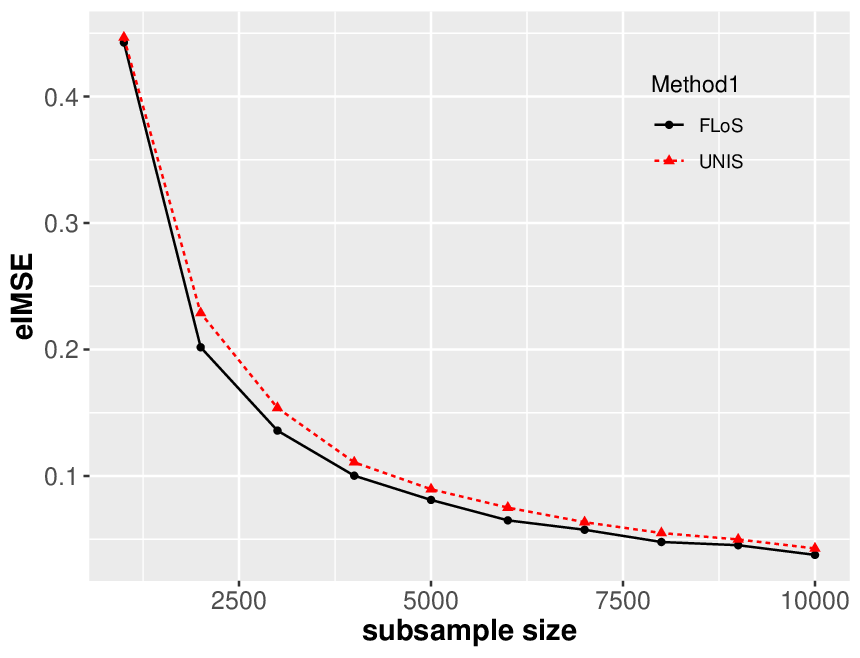}}
  \subfigure[Year = 2020]{\includegraphics[width=4.8cm]{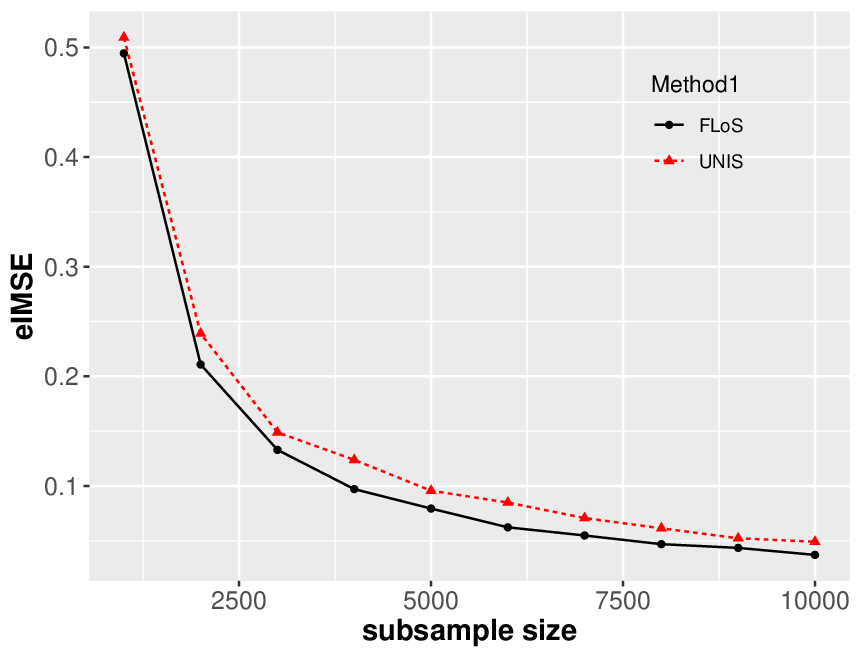}}
  \subfigure[Year = 2100]{\includegraphics[width=4.8cm]{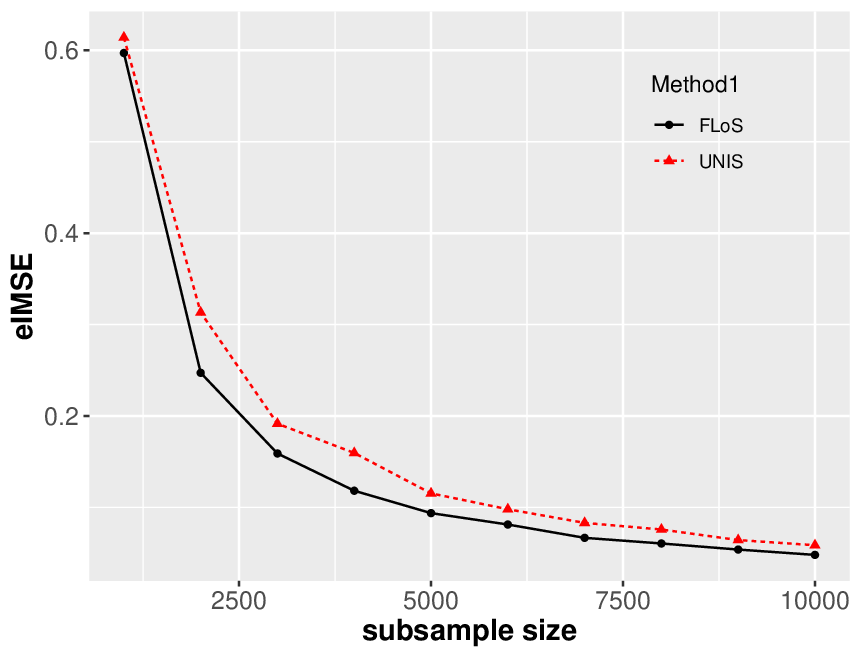}}\\
  \caption{The empirical integrated mean square error (eIMSE) of the estimated functional coefficient from the global climate data set in three distinct years 1950, 2020 and 2100 when using the functional L-optimality subsampling (FLoS) method and the uniform subsampling (UNIS) approach. }\label{f21}
\end{figure}
Our objective is to estimate the effect of the daily average temperature on the log annual precipitation in three distinct years: 1950, 2020 and 2100 and to study whether the temperature effect changes over 150 years.  For each year, we use our proposed functional L-optimality subsampling method and the uniform subsampling approach to estimate the functional linear model:
\begin{equation}\label{pre}
    log(\mathrm{Preciptation}_i) = \alpha+\int_{1}^{365} \mathrm{Temp}_i(t)\cdot \beta(t)dt +\varepsilon_i, \quad i=1,2,\ldots, n,
\end{equation}
where the functional coefficient $\beta(t)$ represents the cumulative effect of the daily temperature on the log annual precipitation. Because we do not know the true functional coefficient, we adopt the empirical integrated mean square error (eIMSE) as the criterion for comparing two subsampling methods, which is defined as 
\begin{equation}\label{eIMSE}
    \mathrm{eIMSE} = \frac{1}{S}\sum_{s=1}^S \int (\widetilde{\beta}^{(s)}(t)-\widehat{\beta}(t))^2dt,
\end{equation}
where $\widetilde{\beta}^{(s)}(t)$ is the estimated functional coefficient using the s-th subsample data set, and $\widehat{\beta}(t)$
is the estimator using the full data.

\figref{f23} (a) and (b) show the effects of the two 
terms 
related to the optimal subsampling probabilities (\ref{eq12}) in the functional L-optimality subsampling method. The heatmap of 
$\|\bm{N}_i\|_2$
has large values in Arctic, Antarctica and the area around the equator. On the other hand, the heatmap of the fitted residual $|y_i-\bm{N}_i^{T}\widehat{\bm{c}}|$ only has large values around the equator in the Atlastic Ocean and the Indian Ocean.
\figref{f23} (c) and (d) display the selected samples with two subsampling methods from the full data in 2020. It shows that the samples with the functional L-optimality subsampling method are concentrated in Arctic, Antarctica and the area around the equator.
\figref{f21} displays the eIMSEs of the estimated functional coefficient when using the two subsampling methods.
It shows that the functional L-optimality subsampling method is better than the uniform subsampling approach in all three distinct years.



\begin{figure}[htbp]
  \centering
  \includegraphics[width=14cm]{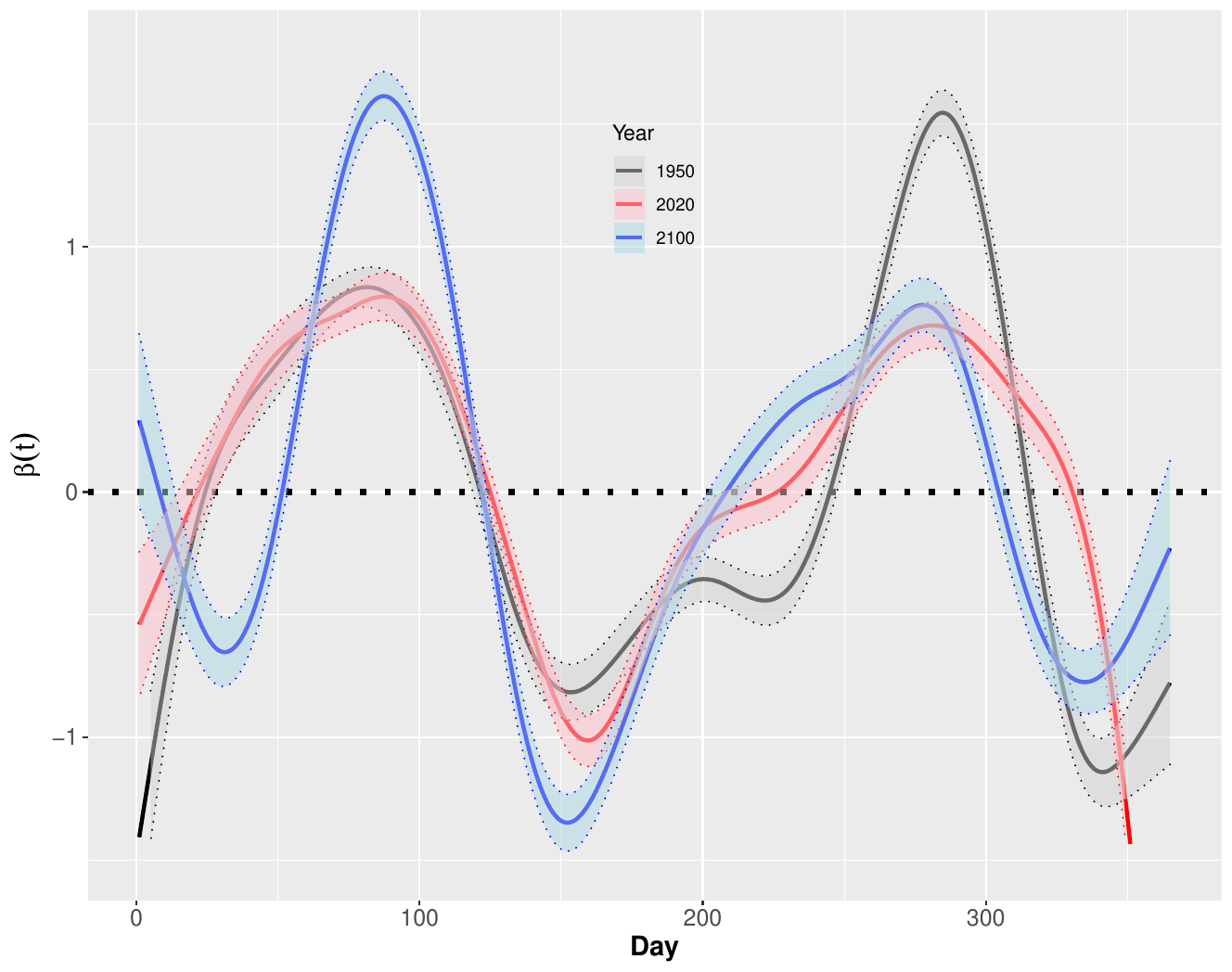}\\
  \caption{The average of the estimated functional coefficient from the global climate data set in three distinct years 1950, 2020 and 2100 using the functional L-optimality subsampling method based on 1000 subsampling datasets with the subsample size $L=10^4$. The shaded areas are the corresponding 95\% point-wise confidence intervals based on 1000 subsampling datasets.}\label{f22}
\end{figure}

\figref{f22} displays the average of the estimated functional coefficient using the functional L-optimality subsampling method based on 1000 subsampling datasets with the subsample size $L=10000$. It shows that the effect of daily temperature on annual precipitation is very different in three distinct years: 1950, 2020 and 2100. In 1950, there is a strong peak in the late fall. In 2020, there are two similar peaks in the late spring and late fall. In 2100, the functional coefficient peaks in the late spring and becomes very negative in May, which may be interpreted as that the contrast between spring and summer temperatures would have a larger effect on the annual precipitation in 2100 than in 1950 and 2020. \figref{f22} also displays the corresponding 95\% point-wise confidence intervals for the estimated functional coefficient, which indicates that 
the daily temperature in almost the whole year has a significant impact on the annual precipitation.



\section{Conclusions and Discussion}
\label{sec::discuss}
We propose the functional L-optimality subsampling method for estimating the functional linear model and the functional generalized linear model to tackle the challenges brought from the extraordinary amount of functional data. 
The asymptotic results of the subsample estimators have also been established.
Several simulation 
studies show that our proposed method is computationally feasible and outperforms the uniform subsampling method for massive data. The proposed subsampling methods are also demonstrated by analyzing the kidney transplant data and the global climate data.
For the kidney transplant data, we find that
the eGFR trajectories during the 4th to the 5.5th year has a significant effect on a recipient's lifespan. The subsample estimators can well approximate the results obtained from the full data. The analysis of the global climate data shows the selected data by the proposed FLoS method is more concentrated in Arctic, Antarctica and the area around the equator. In addition, we also find that the effect of daily temperature on annual precipitation in Year 2100 have very different patterns from Year 1950 and 2020.

In this paper, we consider the subsampling for the scalar on function regressions. There are other functional regressions, such as, function on scalar regressions \citep{zhu2012multivariate,luo2016,Li2017} and function on function regressions \citep{sun2018optimal,cai2021}. For these two types of regressions, how to subsample is still an open problem.
Besides, massive functional data often presents heterogeneity \citep{delaigle2015nonparametric,delaigle2019clustering}
and sometimes part of the massive functional data may be incomplete\citep{delaigle2020estimating}. 
We may efficiently tackle these issues by virtue of the subsampling idea. We will pursue these problems in our future research.


\section*{Acknowledgments}
The kidney transplant data set was supported in part by Health Resources and Services Administration contract 234-2005-370011C. The content about this data set is the responsibility of the authors alone and does not necessarily reflect the views or policies of the Department of Health and Human Services, nor does mention of trade names, commercial products, or organizations imply endorsement by the U.S. Government.

\bibliographystyle{apalike}      
\bibliography{bibfile,References}   

\CJKindent
\end{CJK}

\end{document}